\documentclass[aps,prd,a4paper,onecolumn,amsmath,showpacs,superscriptaddress,nofootinbib,preprintnumbers,notitlepage]{scrartcl}

\usepackage{amsmath,amssymb,latexsym,mathrsfs}
\usepackage{graphicx}
\usepackage{xcolor}
\usepackage{hyperref}
\usepackage{enumitem,booktabs,cfr-lm}
\usepackage[referable]{threeparttablex}
\usepackage{slashed}
\usepackage{bm}
\renewlist{tablenotes}{enumerate}{1}
\makeatletter
\setlist[tablenotes]{label=\tnote{\alph*},ref=\alph*,itemsep=\z@,topsep=\z@skip,partopsep=\z@skip,parsep=\z@,itemindent=\z@,labelindent=\tabcolsep,labelsep=.2em,leftmargin=*,align=left,before={\footnotesize}}
\makeatother

\DeclareMathOperator{\Tr}{Tr}
\DeclareMathOperator*{\argmax}{arg\,max}
\DeclareOldFontCommand{\bf}{\normalfont\bfseries}{\textbf}
\DeclareOldFontCommand{\it}{\normalfont\itshape}{\textit}

\newcommand{\btheta}{\bm{\theta}}
\newcommand{\bd}{\bm{d}}

\newcommand{\der}{\mathrm{d}}
\newcommand{\Ylm}{Y_{\ell m}}
\newcommand{\alm}{a_{\ell m}}
\newcommand{\Cl}{C_\ell}
\newcommand{\ClTT}{\Cl^{TT}}
\newcommand{\ClTE}{\Cl^{TE}}
\newcommand{\ClEE}{\Cl^{EE}}
\newcommand{\ClBB}{\Cl^{BB}}
\newcommand{\ClTB}{\Cl^{TB}}
\newcommand{\ClEB}{\Cl^{EB}}

\newcommand{\fsky}{f_\mathrm{sky}}
\newcommand{\MeV}{\mathrm{MeV}}
\newcommand{\eV}{\mathrm{eV}}

\begin{document}

\title{Likelihood methods for CMB experiments}

\author{Martina Gerbino\thanks{Istituto Nazionale di Fisica Nucleare (INFN), Sezione di Ferrara, Via Giuseppe Saragat 1, I-44122 Ferrara, Italy; HEP Division, Argonne National Laboratory, Lemont, IL 60439, USA}\and Massimiliano Lattanzi\thanks{Istituto Nazionale di Fisica Nucleare (INFN), Sezione di Ferrara, Via Giuseppe Saragat 1, I-44122 Ferrara, Italy} \and Marina Migliaccio\thanks{Dipartimento di Fisica, Universit\`a di Roma Tor Vergata, Via della Ricerca Scientifica 1, 00133, Roma, Italy and Istituto Nazionale di Fisica Nucleare (INFN), Sezione di Roma 2, Via della Ricerca Scientifica 1, 00133, Roma, Italy} \and Luca Pagano\thanks{Dipartimento di Fisica e Scienze della Terra, Universit\`a degli Studi di Ferrara, via Giuseppe Saragat 1, I-44122 Ferrara, Italy; Istituto Nazionale di Fisica Nucleare (INFN), Sezione di Ferrara, Via Giuseppe Saragat 1, I-44122 Ferrara, Italy} \and Laura Salvati\thanks{IFPU - Institute for Fundamental Physics of the Universe, Via Beirut 2, 34014 Trieste, Italy; INAF - Osservatorio Astronomico di Trieste, via G. B. Tiepolo 11, I-34143 Trieste, Italy; Institut d'Astrophysique Spatiale, CNRS (UMR 8617) Universit\'e Paris-Sud, B\^atiment 121, Orsay, France} \and Loris Colombo\thanks{Dipartimento di Fisica, Universit\`a degli Studi di Milano, Via Celoria, 16, Milano, Italy} \and Alessandro Gruppuso\thanks{INAF-OAS Bologna, Osservatorio di Astrofisica e Scienza dello Spazio di Bologna, Istituto Nazionale di Astrofisica, via Gobetti 101, I-40129 Bologna, Italy; Istituto Nazionale di Fisica Nucleare (INFN), Sezione di Bologna, Viale Berti-Pichat 9/2, 40127 Bologna, Italy} \and Paolo Natoli\thanks{Dipartimento di Fisica e Scienze della Terra, Universit\`a degli Studi di Ferrara, via Giuseppe Saragat 1, I-44122 Ferrara, Italy, and Istituto Nazionale di Fisica Nucleare (INFN), Sezione di Ferrara, Via Giuseppe Saragat 1, I-44122 Ferrara, Italy} \and Gianluca Polenta\thanks{Space Science Data Center - Agenzia Spaziale Italiana, Via del Politecnico snc, 00133, Roma, Italy}}

\maketitle

\begin{abstract}
A great deal of experimental effort is currently being devoted to the precise measurements of the cosmic microwave background (CMB) sky in temperature and polarisation. Satellites, balloon-borne, and ground-based experiments scrutinize the CMB sky at multiple scales, and therefore enable to investigate not only the evolution of the early Universe, but also its late-time physics with unprecedented accuracy. The pipeline leading from time ordered data as collected by the instrument to the final product is highly structured. Moreover, it has also to provide accurate estimates of statistical and systematic uncertainties connected to the specific experiment.
In this paper, we review likelihood approaches targeted to the analysis of the CMB signal at different scales, and to the estimation of key cosmological parameters. We consider methods that analyze the data in the spatial (i.e., pixel-based) or harmonic domain. We highlight the most relevant aspects of each approach and compare their performance. 
\end{abstract}

\tableofcontents


\newpage
\section{Introduction}
After 71 years from its first predictions, and after 55 years from its first observational evidences, the cosmic microwave background (CMB) is nowadays one of the most important probes in cosmology. During the past decades, theoretical efforts have elucidated the physics leading to the pattern of anisotropies in temperature and polarisation (see e.g.~\cite{Hu:2001bc} for a review on CMB physics and~\cite{Kolb:1990vq} for an exhaustive review on early Universe physics). Quantum fluctuations in the early universe generate metric perturbations. Scalar perturbations are converted into matter perturbations and radiation anisotropies that evolve in the expanding universe according to a set of coupled Einstein, Boltzmann and fluid equations. Matter perturbations eventually grow into galaxies and galaxy clusters. Primary CMB anisotropies are frozen at the time of matter-radiation decoupling, and subsequently modified during the propagation through evolving structures from the last scattering surface to the observer. Scattering between free electrons and CMB photons in two distinct epochs (recombination and reionization) further enriches the CMB structure with the addition of a polarisation ``curl-free'' ($E$-mode) pattern in the CMB radiation. Gravitational lensing of the CMB due to the propagation of CMB photons throughout large-scale structures (LSS) generate a polarised ``divergence-free'' ($B$-mode) patter from the distortion of the CMB $E$-modes. Perhaps more elusive, though of paramount importance, is the primordial $B$-mode signal sourced by tensor perturbations to the metric (gravitational waves). 

On the other hand, observational efforts have progressively lead the field to the current stage of precision cosmology. Observations of the CMB sky from space missions~\cite{Aghanim:2018eyx,WMAP,Bennett:1996ce}, balloon-borne experiments~\cite{spider,2018ApJS..239....7T,deBernardis:2000sbo}, and from ground-based telescopes~\cite{Louis:2016ahn,Henning:2017nuy,Ade:2018gkx,Ade:2017uvt,Essinger-Hileman:2014pja} provided measurements of CMB anisotropies in temperature and polarisation over a wide range of angular scales. While we are writing this review, the Planck collaboration~\cite{Akrami:2018vks} is preparing the final public release of the Planck legacy products, which will likely represent the state of the art of CMB measurements from a single experiment for the next decade and more. Current observations~\cite{Aghanim:2018eyx} are in agreement with the standard cosmological model of a homogeneous and isotropic Universe at large scales, based on General Relativity and on the standard model (SM) of particle physics, complemented with a mechanism for the generation of primordial perturbations, i.e., the inflationary paradigm. When interpreted in this $\Lambda\mathrm{CDM}$ framework, cosmological data point to a spatially flat Universe composed by baryons ($\Omega_b h^2=0.02237\pm0.00015$, $\sim5\%$ of the total density), dark matter ($\Omega_c h^2=0.1200\pm0.0012$, $\sim25\%$), and dark energy ($\Omega_\Lambda=0.6847\pm0.0073$, $\sim70\%$), a component that behaves like a cosmological constant, and is responsible for the present accelerated expansion, plus photons (a few parts in $10^5$) and light neutrinos. Further advances in CMB observations are still to come. Planned upgrades of existing ground-based and balloon experiments are ongoing~\cite{Henderson:2015nzj,Benson:2014qhw,SA,Bergman:2018rar}. The next generation of CMB observatories is under construction and is paving the way to the ``stage IV'' experiments targeting the ultimate measurements of the CMB polarisation field~\cite{S4,Ade:2018sbj,Matsumura:2013aja,Delabrouille:2017rct,Hanany:2019lle}.

The long run that lead from the pivotal observations of Penzias and Wilson~\cite{Penzias:1965wn} to the Planck legacy release has seen the dramatic improvement of the sensitivity to key cosmological parameters. Planck 2018 data provides sub-percent constraints on the base-$\Lambda$CDM parameters\footnote{The base $\Lambda$CDM parameters are: the angular size of the sound horizon at recombination $\theta_*$, the amplitude $A_S$ and tilt $n_s$ of the spectrum of primordial scalar perturbations, the reionization optical depth $\tau$, the energy density in baryonic matter $\Omega_b h^2$ and in cold dark matter $\Omega_c h^2$.}~\cite{Aghanim:2018eyx}.
Moreover, advances in experimental cosmology over the past decades made cosmology itself a new avenue to the investigation of fundamental physics properties complementary to laboratory searches. A clear example is given by the possibility to constrain neutrino properties, such as their number $N_\mathrm{eff}$ and the sum of their masses $\sum m_\nu$. Indeed, the combination of Planck 2018 data and LSS information (in the form of measurements of baryon acoustic oscillations, BAO) can exclude at 95\% c.l. the presence of light thermal relics decoupling after the QCD phase transition ($T<100\,\MeV$) and provides a bound on the sum of the neutrino masses of $\sum m_\nu<0.12\,\eV$ at 95\% c.l.\footnote{Constraints derived in the context of minimal extensions of the standard $\Lambda$CDM model.}~\cite{Aghanim:2018eyx}. 

In this context, a key ingredient is the suitable choice of the likelihood function to compare observed data with theoretical predictions in order to constrain the model parameters. In the standard cosmological model of the early universe, primordial perturbations are Gaussian distributed, and so are CMB fluctuations. Therefore, all relevant physical information in the CMB field are contained in the variance of the distribution. This is the reason why the full-sky power spectra of CMB fluctuations are a sufficient statistics. The power spectra of observed data also provide an unbiased estimator of the ensemble averaged variance of the CMB fluctuations. In the simple case of full-sky observations and isotropic and mode-uncorrelated experimental noise, the likelihood function can be derived analytically. In particular, for correlated temperature and polarisation field, the probability of the data given the theoretical model (i.e., the likelihood $\mathcal{L}$) is given by a Wishart distribution.

However, this simple case does not capture the properties of realistic observations. Depending on the experimental platform (satellite, balloon, ground), each telescope has access to fractions of the sky $\fsky$ of different size. As an example, compare the almost full-sky observations of the Planck satellite~\cite{Akrami:2018vks} with the $\fsky\sim1\%$ sky coverage of the ground-based BICEP-2 experiment~\cite{Ade:2018gkx}. Even in the case of full-sky observations, only a certain fraction of the sky can be retained for cosmological analysis. Foreground emissions from astrophysical and galactic sources should be masked if particularly bright contaminants. In addition to limited access to the sky coverage, a particular choice of the observational (or scanning) strategies of the sky can break the assumption of isotropic noise, due to repeated visits to the same part of the sky. As an example, consider the Planck scanning strategy featuring a longer integration time in the proximity of the Ecliptic poles (i.e., at lower Galactic latitudes, where galactic foreground contaminations are smaller).

In general, complications to the simple case of full-sky and isotropic noise arising from realistic experimental conditions require a different likelihood analysis. First of all, specific estimators of the power spectra should be defined in the partial-sky regime, which take into account spurious correlations between fields induced by the incomplete sky coverage. Secondly, the use of a Wishart distribution as a likelihood function is no longer possible. Either the new estimators are no longer distributed according to a Wishart, and therefore this choice is not exact anymore. Or, the use of the exact likelihood is unfeasible as one moves to the analysis of smaller scales (larger multipoles) and higher-resolution maps, due to the huge computational cost of inverting large covariance matrices. At large scales and for low-enough angular resolutions, the exact likelihood in pixel space can still be adopted.

In all the above situations, approximate forms of the likelihood functions have been developed. At small scales, the central limit theorem allows to approximate the Wishart distribution as a Gaussian in the power spectra. In general, quadratic forms in some functions of the CMB spectra have been adopted as approximate likelihood functions, with various choices of the covariance matrix. To conclude this long introduction, it has to be stressed that the choice of the likelihood strongly depends on the characteristics of the experiment at large, i.e., on the observational strategy, on the range of scales probed, on the noise properties, etc.

The aim of this manuscript is to review the basics of likelihood analysis in CMB experiments. This goal is motivated by the fact the we are at a crucial point in the history of observational CMB cosmology. The level of maturity and complexity reached by current CMB experiments boosted the theoretical efforts in finding smart solutions to the issue of identifying a suitable likelihood choice. A rich literature has been produced in this sense, although an exhaustive overview of the topic is not available, to the best of our knowledge. This review would fill the gap. Such a review could also serve as a good starting point for those who are approaching the field of CMB data analysis today or in the next future, and would be ideally contributing to the advances of CMB science in the next decades.

The structure of the manuscript is as follows. Section~\ref{sec:CMBstat} is devoted to the statistics of the CMB fields. The approach to the topic is pedagogical, in a sense that we begin with a discussion in the single-field, temperature-only regime and introduce the basic statistical properties of CMB fluctuations. Then, we move to the more general case of correlated $T,E,B$ fields. The discussion is carried over in the full-sky regime, with no distinction made between applications to large- and small-scales. We conclude Sec.~\ref{sec:CMBstat} with the introduction of the exact likelihood in full sky. Specific approximations to the exact likelihood are presented in Sec.~\ref{sec:highl} (applications to the small-scale regime) and in Sec.~\ref{sec:lowl} (applications to the large-scale regime). In both Sections, attention is devoted to complications due to partial-sky coverage and noise contamination. The inclusion of physical late-time Universe effects on the CMB photons in terms of gravitational lensing is detailed in Sec.~\ref{sec:lensing}, whereas the important issues related to the presence of foreground emissions are described in Sec.~\ref{sec:foreground}. The discussion of the various likelihood approaches in terms of computational cost (where applicable) and robustness with respect to the ability to provide unbiased estimations of cosmological parameters is detailed in Sec.~\ref{sec:comparison}. Our conclusions are summarised in Sec.~\ref{sec:conclusion}. Some useful tools that will be mentioned throughout the main text are further discussed in Appendix. In particular, in Appendix~\ref{sec:basic}, we review the basic notions of statistics needed to develop the formalism of CMB statistics. In Appendix~\ref{sec:appendixB}, we discuss the construction of power spectrum estimators, including pseudo-$C_\ell$, the ``pure'' formalism, and quadratic maximum likelihood (QML) estimator.


\newpage
\section{Statistics of the Cosmic Microwave Background}\label{sec:CMBstat}
We now introduce some basic aspects of the statistics of the CMB. The basic object that we are interested in is the \emph{likelihood function} $\mathcal{L}$, i.e., the probability of the observed data $\mathbf d$ given a model, regarded as a function of the model itself. If the model is defined in terms of a vector of parameters $\btheta$, we thus have:
\begin{equation}
{\mathcal L}(\btheta) = p(\bd|\btheta) \, .
\end{equation}
The notation used throughout this review is presented in Appendix~\ref{sec:basic}, were we also recall some basic notion of probability and statistics.

We first derive the exact likelihood function for the CMB fields in \textit{harmonic} space. The main statistical concepts are introduced in the limit of single field (Sec.~\ref{sec:CMBstat_T}), i.e., temperature only, for the sake of simplicity. We then generalise these main findings in the case of joint temperature and polarization analysis (Sec.~\ref{sec:CMBstat_TEB}). The exact likelihood in \textit{real} space are derived in Sec.~\ref{sec:CMBstat_real}.

We assume an ideal scenario of full-sky observations with infinite angular resolution and absence of noise and foreground contaminations. Obviously, this scenario is highly idealized. Nevertheless, it allows to easily derive the basic concepts of CMB statistics. Modifications to this picture arising from realistic observational issues (limited sky coverage, masked sky, experimental noise, finite angular resolution) are introduced in Sec.~\ref{sec:highl_noise}. Foregrounds are briefly discussed in Sec.~\ref{sec:foreground}. We also assume that the temperature and polarization fluctuations are Gaussian, thus neglecting any non-Gaussianity, either of primordial origin (which are anyway bounded to be small~\cite{Akrami:2019izv}), or coming from unresolved systematics (e.g., foreground residuals).

A final remark concerns the physical, late-time-Universe effects on the CMB fields due to the propagation of CMB photons from the last-scattering surface to the observer  throughout the evolving large-scale structures. Weak gravitational lensing due to the gravitational potential of cosmological structures deflects CMB photons and modifies the observed statistics of CMB anisotropies with respect to the pattern arising at decoupling. In what follows, we will implicitly consider \textit{unlensed} CMB fields, i.e., we will ignore the effects of gravitational lensing for the sake of simplicity. The non-trivial modifications induced by the gravitational potential will be discussed later in Sec.~\ref{sec:lensing}.

\subsection{Statistics of CMB temperature field -- Exact likelihood in harmonic space}\label{sec:CMBstat_T}
The CMB temperature field $T(\vec{x},\hat{n},\tau)\!=\!\bar{T}(\tau)\left[1+\Theta(\vec{x},\hat{n},\tau)\right]$ 
observed\footnote{In this section, ``observed'' stands for what would be observed in the idealised situation considered here, i.e., negligible instrumental noise and absence of contaminants of any kind. A more proper term would be ``realised'', but in this context we avoid to use it as we reserve it for simulated CMB sky signals.} in a given direction $\hat n$ is defined at every point $(\vec x, \tau)$ in space and time. The field has been decomposed in an isotropic background value $\bar{T}(\tau)$ and a small perturbation, the anisotropy field $\Theta(\vec{x},\hat{n},\tau)=(T-\bar{T})/\bar{T}$. Anisotropies are assumed to be the result of a Gaussian random process originated from quantum fluctuations in the early Universe. The observed temperature field generated by scalar fluctuations is a linear operator acting on three-dimensional perturbation fields:
\begin{equation}
\Theta(\hat{n}) = \int d^3 \vec{k} \,\xi(\vec{k})\, \int_0^{\tau_0} d\tau\,e^{i (\hat{k}\cdot\hat{n})(\tau_0-\tau)} S_T^{(s)}(k,\,\tau)\,
\label{eq:linearop}
\end{equation}
where $\xi(\vec{k})$ is the primordial curvature perturbation, and the source function $S_T^{(s)}$ for scalar temperature fluctuations is a linear combination of the cosmological perturbation fields (see Ref. \cite{Zaldarriaga:1996xe} for the explicit expression). A similar expression holds for temperature fluctuations generated by tensor perturbations \cite{Zaldarriaga:1996xe}.  In Eq.~(\ref{eq:linearop}), we have suppressed the $\tau$ and $\vec{x}$ dependence in $\Theta$ as we are implicitly assuming that the temperature field is observed in a given position at a fixed point in time.

Assuming a given cosmological model, we cannot directly predict the particular realisation of the temperature field. Instead, we shall infer statistical properties of the observed perturbation field. It is useful, to this purpose, to decompose the angular dependence of the temperature anisotropy field in spherical harmonics $\Ylm(\hat{p})$

\begin{equation}\label{eq:Spherical harmonics decomposition}
\Theta(\vec{x},\hat{p},\tau)=\overset{\infty}{\underset{\ell=1}{\sum}}\overset{\ell}{\underset{m=-\ell}{\sum}}\alm(\vec{x},\tau)\Ylm(\hat{p}) 
\end{equation}
where the harmonic $\Ylm$ corresponds to an angular scale $\theta\sim\pi/\ell$ with $(2\ell+1)$ $m$-modes for each multipole $\ell$. Low multipoles (low-$\ell$) in the expansion correspond to large angular scales in the sky, whereas high multipoles (high-$\ell$) correspond to small scales. Since $\Theta$ is real, the decomposition coefficients $\alm$
have to satisfy the reality condition

\begin{equation}
\alm^{*}\!=\!a_{\ell\,-m}\label{eq:Reality condition of alm}
\end{equation}

\noindent All the information about the $\vec{x}$ and $\tau$ dependence of
$\Theta$ is now encoded in the $\alm$'s. 

We are interested in extracting information about the statistical properties of the $\alm$'s from the observations. In the standard cosmological model, $\alm$s follow a Gaussian distribution, with vanishing average ($\left\langle \alm\right\rangle =0$, since the $\alm$ are expansion coefficients of the anisotropy field, whose mean vanishes),
and covariance 
\begin{equation}\label{eq:cl}
\langle\alm a_{\ell'm'}^* \rangle = \delta_{\ell\ell^{'}}\delta_{mm^{'}}\Cl
\end{equation}
where the constraints imposed by the two Dirac delta functions follow from the $\alm$ being independent random variables (diagonal covariance). Moreover, statistical isotropy ensures that the variance does not depend on $m$ (rotational invariance of $\Cl$). The $\Cl$'s are the angular power spectrum of the CMB temperature field. The power spectrum is related to the two-point correlation function of the field $C(\theta)=\langle\Theta(\hat{n}_1)\Theta(\hat{n}_2)\rangle$ observed at two directions $\hat{n}_{1}$ and $\hat{n}_2$ in the sky such that $\hat{n}_1\cdot\hat{n}_2=\cos\theta$:
\begin{equation}
C(\theta)={\underset{\ell}{\sum}}\frac{2\ell+1}{4\pi}\Cl P_\ell(\cos\theta),
\end{equation}
where $P_\ell$ is the Legendre polynomial of order $\ell$.

If a random variable is Gaussian distributed, all the statistical properties are encoded in its mean and variance, which are the only momenta of the distribution we need to know. In fact, for a Gaussian distribution, odd momenta vanish and even momenta beyond the second can be recast as a function of the variance (Wick's theorem). Thus, the power spectrum $\Cl$, or equivalently the two-point correlation function $C(\theta)$, completely characterizes the statistical properties of the anisotropy field.

Since the $\alm$'s follow a Gaussian distribution with zero mean and variance $\Cl$, we can readily write the probability density function $p(\alm|\Cl)$ of the $\alm$'s conditioned by the $\Cl$'s:
\begin{equation} \label{eq:likealm}
p(\alm|\Cl)=\frac{1}{\sqrt{2\pi \Cl}} \exp{\left(-\frac{|\alm|^2}{2\Cl}\right)}.
\end{equation}
Given the observed temperature field and the corresponding $\alm$'s, this expression already provides the likelihood function for the theoretical (model) $\Cl$'s. However the information contained in the $\alm$'s can be further compressed, as we shall see in the following.

Statistical isotropy of the $\Cl$'s allows us to rewrite Eq.~\ref{eq:cl} as:
\begin{equation} \label{eq:cl2}
\Cl=\frac{1}{2\ell+1}\overset{\ell}{\underset{m=-\ell}{\sum}}\langle|\alm|^2\rangle
\end{equation}
Some considerations are in order at this point. The average operation defined with the symbol $\langle...\rangle$ in Eqs.~\ref{eq:cl} and~\ref{eq:cl2} is an \textit{ensemble} average. As noted above, the CMB field is a realization of a random process and statistical information about the outcome of such a process should be obtained by averaging over all possible realizations. In practice, however, we can only observe a single realization of the CMB field. A way out is provided by the statistical omogeneity and isotropy of the CMB fluctuations, that in principle allows to substitute the ensemble average in Eq.~\ref{eq:cl} with an average over different positions and directions. According to this \emph{ergodic hypothesis}, different regions that are widely separated in the sky are statistically independent from each other and can be considered as different statistical realizations of the same stochastic process.
Since we only have access to the CMB field observed at $\vec{x}_{0}$ and $\tau_{0}$, i.e., the CMB field here and now, what we are really left is the average over different directions, or equivalently over different values of $m$. 
In other words, for a given $\ell$, all the $\alm$ are drawn from the same distribution, which
can be therefore sampled by measuring all the $2\ell+1$ coefficients. We are thus led to define an estimator of the observed power spectrum 
\begin{equation}\label{eq:clmeas}
\hat{\Cl}=\frac{1}{2\ell+1}\sum_m |\alm|^2 \, ,
\end{equation}
with the property $\langle {\hat C}_\ell\rangle = C_\ell$. Note that in Eq.~\ref{eq:clmeas} the ensemble average does not appear: we are forced to measure $\Cl$ only with a limited number of values. This induces an intrinsic source of inaccuracy due to replacing the true variance $\Cl$ with the observed power $\hat{\Cl}$ (i.e., by replacing the ensemble average with the average over directions). This effect is known as \emph{cosmic variance}:
\begin{eqnarray}\label{eq:cv}
\left\langle \left(\frac{\hat{\Cl}-\Cl}{\Cl}\right)^2\right\rangle  &\!=\!& -1+\frac{1}{(2\ell+1)^{2}\Cl^{2}}\underset{mm'}{\sum}\left\langle \alm a_{\ell m}^{*}a_{\ell m'}a_{\ell m'}^{*}\right\rangle \nonumber \\
&=&-1+\frac{1}{(2\ell+1)^{2}\Cl^{2}}\left[\underset{m}{\sum}\left\langle \alm a_{\ell m}^{*}a_{\ell m}a_{\ell m}^{*}\right\rangle+\underset{m,m'\neq m}{\sum}\left\langle \alm a_{\ell m}^{*}a_{\ell m'}a_{\ell m'}^{*}\right\rangle \right] \nonumber \\
&=&-1+\frac{1}{(2\ell+1)^2\Cl^2}\left(3\Cl^2(2\ell+1)+2\ell\Cl^2(2\ell+1)\right) \nonumber \\
&\!=\!&\frac{2}{2\ell+1}\label{eq:Cosmic variance}
\end{eqnarray}
where the third equality follows from Wick's theorem.

Cosmic variance is an irreducible source of uncertainty in cosmological measurements of the CMB power spectrum, and one of the major sources of uncertainties especially at the largest scales (low-$\ell$), where we have only a limited number of coefficients $\alm$ to average over with respect to the small-scale (high-$\ell$) regime.  Eq.\eqref{eq:Cosmic variance} is valid provided full-sky observations. However, in real data analysis, even if we are able to observe the full sky (e.g., with space missions), we are nevertheless forced to mask a certain fraction of the sky, e.g., to avoid foreground contamination. An approximate estimate of the increase is given by a factor $1/f_{\mathrm{sky}}$, where $f_{\mathrm{sky}}$ but see e.g.,~\cite{Efstathiou:2003dj} for a careful counting of the degrees of freedom available in cut-sky regimes. Current experiments like the Planck satellite are ideally cosmic-variance-limited up to very high multipoles, i.e., $\ell\sim1500$.

To derive the distribution of the observed $\Cl$'s, we note that the sum of $\nu=(2\ell+1)$ standard Gaussian variables follows a $\chi^2$ distribution with $\nu$ degrees of freedom. If we define $\hat{Y}_\ell=\sum_m (|\alm/\sqrt{\Cl}|^2)$, this new variable has a $\chi^2$ distribution:

\begin{equation}
p(\hat{Y}_\ell|\Cl)=\frac{\hat{Y}_\ell^{\nu/2-1}}{\Gamma(\nu/2)2^{\nu/2}}\exp{\left(-\frac{\hat{Y}_\ell}{2}\right)}
\end{equation}

The estimator (hereafter \textit{observed}) $\hat{\Cl}$ is a multiple of $\hat{Y}_\ell$: $\hat{\Cl}=\Cl\hat{Y}_\ell/(2\ell+1)$, and multiples of $\chi^2$-distributed variables follow a Gamma distribution:

\begin{equation}\label{eq:like}
p(\hat{\Cl}|\Cl)\propto \Cl^{-1} \left(\frac{\hat{\Cl}}{\Cl}\right)^{\nu/2-1}\exp{\left(-\frac{\nu}{2}\frac{\hat{\Cl}}{\Cl}\right)}
\end{equation}

The previous expression is the probability of the observed power spectrum given the fiducial power, and for fixed data it can be still regarded as a likelihood $\mathcal{L}(C_\ell)$, in which the role of the data is not played by the $\alm$'s as in Eq.~\ref{eq:likealm}, but by the $\hat\Cl$. The mean and variance of the distribution of the $\hat\Cl$'s are $\mathrm{E}[\hat{C}_\ell]=\Cl$ and $\mathrm{Var}[\hat{C}_\ell]=2\Cl^2/\nu$. The maximum of the distribution is in $(\nu-2)/\nu \Cl$, that does not coincide with the mean of the distribution. As such, the distribution of observed $\hat{\Cl}$ is skewed. However, in the limit $\nu\rightarrow \infty$, the distribution in Eq.~\ref{eq:like} tends to a Gaussian distribution with same mean and variance, according to the central limit theorem. Note that the variance of the distribution is exactly the cosmic variance introduced in Eq.~\ref{eq:cv}. This further stresses the meaning of the cosmic variance as an irreducible source of uncertainty due to the limitation of having access to a single realisation of the Universe (i.e., the limitation due to estimating the true power spectrum $\Cl$ with the observed power spectrum $\hat{C}_\ell$).

\subsection{Statistics of joint CMB temperature and polarization fields -- Exact likelihood in harmonic space}\label{sec:CMBstat_TEB}
The above treatment has to be generalised in the case of the joint analysis of temperature and polarization fields $T,E,B$. In analogy to the temperature case, we can define two sets of spherical harmonics coefficients for $E$ and $B$: 
\begin{subequations}
\begin{eqnarray}\label{eq:almEB}
\alm^E&\equiv&-\frac{1}{2}\left(_{+2}a_{\ell m}+ _{-2}a_{\ell m}\right)\\
\alm^B&\equiv&\frac{i}{2}\left(_{+2}a_{\ell m}-_{-2}a_{\ell m}\right)
\end{eqnarray}
\end{subequations}
where $_{\pm2}a_{\ell m}$ are the expansion coefficients of the combinations of Stokes parameters describing the polarisation state of the CMB signal -- $(Q\pm iU)$ -- in spin-2 spherical harmonics $_{\pm2}Y_{\ell m}$ (see e.g.~\cite{Zaldarriaga:2003bb} for a derivation of the formalism). 

The variable $\textbf{X}_a=(\alm^T,\alm^E,\alm^B)$ is distributed according to a Gaussian multivariate distribution with covariance matrix
\begin{subequations}
\begin{eqnarray}
\mathrm{cov}\left[\textbf{X}_a,\textbf{X}_a\right]&\equiv& \langle\textbf{X}_a \textbf{X}_a^\dag\rangle-\langle|\textbf{X}_a|\rangle^2\\
&\equiv& \textbf{V}_\ell=\left(
\begin{array}{ccc}
\ClTT & \ClTE & 0\\
\ClTE & \ClEE & 0\\
0 & 0 & \ClBB
\end{array}
\right)\, ,
\end{eqnarray}
\end{subequations}
where it is explicitly seen that the temperature and the $E$-polarisation fields are correlated, whereas the parity-even fields ($T$ and $E$) are uncorrelated with the parity-odd field $B$ (although this is strictly true only in the standard cosmological model when parity violation processes are forbidden in the early Universe).

In analogy to Eq.~\ref{eq:clmeas}, the estimators for the observed power spectra are given by the following matrix:
\begin{equation}\label{eq:clall}
\textbf{S}_\ell=\frac{1}{2\ell+1}\sum_m \textbf{X}_a \textbf{X}_a^\dag=\left(
\begin{array}{ccc}
\hat{\ClTT} & \hat{\ClTE}  & \hat{\ClTB} \\
\hat{\ClTE}  & \hat{\ClEE}  & \hat{\ClEB} \\
\hat{\ClTB}  & \hat{\ClEB}  & \hat{\ClBB} 
\end{array}
\right)\, ,
\end{equation}
where the observed cross-correlations $TB$ and $EB$ may be non-vanishing as well. 

The probability of $\textbf{X}_a$ at each $\ell$ can therefore be expressed as:
\begin{subequations}
\begin{eqnarray}\label{eq:pXa}
-2\ln[p(\textbf{X}_a|\textbf{V}_\ell])&=&\textbf{X}_a^\dag \textbf{V}_\ell^{-1} \textbf{X}_a + \ln\mathrm{det}[2\pi\textbf{V}_\ell]\\
&=&(2\ell+1)\mathrm{trace}[\textbf{S}_\ell \textbf{V}_\ell^{-1}]+\ln\mathrm{det}[\textbf{V}_\ell]+\mathrm{const.}
\end{eqnarray}
\end{subequations}
Note that $\textbf{S}_\ell$ represents sufficient statistics for this likelihood function: in the full-sky regime, Eq.~\ref{eq:pXa} only depends on the data through $\textbf{S}_\ell$ and therefore information on the CMB sky can be losslessly compressed to a set of power spectrum estimators $\textbf{S}_\ell=\textbf{S}_\ell(\hat{C}_\ell^{XY})$, $X,Y=T,E,B$. 

The probability of $\textbf{S}_\ell$ given $\textbf{V}_\ell=\textbf{V}_\ell(C_\ell^{XY})$ is obtained by properly normalizing Eq.~\ref{eq:pXa}.
In the previous section, we have seen that the single-field $\hat{\Cl}$ is Gamma-distributed. It is easy to understand that the full set of observed power spectra $\textbf{S}_\ell$ has a Wishart distribution, i.e., a multi-dimensional generalisation of the Gamma distribution, with $\nu=(2\ell+1)$ degrees of freedom in $p=3$ dimensions:

\begin{equation}\label{eq:Wishart}
p(\textbf{S}_\ell|\textbf{W}_\ell)=\mathcal{L}(\mathbf{W}_\ell)=\frac{|\textbf{S}_\ell|^{(\nu-p-1)/2} \exp{\left[-\mathrm{trace}(\textbf{W}_\ell^{-1}\textbf{S}_\ell /2)\right]}}{2^{p\nu/2}|\textbf{W}_\ell|^{\nu/2}\Gamma_p(\nu/2)}
\end{equation}
where $\textbf{W}_\ell=\textbf{V}_\ell/\nu$. For given $\hat{C}_\ell^{XY}$'s, Eq.~\ref{eq:Wishart} represents the exact expression of the likelihood function of the $C_\ell^{XY}$.

Since $\textbf{V}_\ell$ is separable in the two blocks $TE$ and $B$, we can simplify the problem and consider two separate Wishart distributions for the block $TE$ and for the block $B$: 
\begin{equation}
\mathcal{L}(\mathbf{W}_\ell)=\mathcal{L}(\mathbf{W}_\ell^\mathrm{TE})\mathcal{L}(\mathbf{W}_\ell^\mathrm{B})
\end{equation}
The latter is further simplified since it reduces to the one-dimensional Gamma distribution, as described in details in the previous section. The distribution for the $TE$ block can be fully expanded as:

\begin{subequations}
\begin{eqnarray}\label{eq:Wte}
\mathcal{L}(\mathbf{W}_\ell^\mathrm{TE})&\propto&\frac{\left(\hat{\ClTT}\hat{\ClEE}-(\hat{\ClTE})^2\right)^{(\nu-3)/2}}{\left(\ClTT\ClEE-(\ClTE)^2\right)^{\nu/2}}\\\nonumber
\quad&\times&\exp{\left\{-\frac{\nu}{2}\left[\frac{\ClTT\hat{\ClEE}+\hat{\ClTT}\ClEE-2\hat{\ClTE}\ClTE}{\ClTT\ClEE-(\ClTE)^2}\right]\right\}}
\end{eqnarray}
\end{subequations}

The marginal distribution of each individual diagonal element of $\textbf{S}_\ell^{TE}$ can be obtained by integrating $p(\textbf{S}_\ell|\textbf{W}_\ell)^{TE}$ over $\hat{\ClTE}$ and the other diagonal element, and it is again a Gamma distribution as we expect it to be, in analogy to discussion in the previous section. However, the marginal distribution of the off-diagonal terms $\hat{\ClTE}$ is not a Gamma distribution, and it is interesting to note that it depends on $\ClTT$ and $\ClEE$ in addition to $\ClTE$ (see~\cite{Percival:2006ss} for a detailed calculation).

In the limit $\nu\rightarrow \infty$, the Wishart distribution of $\hat{\textbf{X}}_C=(\hat{\Cl}^{TT},\hat{\Cl}^{TE},\hat{\Cl}^{EE})$ tends to a multivariate Gaussian distribution with covariance matrix:

\begin{subequations}
\begin{eqnarray}\label{eq:covW}
\mathrm{cov}[\textbf{X}_C,\textbf{X}_C]&\equiv&\langle\textbf{X}_C \textbf{X}^\dagger_C\rangle-\langle\textbf{X}_C\rangle^2\equiv\mathbf{C}\\
&=&\frac{1}{2\ell+1} \left(
\begin{array}{ccc}
2(\ClTT)^2 & 2\ClTT\ClTE & 2(\ClTE)^2\\
2\ClTT\ClTE & \ClTT\ClEE+(\ClTE)^2 &2\ClTE\ClEE \\
2(\ClTE)^2 &2\ClTE\ClEE  &2(\ClEE)^2
\end{array}
\right)\, ,
\end{eqnarray}
\end{subequations}

The variance of $\hat{\Cl}^{TT}$ and $\hat{\Cl}^{EE}$ is the same of the single-field limit, whereas the variance of the cross-correlation $\hat{\Cl}^{TE}$ reflects the different marginalised distribution of $\hat{\Cl}^{TE}$ itself.

\subsection{Statistics of joint CMB temperature and polarisation fields -- Exact likelihood in real space}\label{sec:CMBstat_real}
The discussion in Secs.~\ref{sec:CMBstat_T} and~\ref{sec:CMBstat_TEB} refers to the CMB statistics in \textit{harmonic} space, i.e., the space in which the CMB fields are expanded in spherical harmonics and the physical information are encoded in the expansion coefficients $a_{\ell m}$. In this subsection, we will review the basics of CMB statistics in \textit{real} space.

The starting point are the observed CMB maps of the three Stokes parameters $T$, $Q$ and $U$. These maps can be discretized into $N$ pixels and arranged in $N$-dimensional vectors $\mathbf{T}$, $\mathbf{Q}$ and $\mathbf{U}$. As discussed in the previous section, the statistical properties of these objects are fully encoded in the auto- and cross- power spectra $C_{\ell}^{XY}$, with $X,Y=\{ T,E,B \}$ for temperature, $E$-mode and $B$-mode polarization. 

The exact likelihood function in real space (also called the pixel-based likelihood) is defined as
\begin{equation}\label{eq:pixel_like}
\mathcal{L}(C_{\ell}) = p(\mathbf{m}|C_{\ell}) = \dfrac{1}{2 \pi |M| ^{1/2}} \exp{\left( -\dfrac{1}{2} \mathbf{m}^{\text{T}} M ^{-1} \mathbf{m} \right)} \, ,
\end{equation}
where $\mathbf{m}$ is the vector with $3N$ elements built from the justaxposition of $\mathbf{T}$, $\mathbf{Q}$ and $\mathbf{U}$, and $M$ is the total covariance matrix. The matrix $\mathbf{M}$ depends only on the angle between two directions in the sky $\hat{n}_{i,j}$
\begin{equation}\label{eq:M1}
\mathbf{M}(\hat{n}_i\cdot\hat{n}_j)=\left(\begin{array}{ccc}
\langle T_iT_j\rangle  &\langle T_iQ_j\rangle  &\langle T_iU_j\rangle\\
\langle T_iQ_j\rangle  &\langle Q_iQ_j\rangle  &\langle Q_iU_j\rangle\\
\langle T_iU_j\rangle  &\langle Q_iU_j\rangle  &\langle U_iU_j\rangle
\end{array}\right).
\end{equation}

The $(3\times3)$ entries in Eq.~\ref{eq:M1} for any given pair of pixels $ij$ depend on the Legendre polynomial $P_\ell$ and the fiducial power spectra. As a straightforward example, the entry $\langle T_iT_j\rangle$ is the expression
\begin{equation}\label{eq:covmat}
\langle T_iT_j\rangle=\sum_\ell\frac{2\ell+1}{4\pi}P_\ell(\hat{r}_i\cdot\hat{r}_j)C_\ell^{TT}.
\end{equation}
where $P_\ell(\hat{r}_i\cdot\hat{r}_j)=\frac{4\pi}{2\ell+1}\sum_m Y_{\ell m}(\hat{r}_i)Y^*_{\ell m}(\hat{r}_j)$. A detailed description of the full procedure to obtain the covariance matrix, together with the expressions of the $(3\times3)$ entries, can be found in Appendix A of Ref.~\cite{Tegmark:2001zv}.

It should be noted that the pixel-based likelihood in Eq.~(\ref{eq:pixel_like}) is exact even in the case of partial sky coverage; this not the case for the likelihood in harmonic space in Eq.~(\ref{eq:Wishart}). Note however that it is still possible to derive an exact form for the harmonic-space likelihood even for partial sky coverage \cite{Upham:2019ruv}. The pixel-based approach ensures mathematical rigour in the evaluation of the likelihood function. Nevertheless, it is highly expensive from a computational point of view. 
Indeed, the number of pixels needed to retain the information in the first $\ell_{max}$ multipoles of the power spectrum scales as $\ell _{\text{max}}^2$ and therefore the Cholesky decomposition required to evaluate the inverse of the covariance matrix in Eq.~\ref{eq:pixel_like} scales roughly as $\ell_{\text{max}}^6$, where $\ell_\mathrm{max}$ is the highest multipole retained in the analysis. The computational cost is therefore driven by the evaluation of inverse matrix and determinants and becomes prohibitive for $\ell_{\text{max}}$ larger than few hundreds. For this reason, this exact approach is feasible only to study large angular scales, where the information is contained in a relatively small number of multipoles.


\newpage
\section{Likelihood approaches -- Small-scale regime}\label{sec:highl}
The exact likelihood of the observed CMB $\hat{C}_\ell^{XY}$ as a function of the underlying fiducial CMB $C_\ell^{XY}$ is given by Eq.~\ref{eq:Wishart} in case of full-sky observations:
\begin{equation}
p(\textbf{S}_\ell|\textbf{W}_\ell)=\mathcal{L}(\mathbf{W}_\ell)=\frac{|\textbf{S}_\ell|^{(\nu-p-1)/2} \exp{\left[-\mathrm{trace}(\textbf{W}_\ell^{-1}\textbf{S}_\ell /2)\right]}}{2^{p\nu/2}|\textbf{W}_\ell|^{\nu/2}\Gamma_p(\nu/2)}.
\end{equation}

However, complications arise in real analysis that make it necessary to replace Eq.~\ref{eq:Wishart} with a suitable approximation. Complications usually include time-consuming evaluations of Eq.~\ref{eq:Wishart} due to the inversion of large covariance matrices for each theoretical model.

A standard approach is to develop an approximation of Eq.~\ref{eq:Wishart} in the full-sky regime that is quadratic in some function of $C_\ell^{XY}$, and that can be easily generalised to the cut-sky regime with a proper estimate of the covariance matrix:

\begin{equation}\label{eq:quad}
-2\ln p(\hat{\textbf{X}}_C|\textbf{X}_C)\equiv-2\ln\mathcal{L}(\textbf{X}_C)\propto\left[(\textbf{Z}_C-\hat{\textbf{Z}}_C)^T \textbf{Y}^{-1} (\textbf{Z}_C-\hat{\textbf{Z}}_C)+\ln|\mathbf{Y}|\right]
\end{equation}

where $\textbf{Z}_C$ ($\hat{\textbf{Z}}_C$) is the vector containing functions of $\Cl$s ($\hat{\Cl}$s) and $\textbf{Y}$ is a suitable choice of the covariance matrix. 

In what follows, we introduce a list of the most common approximate forms among those proposed in the literature (see e.g.,~\cite{Percival:2006ss,Hamimeche:2008ai,Bond:1998qg,Smith:PRD732006,Lewis:PRD652001,Slosar:2004fr}). We further quantify the goodness of the approximation in the full-sky regime following the approach in Ref.~\cite{Percival:2006ss}: we expand the exact likelihood and the approximate forms along the standard axes $(TT,\,TE,\,EE)$ around the maximum $\textbf{X}_C=\hat{\textbf{X}}_C$, and compare the expansion coefficients up to a certain order.

As already commented in the previous section, the analytic comparison of the various approximations is carried in absence of noise contaminations and in the limit of infinite angular resolution. We also implicitly assume that the CMB spectra are \textit{unlensed}, i.e., they are the spectra as they would be observed in absence of gravitational lensing effects on the CMB photons. The inclusion of experimental noise, experimental angular resolution, and gravitational lensing effects will be discussed in Sec.~\ref{sec:highl_noise} and Sec.~\ref{sec:lensing}.

Before moving to the list of the most common approximate likelihood functions, we would like to mention that it is not trivial to construct an unbiased estimator of the true $C_\ell$ in the cut-sky regime. We don't have access to the full-sky set of $a_{\ell m}$ and therefore we cannot directly construct $\hat{C}_\ell$. In the case of cut-sky maps, appropriate algorithms have been developed to derive the unbiased estimator $\hat{C}_\ell$ to be used in the likelihood analysis. For example, pseudo-$C_\ell$ power spectra $\tilde{C}_\ell$ can be defined from cut-sky harmonic coefficients $\tilde{a}_{\ell m}$, see Sec.~\ref{sec:appendixB} for further details and references. The pseudo-$C_\ell$ are related to the \textit{true} $C_\ell$ in ensemble average as

\begin{equation}\label{eq:cutsky}
\langle \tilde{C}_{\ell'}\rangle=\sum_{\ell}M_{\ell'\ell}\langle C_{\ell}\rangle
\end{equation}

where $M_{\ell_1\ell_2}$ is a coupling matrix that encodes the geometrical effects of cut-sky observations. From Eq.~\ref{eq:cutsky}, it is possible to operatively define an estimator for the $C_\ell$ in the cut-sky regime as:

\begin{equation}
\hat{C}_\ell=\sum_{\ell'} M_{\ell \ell'}^{-1} \tilde{C}_{\ell'}\;.
\end{equation}

The interested reader can find a detailed discussion in Appendix~\ref{sec:appendixB}, where we also report alternative methods adopted to construct estimators for the $BB$ spectrum~\ref{sec:pureEB}, and for power spectra at large scales via the quadratic maximum likelihood (QML) approach~\ref{sec:qml}. A final remark on the cut-sky case: the compression of information from CMB maps ($\sim (N_\mathrm{pix}\times N_\mathrm{pix})$ pixels) to CMB spectra ($\sim (\ell_\mathrm{max}-\ell_\mathrm{min})$ bandpowers) is lossless in the full-sky regime, i.e., the power spectra represent sufficient statistics. In the cut-sky regime, the compression is partly lossy, as the masked regions induce correlations between multipoles which have to be taken into account (see e.g., discussion in Appendix~\ref{sec:appendixB}).

\subsection{Approximate forms}\label{sec:approxtheo}
The most common approximations are given by quadratic/Gaussian expressions in $C_\ell^{XY}$ with different choices for the covariance matrix. Alternatively, quadratic expressions involving more complicated functions $f=f(C_\ell^{XY})\equiv\mathbf{Z}_C$, as well as \textit{ad-hoc} combinations of various approximations have been developed to match the exact likelihood up to a certain order in the perturbative regime (see Sec.~\ref{subsec:compexac}).

\begin{itemize}
\item \textbf{Symmetric Gaussian}. This approximation is quadratic in $\textbf{Z}_C=C_\ell^{XY}$, with covariance matrix given by the curvature of the Wishart, see Eq.~\ref{eq:Ym1}:

\begin{equation}\label{eq:gaussian}
-2\ln\mathcal{L}(\textbf{X}_C)\propto (\hat{\textbf{X}}_C-\textbf{X}_C)^T \textbf{Y}^{-1}_C(\hat{\textbf{X}}_C) (\hat{\textbf{X}}_C-\textbf{X}_C)+\mathrm{const.}
\end{equation}

The inverse of the covariance matrix $\mathbf{Y}_C$ is the curvature of the Wishart distribution in Eq.~\ref{eq:Wte}, i.e., $\textbf{Y}^{-1}_{ij}=\der^2(-2\ln p)_{ij}|_{\Cl=\hat{\Cl}}$, computed in $\textbf{X}_C=\hat{\textbf{X}}_C$:

\begin{eqnarray}\label{eq:Ym1}
\textbf{Y}^{-1}_\ell&=&\frac{\nu}{2\left[\hat{\Cl}^{TT}\hat{\Cl}^{EE}-(\hat{\Cl}^{TE})^2\right]^2}\\\nonumber
&\times&\left(
\begin{array}{ccc}
(\hat{\Cl}^{EE})^2 & -2\hat{\Cl}^{TE}\hat{\Cl}^{EE} & (\hat{\Cl}^{TE})^2\\
-2\hat{\Cl}^{TE}\hat{\Cl}^{EE} & 2[\hat{\Cl}^{TT}\hat{\Cl}^{EE}+(\hat{\Cl}^{TE})^2] & -2\hat{\Cl}^{TE}\hat{\Cl}^{TT}\\
(\hat{\Cl}^{TE})^2 & -2\hat{\Cl}^{TE}\hat{\Cl}^{TT} & (\hat{\Cl}^{TT})^2
\end{array}
\right)\, ,
\end{eqnarray}

\item \textbf{Improper Gaussian}. This approximation is similar to the Symmetric Gaussian in Eq.~\ref{eq:gaussian}, with the covariance matrix that appears in the first term replaced by $\textbf{Y}=\textbf{Y}(C_\ell^{XY})$; i.e., the covariance matrix is given in terms of the model $C_\ell^{XY}$. This approximation is an improper Gaussian in a sense that there is no determinant term: 

\begin{equation}\label{eq:improper}
-2\ln\mathcal{L}(\textbf{X}_C)\propto (\hat{\textbf{X}}_C-\textbf{X}_C)^T \textbf{Y}^{-1}_C(\textbf{X}_C) (\hat{\textbf{X}}_C-\textbf{X}_C)
\end{equation}

\item \textbf{Determinant Gaussian}. The expression in Eq.~\ref{eq:improper} can be slightly modified to provide a better fit to the exact likelihood approach (see Sec.~\ref{subsec:compexac}). The modification consists in the addition of a $C_\ell^{XY}$-dependent determinant term:

\begin{equation}\label{eq:determinant}
-2\ln\mathcal{L}(\textbf{X}_C)\propto (\hat{\textbf{X}}_C-\textbf{X}_C)^T \textbf{Y}^{-1}_C(\textbf{X}_C) (\hat{\textbf{X}}_C-\textbf{X}_C)+\ln|\textbf{Y}_C(\textbf{X}_C)|
\end{equation}

\item \textbf{Fiducial Gaussian}. This approximation is similar to Eq.~\ref{eq:gaussian} and Eq.~\ref{eq:improper}, with a constant determinant term (as in Eq.~\ref{eq:gaussian}) and the covariance matrix computed for a given fiducial model (as in Eq.~\ref{eq:improper}). The fiducial model for the covariance matrix is however kept fixed, and assumed to be smooth and a close approximation to the underlying model under scrutiny:

\begin{equation}\label{eq:fiducial}
-2\ln\mathcal{L}(\textbf{X}_C)\propto (\hat{\textbf{X}}_C-\textbf{X}_C)^T \textbf{Y}^{-1}_C(\textbf{X}_{C,\mathrm{fid}}) (\hat{\textbf{X}}_C-\textbf{X}_C)+\mathrm{const.}
\end{equation}

The fiducial Gaussian approximation is used in the official analyses of the Planck \cite{Aghanim:2015xee,planckL05}, ACT \cite{Louis:2016ahn} and SPT \cite{Henning:2017nuy} collaborations. 

\item \textbf{Log-normal}. This approximation is quadratic in a peculiar function of theoretical and observed spectra, i.e., $\textbf{Z}_C=\hat{C}_\ell^{XY}\ln(C_\ell^{XY})$, with fixed covariance matrix $\textbf{Y}_C=\textbf{Y}_C(\hat{C}_\ell^{XY})$:

\begin{equation}\label{eq:lognorm}
-2\ln\mathcal{L}(\textbf{X}_C)\propto (\hat{\textbf{Z}}_C-\textbf{Z}_C)^T \textbf{Y}^{-1}_C(\hat{\textbf{X}}_C) (\hat{\textbf{Z}}_C-\textbf{Z}_C)+\mathrm{const.}
\end{equation}

\item \textbf{Offset log-normal}. This approximation is a generalisation of Eq.~\ref{eq:lognorm}. The data vector is generalised to $\textbf{Z}_C=(1+a_{XY})\hat{C}_\ell^{XY}\ln(C_\ell^{XY}+a_{XY}\hat{C}_\ell^{XY})$, with $a_{XY}$ a suitable real offset coefficient that may or may be not be the same for every $XY$ pair. The covariance matrix is again as in Eq.~\ref{eq:lognorm} ($\textbf{Y}_C=\textbf{Y}_C(\hat{C}_\ell^{XY})$):

\begin{equation}\label{eq:offlognorm}
-2\ln\mathcal{L}(\textbf{X}_C)\propto (\hat{\textbf{Z}}_C-\textbf{Z}_C)^T \textbf{Y}^{-1}_C(\hat{\textbf{X}}_C) (\hat{\textbf{Z}}_C-\textbf{Z}_C)+\mathrm{const.}
\end{equation}

\item \textbf{One-third-two-thirds}. We briefly mention this approximation as an example of combined likelihood appositely built to match the exact likelihood up to the third order. It is a weighted combination of the improper Gaussian in Eq.~\ref{eq:improper} (with weight $1/3$) and of the log-normal approximation in Eq.~\ref{eq:lognorm} (with weight $2/3$). Note that the approximation was explicitly built for the single-field TT-only WMAP analysis~\cite{Verde:2003ey}:

\begin{equation}\label{eq:onetwothird}
\ln\mathcal{L}(C_\ell^{TT})\propto \frac{1}{3}\ln\left[\mathcal{L}_\mathrm{improper}(C_\ell^{TT})\right]+\frac{2}{3}\ln \left[\mathcal{L}_\mathrm{lognorm}(C_\ell^{TT})\right]
\end{equation}

\item \textbf{Hamimeche-Lewis}. In Ref.~\cite{Hamimeche:2008ai}, Hamimeche \& Lewis (HL) have developed a form of the likelihood for correlated Gaussian fields (CMB temperature and polarisation) that coincides with the exact likelihood in full sky. The authors show with simulations that it provides a very good approximation to the exact likelihood in the cut-sky regime at small scales\footnote{One of the assumptions is that the matrix of the estimators $\hat{C}_\ell$ is positive definite. This assumption may break up at large scales.} ($\ell\ge30$). The form of the likelihood is quadratic in some peculiar function of the observed, fiducial, and theoretical $C_\ell$, as we shall see in Sec.~\ref{subsec:compexac}. The covariance matrix is precomputed for a fixed fiducial model. The dependence on the fiducial model is negligible. Moreover, should the fiducial fail in matching the true sky, the likelihood is still exact in full sky. The HL likelihood was used in the analysis of the BICEP2/KECK data \cite{Ade:2018gkx}. 

In the HL formalism, the likelihood in cut-sky can be approximated as:
\begin{equation}
-2 \ln  \mathcal{L} (\mathbf{X}_C) \simeq \mathbf{X}_{g\ell} ^{T}\mathbf{M}_{f\ell}^{-1}\mathbf{X}_{g\ell} \label{eq:HL4}
\end{equation}

where $\mathbf{X}_g$ is a vector of a specific function of the observed, fiducial, and theoretical $C_\ell$, and $\mathbf{M}_f$ is the fiducial model covariance block matrix 
\begin{equation}\label{eq:HL5}
[\mathbf{M}_f]_{\ell \ell'} = \langle  (\hat{\mathbf{X}}_{\ell} - \mathbf{X}_{\ell})  (\hat{\mathbf{X}}_{\ell '} - \mathbf{X}_{\ell '})^T  \rangle _f 
\end{equation}
with $n(n+1)/2 \times n(n+1)/2$ blocks ($n$ is the number of fields), labeled by $\ell$ and $\ell '$, i.e., we explicitly take into account the possibility that either the cut-sky or anisotropic noise can induce correlations between different multipoles (non-diagonal covariance).

The derivation of Eq.~\ref{eq:HL4} is provided in Sec.~\ref{subsec:compexac}, where we will show that it is equivalent to the exact likelihood in the full-sky regime. For more details and a thorough definition of the notation, see Ref.~\cite{Hamimeche:2008ai}
\end{itemize}

\subsection{Comparison with the exact likelihood in the full-sky regime}\label{subsec:compexac}

In this section, we comment on the goodness of the approximations listed in the previous section. The goodness is defined in terms of the ability to match the exact likelihood in the full-sky regime up to a certain order, when both the exact likelihood and the approximate form are expanded around the maximum. This approach is described in Ref.~\cite{Percival:2006ss}.

Let's start by expanding the Wishart distribution in Eq.~\ref{eq:Wte} along the $TT$ direction. In particular, we write Eq.~\ref{eq:Wte} with the following substitutions: $\ClTE\rightarrow \hat{\Cl}^{TE}$, $\ClEE\rightarrow \hat{\Cl}^{EE}$, $\ClTT\rightarrow (1+\epsilon)\hat{\Cl}^{TT}$. We further expand in $\epsilon$. We obtain:

\begin{equation}\label{eq:Wtteps}
-2 \ln\mathcal{L}(\ClTT) = \frac{\nu}{(\hat{r}^2-1)^2}\left[\frac{\epsilon^2}{2}+\frac{2}{3(\hat{r}^2-1)}\epsilon^3+\mathcal{O}(\epsilon^4)\right]
\end{equation}

where $\hat{r}=\hat{\Cl}^{TE}/\sqrt{\hat{\Cl}^{TT}\hat{\Cl}^{EE}}$ and $\nu=2\ell+1$. It is straightforward to show that the expansion along the $EE$ axis provides the same form of Eq.~\ref{eq:Wtteps} for $-2 \ln[\mathcal{L}(\ClEE)]$. 

Let's now expand along the $TE$ axis with the following substitutions: $\ClTT\rightarrow \hat{\Cl}^{TT}$, $\ClEE\rightarrow \hat{\Cl}^{EE}$, $\ClTE\rightarrow (1+\epsilon)\hat{\Cl}^{TE}$. The expansion is:

\begin{equation}\label{eq:Wteeps}
-2 \ln\mathcal{L}(\ClTE) = \frac{2\hat{r}^2 (\hat{r}^2+1) \nu}{(\hat{r}^2-1)^2}\left[\frac{\epsilon^2}{2}-\frac{2 \hat{r}^2}{3(\hat{r}^2-1)}\epsilon^3+\mathcal{O}(\epsilon^4)\right]
\end{equation}

Equation \ref{eq:Wteeps} reflects again the difference in the marginal distribution of the cross-correlation spectrum $TE$ with respect to the distributions of the auto-spectra $TT$ and $EE$ (see discussion in Sec.~\ref{sec:CMBstat_TEB}). 

Let's now move to expand each of the approximations reported in Sec.~\ref{sec:approxtheo}.

\begin{itemize}

\item \textbf{Symmetric Gaussian}. The expansion of Eq.~\ref{eq:gaussian} is of the same form along each of the standard axes $TT$, $TE$, $EE$, with a different normalisation factor in the case of the expansion along $TE$:

\begin{subequations}
\begin{eqnarray}\label{eq:gaussexp}
-2\ln\mathcal{L}(\ClTT) &\propto& \frac{\nu}{(\hat{r}^2-1)^2}\left(\frac{\epsilon^2}{2}\right)\\
-2\ln\mathcal{L}(\ClEE) &\propto& \frac{\nu}{(\hat{r}^2-1)^2}\left(\frac{\epsilon^2}{2}\right)\\
-2\ln\mathcal{L}(\ClTE) &\propto& \frac{2\hat{r}^2(\hat{r}^2+1)\nu}{(\hat{r}^2-1)^2}\left(\frac{\epsilon^2}{2}\right)
\end{eqnarray} 
\end{subequations}

Note that the expansion is truncated at the second order on $\epsilon$. This result is an exact expansion and it is expected, given the initial form (symmetric Gaussian) of the approximate likelihood. If we compare Eq.~\ref{eq:gaussexp} with the expansion of the Wishart in Eqs.~\ref{eq:Wtteps}-\ref{eq:Wteeps}, we observe what follows. Firstly, the Symmetric Gaussian matches the Wishart only up to the second order on $\epsilon$. Secondly, the approximate form is, by definition, symmetric in $\epsilon$ and therefore fails in capturing the skewness of the exact likelihood. Finally, it is biased low along the $TT$, $EE$ axes in a sense that $-2\ln(\mathcal{L}_\mathrm{gaussian}(\epsilon))<-2\ln(\mathcal{L}_\mathrm{exact}(\epsilon))$ for $\epsilon>0$ (i.e., for $C_\ell>\hat{C}_\ell$). For the opposite reason, it is biased high along the $TE$ axis. 

\item \textbf{Improper Gaussian}. The expansion of Eq.~\ref{eq:improper} along the standard axes are as follows:

\begin{subequations}
\begin{eqnarray}\label{eq:improperexp}
-2\ln\mathcal{L}(\ClTT) &\propto& \frac{\nu}{(\hat{r}^2-1)^2}\left(\frac{\epsilon^2}{2}+\frac{\epsilon^3}{(\hat{r}^2-1)}+\mathcal{O}(\epsilon^4)\right)\\
-2\ln\mathcal{L}(\ClEE) &\propto& \frac{\nu}{(\hat{r}^2-1)^2}\left(\frac{\epsilon^2}{2}+\frac{\epsilon^3}{(\hat{r}^2-1)}+\mathcal{O}(\epsilon^4)\right)\\
-2\ln\mathcal{L}(\ClTE) &\propto& \frac{2\hat{r}^2(\hat{r}^2+1)\nu}{(\hat{r}^2-1)^2}\left(\frac{\epsilon^2}{2}-\frac{\hat{r}^2 (\hat{r}^2+3)\epsilon^3}{(\hat{r}^2-1)}+\mathcal{O}(\epsilon^4)\right)
\end{eqnarray}
\end{subequations}

With respect to the symmetric Gaussian, the improper Gaussian approximation is skewed in the same direction of the Wishart. However, it is still a correct match only up to the second order. With respect to the exact likelihood, Eqs.~\ref{eq:improperexp} show that the improper Gaussian is biased high along the $TT$, $EE$ directions, where $-2\ln(\mathcal{L}_\mathrm{improper}(\epsilon))>-2\ln(\mathcal{L}_\mathrm{exact}(\epsilon))$ for $\epsilon>0$ (i.e., for $C_\ell>\hat{C}_\ell$). For the opposite reason, it is biased low along the $TE$ direction.

\item \textbf{Determinant Gaussian}. In this case, it is clear from Eq.~\ref{eq:determinant} that the likelihood is biased at each multipole $\ell$ as the $C_\ell$-dependent determinant term implies that the minimum value for this approximate form is not in $\epsilon=0$. Indeed, the expansions of Eq.~\ref{eq:determinant} along the standard axes include a term of order $\epsilon$:

\begin{subequations}
\begin{eqnarray}\label{eq:determinantexp}
-2\ln\mathcal{L}(\ClTT) &\propto& \frac{1}{\left(\hat{r}^2-1\right)^2}\left[-3\epsilon(\hat{r}^2-1)+\frac{(\nu-3)\epsilon^2}{2}+\frac{\epsilon^3}{(\hat{r}^2-1)}+\mathcal{O}(\epsilon^4)\right]\\
-2\ln\mathcal{L}(\ClEE) &\propto& \frac{1}{\left(\hat{r}^2-1\right)^2}\left[-3\epsilon(\hat{r}^2-1)+\frac{(\nu-3)\epsilon^2}{2}+\frac{\epsilon^3}{(\hat{r}^2-1)}+\mathcal{O}(\epsilon^4)\right]\\
-2\ln\mathcal{L}(\ClTE) &\propto& \frac{2\hat{r}^2}{\left(\hat{r}^2-1\right)^2}\\\nonumber
&&\left[3\epsilon(\hat{r}^2-1)+\frac{(\nu-3)(\hat{r}^2+1)\epsilon^2}{2}-\frac{(\nu-1)\hat{r}^2(\hat{r}^2+1)\epsilon^3}{(\hat{r}^2-1)}+\mathcal{O}(\epsilon^4)\right]
\end{eqnarray}
\end{subequations}

However, it can be shown that this approximation, although biased at each individual $\ell$, is unbiased ``on average'', i.e., reproduces the correct result with reasonable accuracy when summed over a wide-enough range of multiples (see HL).

\item \textbf{Fiducial Gaussian}. The expansion of Eq.~\ref{eq:fiducial} is equivalent to the expansion in Eq.~\ref{eq:gaussexp}, only with a different normalization factor. Indeed, the covariance matrix in Eq.~\ref{eq:fiducial} is fixed to that of a given fiducial model, and therefore the approximation is quadratic in $\epsilon$. Note however that, although the form of the expansion is similar at any $\ell$ between the symmetric Gaussian and the fiducial Gaussian, the latter provides a better approximation of the exact likelihood when summed over a range of multipoles (see discussion in Ref.~\cite{Hamimeche:2008ai}). 

\item \textbf{Log-normal}. In this case, we have:

\begin{equation}
Z_C=\left(
\begin{array}{c}
\hat{\Cl}^{TT}\ln[\ClTT]\\
\hat{\Cl}^{TE}\ln[\ClTE]\\
\hat{\Cl}^{EE}\ln[\ClEE]
\end{array}
\right)\, ,
\end{equation}

in Eq.~\ref{eq:quad}, with $Y_C=Y_C(\hat{C}_\ell^{XY})$ being the curvature matrix. The expansions along the auto- and cross-spectra directions can be easily obtained up to normalization factors:

\begin{subequations}
\begin{eqnarray}\label{eq:lognormexp}
-2\ln\mathcal{L}(\ClTT) &\propto& \frac{\nu}{(\hat{r}^2-1)^2}\left(\frac{\epsilon^2}{2}-\frac{\epsilon^3}{2}+\mathcal{O}(\epsilon^4)\right)\label{eq:lognormexp1}\\
-2\ln\mathcal{L}(\ClEE) &\propto& \frac{\nu}{(\hat{r}^2-1)^2}\left(\frac{\epsilon^2}{2}-\frac{\epsilon^3}{2}+\mathcal{O}(\epsilon^4)\right)\label{eq:lognormexp2}\\
-2\ln\mathcal{L}(\ClTE) &\propto& \frac{2\hat{r}^2(\hat{r}^2+1)\nu}{(\hat{r}^2-1)^2}\left(\frac{\epsilon^2}{2}-\frac{\epsilon^3}{2}+\mathcal{O}(\epsilon^4)\right)\label{eq:lognormexp3}
\end{eqnarray}
\end{subequations}

Regardless of the normalization factors, a comparison between Eqs.~(\ref{eq:lognormexp1})-(\ref{eq:lognormexp3}) and Eqs.~(\ref{eq:Wtteps}) and (\ref{eq:Wteeps}) shows that the log-normal distribution provides a good approximation to the exact likelihood up to the second order in the expansion around the maximum. The two distributions have a different shape starting from the third-order term in the series expansion. It is also interesting to note that, normalization factors aside, the expansions along the standard axes are identical. This is a further difference with respect to the case of the Wishart distribution.

\item \textbf{Offset log-normal}. The log-normal distribution can be slightly modified in a way that it could approximate the exact Wishart distribution up to third order. The modified log-normal, or offset log-normal, is quadratic in:

\begin{equation}
Z_C=\left(
\begin{array}{c}
\hat{\Cl}^{TT}(1+a_\mathrm{TT})\ln[\ClTT+a_\mathrm{TT}\hat{\Cl}^{TT}]\\
\hat{\Cl}^{TE}(1+a_\mathrm{TE})\ln[\ClTE+a_\mathrm{TE}\hat{\Cl}^{TE}]\\
\hat{\Cl}^{EE}(1+a_\mathrm{EE})\ln[\ClEE+a_\mathrm{EE}\hat{\Cl}^{EE}]
\end{array}
\right)\, ,
\end{equation}

where the offset factors $a_\mathrm{TT},\,a_\mathrm{TE},\,a_\mathrm{EE}$ can be adjusted to match the Wishart distribution up to the third order. The covariance matrix is again assumed to be $Y_C=Y_C(\hat{C}_\ell^{XY})$. Expanding the offset log-normal in the usual way, one gets:

\begin{subequations}
\begin{eqnarray}\label{eq:offlognormexp}
-2\ln\mathcal{L}(\ClTT) &\propto& \frac{\nu}{(\hat{r}^2-1)^2}\left(\frac{\epsilon^2}{2}-\frac{\epsilon^3}{2 (1+a_\mathrm{TT})}+\mathcal{O}(\epsilon^4)\right)\\\label{eq:offlognormexp1}
-2\ln\mathcal{L}(\ClEE) &\propto& \frac{\nu}{(\hat{r}^2-1)^2}\left(\frac{\epsilon^2}{2}-\frac{\epsilon^3}{2 (1+a_\mathrm{EE})}+\mathcal{O}(\epsilon^4)\right)\\\label{eq:offlognormexp2}
-2\ln\mathcal{L}(\ClTE) &\propto& \frac{2\hat{r}^2(\hat{r}^2+1)\nu}{(\hat{r}^2-1)^2}\left(\frac{\epsilon^2}{2}-\frac{\epsilon^3}{2 (1+a_\mathrm{TE})}+\mathcal{O}(\epsilon^4)\right)\label{eq:offlognormexp3}
\end{eqnarray}
\end{subequations}

A comparison between Eqs.~(\ref{eq:lognormexp1})-(\ref{eq:lognormexp3}) and Eqs.~\ref{eq:Wtteps},\ref{eq:Wteeps} makes it clear that the offset log-normal distributions is a good approximation to the Wishart distribution up to the third order in the expansion, provided that

\begin{equation}\label{eq:aoff}
a_\mathrm{TT}=a_\mathrm{EE}=-\frac{1}{4}(1+3\hat{r}^2),\quad a_\mathrm{TE}=-\frac{1}{2}\left(2+\frac{3(\hat{r}^4-1)}{2\hat{r}^2(\hat{r}^2+3)}\right)
\end{equation}

\item \textbf{One-third two-thirds}. Comparing the TT expansion in Eq.~\ref{eq:improperexp} and in Eq.~\ref{eq:lognormexp} with the TT expansion of the exact likelihood in Eq.~\ref{eq:Wtteps}, it is clear that the weighted sum of the improper Gaussian and the lognormal distribution with weights $1/3$ and $2/3$ respectively matches the Wishart distribution up to the third order in $\epsilon$:

\begin{eqnarray}
-2\ln\mathcal{L}(\ClTT)&=&\left[\frac{1}{3}(-2\ln\mathcal{L}_\mathrm{improper})+\frac{2}{3}(-2\ln\mathcal{L}_\mathrm{lognorm})\right]\\\nonumber&\propto&\nu\left[\frac{\epsilon^2}{2}-\frac{2\epsilon^3}{3}+\mathcal{O}(\epsilon^4)\right]
\end{eqnarray}

\item \textbf{Hamimeche-Lewis}. By construction, this likelihood approximation matches exactly the Wishart distribution in the full sky regime. Indeed, 
the true power of this approximation stands in the fact that the exact quadratic form derived from the full-sky exact likelihood result is assumed to be valid also in the cut-sky regime and at high multipoles, where it is faster to evaluate than the exact calculation. 

We show here the equivalence between the exact likelihood in Eq.~\ref{eq:Wishart} and the Hamimeche-Lewis formalism in full sky.
In what follows, we make use of the matrix notation adopted in Ref.~\cite{Hamimeche:2008ai}. This notation is somehow different from the formalism used in the previous examples. However, it is a more suitable choice to better appreciate the H\&L approximation. We assume the matrix of the estimators $\hat{\mathbf{C}}_{\ell}$ to be positive definite. In the full-sky limit, given $n$ gaussian fields, the likelihood function is defined as in Eq.~\ref{eq:pXa}
\begin{subequations}
\begin{eqnarray}\label{eq:HLpXA}
-2 \ln \mathcal{L} (\mathbf{C}_{\ell}) &=& (2\ell +1) \times  \left[ \text{trace} [ \hat{\mathbf{C}}_{\ell}  \mathbf{C}_{\ell}^{-1}] - \ln \vert \mathbf{C}_{\ell}^{-1} \hat{\mathbf{C}}_{\ell}  \vert - n \right] \label{eq:HL1}\\
&=& (2\ell +1) \times  \left[ \text{trace} [\mathbf{C}_{\ell}^{-1/2} \hat{\mathbf{C}}_{\ell}  \mathbf{C}_{\ell}^{-1/2}] - \ln \vert \mathbf{C}_{\ell}^{-1/2} \hat{\mathbf{C}}_{\ell} \mathbf{C}_{\ell}^{-1/2} \vert - n \right]\label{eq:HL12}\\
& = & (2\ell +1) \sum _i^n \left[ D_{\ell ,ii} - \ln (D_{\ell, ii}) -1 \right] . \label{eq:HL2}
\end{eqnarray}
\end{subequations}

where\footnote{In passing from Eq.~\ref{eq:HL1} to Eq.~\ref{eq:HL12}, we have made use of the properties of the $\text{trace}[]$ operator.}, with respect to Eq.~\ref{eq:pXa}, $\textbf{S}_\ell\rightarrow\hat{\textbf{C}}_\ell$ and $\textbf{V}_\ell\rightarrow\textbf{C}_\ell$. In passing from Eq. \eqref{eq:HL1} to Eq. \eqref{eq:HL2}, we consider that the symmetric form is defined using the Hermitian square root and $\mathbf{C}_{\ell}^{-1/2} \hat{\mathbf{C}}_{\ell} \mathbf{C}_{\ell}^{-1/2} = \mathbf{U}_{\ell} \mathbf{D}_{\ell} \mathbf{U}_{\ell}^{T}$, for orthogonal $ \mathbf{U}_{\ell}$ and diagonal $\mathbf{D}_{\ell}$. In other words, we diagonalise the matrix $\mathbf{C}_{\ell}^{-1/2} \hat{\mathbf{C}}_{\ell} \mathbf{C}_{\ell}^{-1/2}$.

In order to generalise Eq.~\ref{eq:HL2} to the cut sky, we want to reshuffle it in such a way that it resembles a quadratic form:
\begin{subequations}
\begin{eqnarray}
-2 \ln \mathcal{L} (\mathbf{C}_{\ell}) &=& \dfrac{2 \ell +1}{2} \sum _i [ g(D_{\ell,ii}) ] ^2  \\
&=& \dfrac{2 \ell +1}{2}  \Tr [ \mathbf{g}(\mathbf{D}_{\ell})^2 ] \, . \label{eq:HL7}
\end{eqnarray}
\end{subequations}
where the function $g(x)$ is defined as
\begin{equation}\label{eq:HL8}
g(x) \equiv \text{sign}(x-1) \sqrt{2(x - \ln(x) - 1)} \, ,
\end{equation}
and $ [ \mathbf{g}(\mathbf{D}_{\ell})] _{ij} = g(D_{\ell, ii}) \delta _{ij}$.

In order to transform Eq.~\ref{eq:HL7} in a form that is quadratic also in the matrix elements, we exploit the following matrix identity \footnote{Using the invariance of $\Tr[\textbf{A}]$ under diagonalisation of $\textbf{A}$, one has that 
\begin{subequations}
\begin{eqnarray}
\mathrm{Trace}[\mathbf{g(D_\ell)}]&=&\mathrm{Trace}[\mathbf{U\,g(D_\ell)\,U^T}]\\
&=&\mathrm{Trace}[\mathbf{C}_\ell^{-1/2} \hat{\mathbf{C}}_\ell \mathbf{C}_\ell^{-1/2}]\\
&=&\mathrm{Trace}[\mathbf{C}_{f\ell}^{-1/2} \mathbf{C}_{f\ell}^{1/2} \left(\mathbf{C}_\ell^{-1/2} \hat{\mathbf{C}}_\ell \mathbf{C}_\ell^{-1/2}\right) \mathbf{C}_{f\ell}^{1/2} \mathbf{C}_{f\ell}^{-1/2}]\\
&=&\mathrm{Trace}[\mathbf{C}_{f\ell}^{-1/2} \mathbf{C}_{g\ell}  \mathbf{C}_{f\ell}^{-1/2}]
\end{eqnarray}
\end{subequations}
}:

\begin{equation}
\dfrac{2 \ell +1}{2} \text{Trace} [ (\mathbf{C}_{f \ell} ^{-1/2} \mathbf{C}_{g \ell} \mathbf{C}_{f \ell} ^{-1/2}) ^2 ] = \mathbf{X}_{g \ell}^T \mathbf{M}_{f \ell}^{-1} \mathbf{X} _{g \ell} \, ,
\end{equation}
where $\mathbf{X}_{g \ell}$ is the vector of $\mathbf{C}_{g \ell} \equiv \mathbf{C}_{f \ell} ^{1/2} \mathbf{U} _{\ell} \mathbf{g} (\mathbf{D}_{\ell}) \mathbf{U}_{\ell}^T \mathbf{C}_{f \ell} ^{1/2}$ elements for a given fiducial model $\mathbf{C}_{f \ell}$, with dimension $n(n+1)/2$, and $\mathbf{M}_{f \ell}$ is the covariance of $\hat{\mathbf{X}}$ evaluated at $\mathbf{C}_{\ell} = \mathbf{C}_{f \ell}$. Therefore Eq. \eqref{eq:HL7} can be rewritten as
\begin{subequations}
\begin{eqnarray}\label{eq:HLquad}
-2 \ln \mathcal{L} (\mathbf{C}_{\ell}) &=& \dfrac{2 \ell +1}{2} \text{Trace} [ ( \mathbf{C}_{f \ell} ^{-1/2} \mathbf{C}_{g \ell} \mathbf{C}_{f \ell} ^{-1/2} )^2]  \\
&=& \mathbf{X}_{g \ell}^T \mathbf{M}_{f \ell}^{-1} \mathbf{X} _{g \ell} \, .
\end{eqnarray}
\end{subequations}

We stress that this formulation is exact in the full sky regime, as it has been obtained by means of matrix identities and no approximations have been adopted so far.

\end{itemize}

\subsection{Including the effects of noise and beam smearing}\label{sec:highl_noise}
The signal observed with a real CMB experiment is affected by various sources of experimental contaminations. Here, we focus on two main classes of experimental effects: the noise bias and the beam smearing. 

The noise bias is due to the instrumental noise from detectors that adds up to the cosmological signal. It has to be characterised and subtracted from the (overall) signal (an alternative approach posits in cross-correlating different detectors and assuming their individual noise to be uncorrelated, see Sec.~\ref{sec:highl_multinu}). In the simple case of isotropic noise in real space, the noise level is independent from the direction. This translates in a diagonal noise in harmonic space, i.e., the noise power spectrum $N_\ell$ is an additive bias for the CMB power spectrum. A very simple example is the case of isotropic and homogenous noise in real space, i.e., the noise level is the same in each pixel. This translates in a ``white noise'' in harmonic space, i.e., $N_\ell$ is constant in $\ell$. A usual assumption is also to consider the noise in temperature and polarisation to be uncorrelated. If the noise is anisotropic (i.e., it changes from pixel to pixel) for example because of a particular scanning strategy that induces anisotropic sky coverage, it may induce correlations between $a_{\ell m}$, and different considerations apply.

The beam smearing is due to the fact that the instrument has a finite angular response. The signal observed along a certain direction takes contributions from all angular directions. These contributions are weighted with the angular response of the instrument. In real space, this effect is described as a convolution of the observed signal with the angular response (hereafter \textit{beam}) of the instrument $\Theta(\theta,\phi)\propto\int d\Omega' B(\theta-\theta',\phi-\phi') \Theta(\theta',\phi')$. In harmonic space, the convolution becomes a product between the harmonic expansions of the CMB fields and the beam $a'_{\ell m}=b_{\ell m}a_{\ell m}$. In the simple case of gaussian beam of width $\sigma_\mathrm{FWHM}=2\sqrt{2\ln2}\sigma$, the harmonic expression of the beam is independent from $m$ and takes the simple form of $b_{\ell m}\rightarrow B_\ell\equiv \exp[-\ell(\ell+1)\sigma^2]=\exp[-\ell(\ell+1)\sigma^2_\mathrm{FWHM}/(8\ln2)]$.The beam smearing is a multiplicative bias for the CMB power spectrum.

In presence of noise and beam smearing, the observed signal is $d_{\ell m}=B_\ell a_{\ell m}+n_\ell$, and the estimator in Eq.~\ref{eq:clmeas} becomes

\begin{equation}\label{eq:dlmeas}
\hat{C}_\ell\rightarrow\hat{D}_\ell\equiv\frac{1}{2\ell+1}\sum_{m}d_{\ell m} d^*_{\ell m}=B_\ell^2 \hat{C}_\ell+\hat N_\ell
\end{equation}

From Eq.~\ref{eq:dlmeas} it is clear that $\hat{D}_\ell$ is a biased estimator of the true power spectrum $C_\ell$, $\langle \hat D_\ell\rangle=B_\ell^2 C_\ell+N_\ell$. In addition, the variance of the estimator takes an additional contribution. In presence of noise and beam effects, the variance becomes

\begin{equation}
\mathrm{var}[\hat{C}_\ell]\equiv\langle(C_\ell-\hat{C}_\ell)(C_\ell-\hat{C}_\ell)\rangle=\frac{2}{2\ell+1}\left(C_\ell+\frac{N_\ell}{B_\ell^2}\right)^2
\end{equation}

The variable $\hat{D}_\ell\equiv\hat{C}_\ell+\hat N_\ell/B_\ell^2$ still has a $\Gamma$ distribution, and all the considerations made for the $\hat{C}_\ell$ estimator still apply to the (slightly) more general case of noise and beam biases, provided that $\hat{C}_\ell$ is replaced with $\hat{D}_\ell$. More in detail, when both temperature and polarisation are considered, the matrix of estimators $\textbf{S}_\ell$ still has a Wishart distribution (see Eq.~\ref{eq:Wishart}) with a revised $\textbf{W}_\ell=\textbf{V}_\ell /(2\ell+1)$ matrix\footnote{We are dropping the $BB$ part of the distribution, as we have seen that the full Wishart in full-sky is separable into a $T-E$ and $B$ component. For the $B$ component, all considerations in the single-field regime apply.}

\begin{equation}
\textbf{W}_a=\frac{1}{2\ell+1}\left(
\begin{array}{ccc}
C_\ell^{TT}+N_\ell^T/(B_\ell^T)^2 & \ClTE & 0\\
\ClTE & C_\ell^{EE}+N_\ell^P/(B_\ell^P)^2 & 0\\
0 & 0 & C_\ell^{BB}+N_\ell^P/(B_\ell^P)^2
\end{array}
\right)\, ,
\end{equation}

where we have allowed for different (and uncorrelated) noise levels in temperature and polarisation, and for different beam sizes in temperature and polarisation.

The covariance matrix of the estimators $\hat{D}_\ell$ is equivalent to that of $\hat{C}_\ell$ in Eq.~\ref{eq:covW}, provided that $C_\ell^{TT}$, $C_\ell^{EE}$ are replaced with the power spectra corrected for noise and beam.

\subsection{Multi-frequency analysis}\label{sec:highl_multinu}
The results presented so far have been discussed assuming the special viewpoint of a single-frequency experiment. In reality, CMB experiments often rely on multi-frequency observations to better characterise the cosmological signal and extract it from the multi-component sky-signal observed (see discussions in e.g.~\cite{Aghanim:2015xee,Ade:2013kta}). Moreover, multiple detectors sharing the same central frequency are always available, so that the final signal at a certain frequency can be effectively thought as a weighted average of the signals observed with multiple detectors.

In general, if $n$ maps are available, there are $n-1$ combinations that are independent from the signal (if two maps share the same signal and have different noise properties, their difference is independent from the common signal). There is one independent combination defined as the weighted average of the $n$ available maps
\begin{equation}\label{eq:almavg}
a_{\ell m}^{NW}=\sum_i^n w_i a_{\ell m}^i
\end{equation}
where $w_i$ are the noise weights associated to each of the $n$ maps. In the simple scenario of isotropic noise $N_\ell$ for each map, the weights can be defined as $w_i=(N_\ell^i)^{-1}/\sum_i (N_\ell^i)^{-1}$. The noise-weighted map is a sufficient statistics for the CMB field, and therefore all the considerations above about the choice of the likelihood also apply to the combined map. An estimator for the power spectrum can be constructed from the noise-weighted map.

Another possibility is to build estimators $\hat{C}_\ell^{ij}=(1/(2\ell+1))\sum_{m} a_{\ell m}^i (a_{\ell m}^j)^*$ from pairs of maps and then define a weighted estimator $\hat{C}_\ell^{NW}=\sum_{ij}w_{ij}\hat{C}_\ell^{ij}$, where the weights $w_{ij}$ depend on the noise levels in the individual maps. It can be shown that the latter solution is equivalent to estimating the observed power spectrum from the noise-weighted map, and again all the considerations about the likelihood choice apply to this case as well (see discussion in Ref.~\cite{Hamimeche:2008ai}, Appendix C).

Regarding the latter solution, a more robust choice is to build the noise-weighted estimator $\hat{C}_\ell$ from cross-spectra only, i.e., from pairs $(ij)$ with $i\neq j$. If the noise in the individual maps is uncorrelated, to take cross-spectra is safe with respect to the introduction of possible biases in the final estimator due to unaccounted errors in the noise model\footnote{If $a_{\ell m}^i=s_{\ell m}^i+n_{\ell}^i$, with noise $n^i$ uncorrelated for any map $i=1,...,nmaps$ and signal $s^i$, then $\sum_m |a_{\ell m}^i (a_{\ell m}^j)^*|=\sum_m |s_{\ell m}^i (s_{\ell m}^j)^*+n_{\ell m}^i (n_{\ell m}^j)^*|=(2\ell+1)\hat{C}_\ell^{ij}$.}. However, the statistics of the estimator obtained from cross-spectra $\hat{C}_\ell^{CS}$ may differ from that of the generic noise-weighted estimator. In particular, the cross-spectra might not be positive-definite. Therefore, in principle one should use a distribution for $\hat{C}_\ell^{CS}$ other than the Wishart. However, it can be demonstrated (see e.g., Appendix C in Ref.~\cite{Hamimeche:2008ai}) that, in the limit of many maps available, the distribution of $\hat{C}_\ell^{CS}$ approaches that of $\hat{C}_\ell^{NW}$, and hence one can use the same approximations developed in the case of the generic noise-weighted estimator.

Before concluding this subsection, a note on the covariance matrix. When multiple maps are available and the estimators are build from a combination of those maps, the expression for the covariance matrix can be generalised as follows:
\begin{equation}\label{eq:cov}
\mathrm{Cov}\left[ {}^{ij}C_{\ell_1}^{XY}, {}^{ab}C_{\ell_2}^{WZ}\right]=\left\langle {}^{ij}C_{\ell_1}^{XY} \times {}^{ab}C_{\ell_2}^{WZ}\right\rangle-\left\langle {}^{ij}C_{\ell_1}^{XY}\right\rangle\left\langle {}^{ab}C_{\ell_2}^{WZ}\right\rangle
\end{equation}
where $ij,ab$ denote all possible combinations of pairs of maps, $XY,WZ$ are pairs of fields $T,E,B$, and we have explicitly taken into account the possibility of mode-coupling between $\ell_1,\ell_2$ (e.g., in the cut-sky regime). In the simple case of single-map in full-sky, Eq.~\ref{eq:cov} reduces to Eq.~\ref{eq:covW}.


\newpage
\section{Gravitational lensing}\label{sec:lensing}
In the discussions so far, we have implicitly assumed that the CMB fields are unlensed. This is not the case in the reality. CMB photons travelling from the last scattering surface to the observer feel the gravitational effects of the evolving structures in the Universe. This effect is analogous to the weak lensing effect observed in galaxy surveys, where images of source galaxies are distorted and magnified by foreground structures acting as gravitational lenses. In the CMB case, the CMB as emitted at the last scattering surface is the source and the whole distribution of total (cold and baryonic) matter along the line of sight acts as the foreground lens. In practice, this means that the anisotropy fields observed at a certain direction in the sky are displaced with respect to the original direction of emission:
\begin{equation}\label{eq:Xlens}
X(\hat{n})=X^\mathrm{unl}(\hat{n}+\nabla\phi(\hat{n}))
\end{equation}
where $X=T,E,B$ and $\phi$ is the lensing potential. The gradient of the lensing potential gives the deflection angle $\alpha=\nabla\phi$. The typical deflection that CMB anisotropies undergo is of order $2.5\,\mathrm{arcmin}$~\cite{Lewis:2006fu}. The lensing potential is given by the integrated contribution of the gravitational potential along the line of sight\footnote{The unperturbed line of sight in the Born approximation.}:
\begin{equation}
\phi(\hat{n})=-2\int_0^{\chi_*} d\chi W(\chi) \Psi(\hat{n})
\end{equation}
where $\Psi$ is the (Weyl) gravitational potential, $\chi$ is the conformal distance and $W(\chi)$ is a geometrical kernel.

Gravitational lensing preserves the total variance of the CMB fields, being a bare displacement of the anisotropy distribution. Very roughly speaking, if we extracted the CMB power spectra from small patches of the sky\footnote{Small enough that the convergence and shear components of the deflection field can be assumed constant in the patch.}, we would observe the acoustic peaks to shift to either smaller or larger scales with respect to the full sky average (see discussion in e.g.~\cite{Ade:2013tyw}). The net effect is a smoothing of the acoustic peak structure in the CMB power spectra that can be as high as 20\% in the case of the sharper structure in the $EE$ power spectrum with respect to the unlensed case. Another important effect is the generation of spurious (i.e., not primordial) $B$-modes from the lensing of primordial $E$-modes, with a power spectrum that resembles a white noise contribution with noise level of $\sim5\,\mathrm{\mu K\,arcmin}$ at $\ell<100$, representing a serious contaminant for searches of primordial $B$-modes. 

A detailed description of the effects of gravitational lensing on the CMB spectra is beyond the scope of this manuscript, and can be found in the excelent review by Lewis \& Challinor~\cite{Lewis:2006fu}. Here, it is relevant to stress that gravitational lensing modifies the statistical properties of the primary CMB fields in two ways. First, let's consider a \textit{fixed} lensing potential realisation and ensemble average over the CMB realisations. If we Taylor-expand Eq.~\ref{eq:Xlens}, take the harmonic expansion coefficients and compute the covariance of two fields $X,X'=T,E,B$, we get~\cite{Okamoto:2003zw,Hu:2001kj}
\begin{eqnarray}\label{eq:covlensed}
\langle x_{\ell m} x'_{\ell' m'}\rangle |_\mathrm{lens}&=&C_\ell^{xx'} \delta_{\ell\ell'}\delta_{m -m'} (-1)^m \nonumber \\
&+&\sum_{LM}(-1)^M\left(\begin{array}{ccc}
\ell &\ell' &L\\
m &m' &-M
\end{array}\right)f^\alpha_{\ell L\ell'}\phi_{LM}
\end{eqnarray}
where the term in brackets is the Wigner-3j and $f^\alpha$ is a weight of different $xx'$ pairs depending on the unlensed power spectra and on geometrical factors (a full list can be found in Ref.~~\cite{Okamoto:2003zw}). The $C_\ell$ in the first term of the RHS are the lensed power spectra. From Eq.~\ref{eq:covlensed}, it is clear that gravitational lensing not only modifies the structure of the primary CMB spectra by smearing the acoustic peaks, but also induces off-diagonal covariance terms (the second term in the RHS). Therefore, for a fixed lensing realisation, the CMB field becomes anisotropic. This property can be exploited to construct an estimator for the lensing potential $\phi$.

The lensing power spectrum $C^{\phi\phi}_L=(2L+1)^{-1}\sum_M\langle\phi_{LM}\phi_{LM}^*\rangle$ (where we are now taking the ensemble average over the lensing realisations as well) is related to the 4-point correlation function of the primary CMB fields, as it is clear by inspecting Eq.~\ref{eq:covlensed}. In other words, the second modification to the CMB statistics induced by lensing, when the ensemble average is taken over both the CMB and the lensing realisations, is a certain amount of non-Gaussianity measurable from the 4-point function. This property is exploited to construct an estimator for the lensing power spectrum.

The presence of gravitational lensing effects represent a pernicious contaminant for searches for primordial GW. In this case, ``delensing'' procedures (see e.g., ~\cite{Lewis:2006fu} and references therein for a description of delensing procedures) aimed to remove the lensing contamination from the measured CMB sky are crucial to allow for the possibility to detect the primordial tensor BB signal. On the other hand, the presence of gravitational lensing also enriches the amount of information that we can extract from the observations of the CMB sky. As an example, since the gravitational lensing is induced by forming structures, it carries information about the late time evolution of the Universe and can be exploited to constrain those cosmological parameters that govern those stages of the Universe expansion, such as massive neutrinos.

There are two ways in which the information encoded in the gravitational lensing signal can be accessed from CMB measurements. One can exploit the anisotropy and non-Gaussianity properties of the lensed CMB fields to construct estimators for the gravitational potential and for the lensing power spectrum, and use those observables directly in a likelihood analysis. This is what is done e.g., by the Planck collaboration~\cite{Aghanim:2018oex}, ACT~\cite{Sherwin:2016tyf}, SPT~\cite{Omori:2017tae}, POLARBEAR~\cite{Ade:2013hjl}, and this is the goal of future CMB experiments that will be able to reconstruct the lensing signal over a large range of angular scales with exquisite sensitivity (CORE~\cite{Challinor:2017eqy}, Simons Observatory~\cite{Ade:2018sbj}, CMB-S4~\cite{S4}, PICO~\cite{Hanany:2019lle}). Concerning the choice of the likelihood function for the lensing signal itself, it has been shown~\cite{Schmittfull:2013uea} that a quadratic expression in the observed lensing power spectrum with a non-diagonal fiducial covariance matrix provides satisfying results. 

As seen in Eq.~\ref{eq:covlensed}, cosmological information carried by the gravitational lensing signal are also encoded in the CMB power spectra. In fact, the lensing contribution modifies the primary acoustic structure as emerged from last scattering. Current generation of Boltzmann solvers usually employed in cosmological analysis, such as CAMB~\cite{CAMB} and CLASS~\cite{CLASS}, carefully compute the lensing effect when constructing the theoretical spectra to be compared with the measured spectra in the likelihood analysis. In general, when lensed spectra are considered, the likelihood analysis should reflect the different statistical properties of the lensed CMB. First of all, the non-Gaussian distribution of the lensed field may require to build a different likelihood function for the lensed fields. Even if the Gaussian approximation can be retained given the current experimental sensitivity, the covariance matrix may still need to reflect the presence of off-diagonal correlations induced by the lensing signal~\cite{2012PhRvD..86l3008B}. It has been shown that ignoring such correlations can still be safe for the sensitivity level of Planck \cite{2012PhRvD..86l3008B,Lewis:2005tp}, but it will become relevant for the next generation of experiments~\cite{Motloch:2017rlk,Motloch:2016zsl}.

A final remark concerns cosmological analysis employing the combination of CMB power spectra and the lensing power spectrum. These two data sets are usually treated as independent. However, they are extracted from the same map and, as such, there is a certain level of correlation between the two. As pointed out in Ref.~\cite{Schmittfull:2013uea}, there are two sources of correlations: cosmic variance in the CMB field, which may affect the lensing reconstruction; cosmic variance in the lensing field, which affects not only the lensing reconstruction, but can propagate to the lensed CMB power spectra.
An alternative solution that might reduce the correlation between fields is the joint analysis of the lensing power spectrum and unlensed CMB spectra. The reconstructed lensing signal -- either from CMB observations themselves or from tracers of large-scale structure correlated with the lensing signal such as the cosmic infrared background (CIB) -- can be used to delens the primary CMB spectra. This procedure is not only key to unveil the primordial $BB$ tensor spectrum, but it can also improve the sensitivity to those cosmological parameters that are mostly constrained via measurements of the high-$\ell$ damping tail in the $TT,TE,EE$ spectra, such as $N_\mathrm{eff}$~\cite{S4,Green:2016cjr}.

All the above concerns will become much more significant for the next generation of experiments, when more sensitive polarisation-based reconstructions will be available.
\newpage


\section{Likelihood approaches -- Large-scale regime}\label{sec:lowl}
In this section, we review the basics of the likelihood approaches at large scales (low multipoles). We consider two generic classes of likelihood methods: pixel-based and simulation-based. When the focus of the likelihood analysis is the sky at large angular scales, the resolution of the map to be analysed is low enough to make a pixel-based approach computationally feasible. Alternative approaches exploit the information encoded in harmonic space, and build the likelihood function from a simulation-based method or component-separation based method (Blackwell-Rao). For large scale studies the Hamimeche-Lewis likelihood can be also considered,\cite{Vanneste:2018azc,Aghanim:2016yuo}, for completeness we will explore later the performance of such likelihood compared with other approximations.

\subsection{Pixel-based approach}\label{subsec:pixel-based}
The great advantage of the pixel-based approach lies in the fact that the likelihood function so defined is always exact (see Sec.~\ref{sec:CMBstat_real}), including in the cut-sky regime. Equation~\ref{eq:pixel_like} can be easily generalised in the presence of (Gaussian-distributed) noise, by defining the data vector as $\mathbf{m}=\mathbf{s}+\mathbf{n}$, where $\mathbf{s}$ is the signal per pixel in temperature and polarisation ($\mathbf{s}=(T,Q,U)$) and  $\mathbf{n}$ is the instrumental noise. We report here the likelihood function in pixel space for convenience:
\begin{equation}\label{eq:pixel_like2}
\mathcal{L}(C_{\ell}) = \mathcal{P} (\mathbf{m}|C_{\ell}) = \dfrac{1}{2 \pi |M| ^{1/2}} \exp{\left( -\dfrac{1}{2} \mathbf{m}^{\text{T}} M ^{-1} \mathbf{m} \right)} \, ,
\end{equation}
The full covariance matrix $M$ in Eq.~\ref{eq:pixel_like2} is consequently generalised to the sum of the signal and noise covariance matrices $M=\mathbf{S}+\mathbf{N}$. Furthermore, the effect of beam smearing discussed in Sec.~\ref{sec:highl_noise} -- and also relevant for the large-scale data -- is now taken into account when constructing the full covariance matrix in terms of the beam-weighted sum of Legendre polynomial (see Eq.~\ref{eq:covmat}). For example, the explicit expression for the weight $P_{\ell} ^{TT}$ for temperature becomes

\begin{equation}\label{eq:PellBell}
(P_{\ell}^{TT})_{i,j} = \dfrac{2 \ell +1}{4 \pi} \, B_{\ell}^2 \, P_{\ell} (\hat{\mathbf{n}}_i \cdot \hat{\mathbf{n}}_j)
\end{equation}
where $\hat{\mathbf{n}}_i$ is the unit vector pointing towards the $i$th pixel, $B_{\ell}$ is given by the product of the instrumental beam Legendre transform and the (HEALPix~\cite{Gorski:2004by}) pixel window~\footnote{The signal in each pixel is the average over the signal at each point within the surface area of the pixel $\Omega_p$, $f_p=\int d\Omega w_p f(\Omega)$, where $w_p=1/\Omega_p$ is the weight inside the pixel and zero otherwise. The harmonic transform of the weight $w_p$ is the pixel window function. Including the full dependence of the window function on $m,\ell$ can be computationally demanding. However, if the size of the pixel is small compared to the angular resolution of the experiment, the $m$-dependence can be neglected and the $m$-averaged window function $w_\ell$ can be defined. The power spectra computed from the pixelised maps $C_\ell^p$ are related to the unpixelised spectra $C_\ell$ via $C_\ell^p=w_\ell^2 C_\ell$.}, and $P_{\ell}$ is the Legendre polynomial of order $\ell$. 

As already noted above, the actual feasibility of this mathematically rigorous approach only applies to the very large scales. Even so, massive parallel coding and memory requirements could still be a necessary ingredient. As an example, the Planck collaboration employed this approach in the low-$\ell$ likelihood analysis in temperature and polarisation for the 2015 data release~\cite{Aghanim:2015xee}. The map resolution was fixed at $N_\mathrm{side}=16$ (for comparison, the analysis conducted by the WMAP team employed $N_\mathrm{side}=8$ maps) to accommodate the $\ell<47$ multipoles in the analysis, resulting in $N_\mathrm{pix}=3\times3072=9216$ total number of pixels, further reduced by the application of the analysis mask. In evaluating the likelihood function in Eq.~\ref{eq:pixel_like2}, the data vector and the noise covariance matrix are fixed, while the signal covariance matrix is recomputed for any given cosmological model to be compared against data, following Eq.~\ref{eq:covmat}. In practice, only a subsection of $\mathbf{S}$ is recomputed, in particular that subsection corresponding to $\ell<30$. The portion of $\mathbf{S}$ corresponding to multipoles $30\le\ell<47$ is precomputed from a fixed fiducial model. The choice of the fiducial model does not affect the performance of the likelihood. In fact, at the resolution employed in the large-scale analysis, the sensitivity to multipoles above $\ell\sim30$ is strongly suppressed.

In 2013, a hybrid approach coupling a MonteCarlo-based approach in temperature (Blackwell-Rao estimator, see Sec.~\ref{subsec:blackwell-rao}) to a pixel-based approach in polarisation for $\ell<23$ was adopted~\cite{Ade:2013kta}. In order to speed up the evaluation of the fully pixel-based likelihood function for the 2015 release at any given cosmological model, the ``brute-force'' approach described here has been optimised with the implementation of the Sherman-Morrison-Woodbury formula, described below.
 
\subsubsection{Sherman-Morrison-Woodbury formula}
The computational cost required by the pixel-based approach can be dramatically reduced if one consider that only a portion of the signal covariance matrix is reconstructed at any evaluation of the likelihood function. The full covariance matrix can then be decomposed into a varying part, which is function of the theoretical $C_\ell$ (i.e., the power spectra of the theoretical models to be compared against data), and a fixed part given by the fiducial $\mathbf{S}$ and the noise covariance matrix. The following step is to further decompose the varying part of the covariance as $S=V^T A V$, via a transformation $V$ that effectively reduces the dimension of the actual evaluation cost from a $N_\mathrm{pix}\times N_\mathrm{pix}$ inversion to a $n_\lambda\times n_\lambda$ inversion, where $n_\lambda=2\ell+1$ is the dimension of the transformed matrix $A$. The latter is the only matrix that depends on the theoretical $C_\ell$ and, therefore, it is the only matrix to be recomputed and inverted. The fixed portions of the covariance matrix as well as the transformation matrix $V$ can be pre-computed and stored. For the Planck 2015 data release, the application of such mathematical formalisms allowed to speed-up the likelihood evaluation by an order of magnitude. The mathematical details leading to the application of the Sherman-Morrison-Woodbury formula can be found in Ref.~\cite{Aghanim:2015xee} (Appendix B.1) and Ref.~\cite{Hinshaw:2006ia} (Appendix A.1).

\subsection{Blackwell-Rao estimator}\label{subsec:blackwell-rao}
An alternative approach to the likelihood evaluation that overcomes the computational cost of an exact likelihood evaluation in pixel space is provided by the combination of the Gibbs sampling method with the Blackwell-Rao estimator. The Gibbs sampling is a MCMC method applied to the estimation of the observed CMB signal from a raw map containing signal, noise, and foreground contamination in a Bayesian framework~\cite{Eriksen:2004ss,Jewell:2002dz,Wandelt:2003uk}. The crucial output for the subsequent construction of the likelihood estimator is a set of samples of the CMB sky, or more precisely, a set of sample variances of the sky samples obtained with the Gibbs algorithm. The Blackwell-Rao estimator is then built as an average over the set of sample variances.

Let's assume that the observed map $\mathbf{m}$ is composed by the CMB signal $\mathbf{s}$ and noise $\mathbf{n}$ (the following procedure can be generalised to the case in which foreground $\mathbf{f}$ also contribute to the total signal, see e.g.~\cite{Eriksen:2007mx}). What we really want to evaluate is the joint probability of having a certain $\mathbf{s}$ with a certain $C_\ell$ given the observed sky $\mathbf{m}$, i.e., we want to evaluate $\mathcal{P}(\mathbf{s},C_\ell|\mathbf{m})$. A brute force evaluation by computing a grid of $\mathbf{s}$ and $C_\ell$ is computationally prohibitive (it is even more prohibitive once one considers the inclusion of foreground into the equation). Another approach is to sample directly from the distribution via specific algorithms. Although it is again computationally unfeasible to sample directly from the joint distribution, as it would require inverting large and dense covariance matrices, it has been proved that the joint distribution can be reconstructed by iteratively sampling over the \textit{conditional} distributions $\mathcal{P}(\mathbf{s}|C_\ell,\mathbf{m})$ and $\mathcal{P}(C_\ell|\mathbf{s},\mathbf{m})$. In fact, the conditional distributions are known. They are a multivariate Gaussian (posterior distribution of a Wiener filter) for $\mathcal{P}(\mathbf{s}|C_\ell,\mathbf{m})$ and an inverse Gamma distribution for $\mathcal{P}(C_\ell|\mathbf{s},\mathbf{m})$. As for the foregrounds, their marginal distribution does not usually have an analytic representation. However, it can be reconstructed numerically~\cite{Eriksen:2007mx}. The iterative sampling is obtained with the implementation of a specific MCMC sampling algorithm, the Gibbs sampling, and follows these steps (we omit foregrounds for simplicity): 
\begin{itemize}
\item i) start from a guess power spectrum $C_\ell^0$; 
\item ii) draw a sample $\mathbf{s}^1$ from $\mathcal{P}(\mathbf{s}|C_\ell^0,\mathbf{m})$; 
\item iii) draw a sample $C_\ell^1$ from $\mathcal{P}(C_\ell|\mathbf{s}^1,\mathbf{m})$; 
\item iv) cycle over step ii) and iii) until convergence is reached. 
\end{itemize}
This algorithm lays the ground for the subsequent evaluation of the posterior $\mathcal{P}(C_\ell(\theta)|\mathbf{m})$, where $\theta$ is a set of cosmological parameters of a theoretical cosmological model. In principle, one could reconstruct a histogram of the sampled $C_\ell$ from the Gibbs sampling and use the histogram to interpolate for a given theoretical model. However, this procedure is not efficient. Luckily, the Gibbs sampling allows to construct an efficient and arbitrarily exact estimator of the likelihood function, the so called Blackwell-Rao estimator. The basic idea~\cite{Wandelt:2003uk,Chu:2004zp} is to note that the $C_\ell$ only depend on the CMB signal and not on the total sky signal (i.e., once we know the CMB sky, there is no additional information coming from the knowledge of other components), $\mathcal{P}(C_\ell|\mathbf{s},\mathbf{m})\propto\mathcal{P}(C_\ell|\mathbf{s})$. In addition, the $C_\ell$ depend on the CMB signal only through its variance, i.e., $\mathcal{P}(C_\ell|\mathbf{s})\propto\mathcal{P}(C_\ell|\hat{C}_\ell)$, where $\hat{C}_\ell$ is the power spectrum of the sample CMB map $\mathbf{s}$. Note that we are already familiar with the probability distribution $\mathcal{P}(C_\ell|\hat{C}_\ell)$ (see Eq.~\ref{eq:Wishart}). At this point, we can write down the following chain of equivalent probability integrals (we omit the dependence of $C_\ell$ from $\theta$ for brevity):
\begin{subequations}
\begin{eqnarray}
\mathcal{P}(C_\ell|\mathbf{m})&=&\int\mathcal{P}(C_\ell,\mathbf{s}|\mathbf{m})d\mathbf{s}\\
&=&\int\mathcal{P}(C_\ell|\mathbf{s},\mathbf{m})\mathcal{P}(\mathbf{s}|\mathbf{m})d\mathbf{s}\\
&=&\int\mathcal{P}(C_\ell|\hat{C}_\ell)\mathcal{P}(\hat{C}_\ell|\mathbf{m})d\hat{C}_\ell\\
&\sim&\frac{1}{N_\mathrm{Gibbs}}\sum_i^{N_\mathrm{Gibbs}} \mathcal{P}(C_\ell|\hat{C}_\ell^i) 
\end{eqnarray}
\end{subequations}
where $N_\mathrm{Gibbs}$ is the number of Gibbs samples evaluated in the component separation analysis, and $\hat{C}_\ell^i$ is the power spectrum of the i-th sample. In other words, the posterior on the cosmological parameters of interest given the observed data can be obtained by averaging the individual posteriors over the Gibbs samples, that can be stored during the Gibbs implementation. The evaluation can be made more accurate by increasing the number of samples. It has to be noted that one only needs to store the variance of the sampled CMB maps $\mathbf{s}^i$ up to a given multipole $\ell_\mathrm{max}$, i.e., it is not necessary to store the much more memory-demanding samples themselves. 

The accuracy of the Blackwell-Rao estimator depends on the number of samples required to reach convergence. The number of samples grows very steeply with the $\ell_\mathrm{max}$ considered in the analysis~\cite{Chu:2004zp}. This limitation makes the Blackwell-Rao estimator more suited for the likelihood analysis of CMB large scales (low multipoles). Modified version of the Blackwell-Rao estimator have been proposed that allow to reduce the number of samples required for convergence, and therefore allow to extend the viability of this approach to higher $\ell_\mathrm{max}$~\cite{Rudjord:2008vc}.

The Gibbs method has been employed by the Planck collaboration as a component separation method to obtain maps of the CMB temperature signal and foreground contaminants~\cite{Akrami:2018mcd,Ade:2015qkp,Ade:2013crn}. The Blackwell-Rao estimator, based on the Gibbs samples so obtained, has been employed by the Planck collaboration as an alternative likelihood method for the analysis of the temperature data at large scales~\cite{Aghanim:2015xee,Ade:2013kta}. It has been employed by the WMAP collaboration for the likelihood analysis of the large-scale ($\ell<32$) temperature data~\cite{WMAP,Larson:2010gs}.

\subsection{Simulation-based approach}\label{subsec:simulation-based}

In some cases accurate estimate of the noise are not available and/or the probability distribution of residual systematic effects, in map space, is not perfectly Gaussian, and therefore the probability distribution, in map or harmonic space, cannot be expressed analytically and it needs to be learned from simulations.

We describe in this section a simulation-based approach to evaluate the likelihood function in the low-multipole regime. The likelihood distribution is evaluated through realistic simulations of CMB, noise and possible residual systematics. 
We report here the main steps:
\begin{itemize}
\item the initial step is the simulation of $n$ theoretical auto- or cross- power spectra, related to $n$ different cosmological models, represented by a set of cosmological parameters $\theta _j$.

\item For each theoretical power spectrum $C_{\ell}^{\text{th},i}$, $m$ CMB maps are produced, including noise and other residual contaminants.

\item For each map, the corresponding power spectrum is evaluated. Therefore, $k = n \times m$ new simulated power spectra $C_{\ell}^{\text{sim},k}$ are obtained.

\item By histogramming the $k$ simulated power spectra $\ell$-by-$\ell$ and $\theta _j$ by $\theta _j$, the probability $\mathcal{P}(C_{\ell}^{\text{sim}}|C_{\ell}^{\text{th}})$ is built empirically. In evaluating this probability, it is necessary to interpolate it with a suitable function, in order to smooth the scatter due to the limited available number of simulations. In particular, one can define a low-order polynomial interpolation function $f^i_{\ell}(C_{\ell}^{\text{sim}}, C_{\ell}^{\text{th}})$ of the logarithmic of the $C_{\ell}^{\text{sim}}$ histogram for each $i$-th initial power spectrum, such that

\begin{equation}\label{eq:sim1}
f^i_{\ell}(C_{\ell}^{\text{sim}}, C_{\ell}^{\text{th}}) \simeq \log \mathcal{P}(C_{\ell}^{\text{sim}}|C_{\ell}^{\text{th}}) \, .
\end{equation}

\item Starting from this approximation, the likelihood function for the observed power spectra $C_{\ell}^{\text{data}}$ can be finally evaluated. 
The $n$ couples $(C_{\ell}^{\text{th},i},f^i_{\ell}(C_{\ell}^{\text{th}}, C_{\ell}^{\text{data}}))$ can be considered as a tabulated version of the log of the joint probability $\log \mathcal{P}(C_{\ell}^{\text{data}},C_{\ell}^{\text{th}})$. The joint probability can then be interpolated with a suitable low-order polynomial (as done above) for each multipole,  $g_{\ell}(C_{\ell}^{\text{data}}, C_{\ell}^{\text{th}})$, such that 
\begin{equation}\label{eq:sim2}
g_{\ell}(C_{\ell}^{\text{data}}, C_{\ell}^{\text{th}}) \simeq \log \mathcal{P}(C_{\ell}^{\text{data}}|C_{\ell}^{\text{th}}) \, . 
\end{equation}
The sum in the multipole range provides the approximation for the log-likelihood, up to a constant,
\begin{equation}\label{eq:sim3}
\log \mathcal{L} (\textbf{C}^{\text{th}}|\textbf{C}^{\text{data}}) \simeq \sum _{\ell = \ell_{\text{min}}} ^{\ell _{\text{max}}} g_{\ell}(C_{\ell}^{\text{data}}, C_{\ell}^{\text{th}}) + cost
\end{equation}

\end{itemize}

This approach has been used in the latest analysis of the Planck collaboration~\cite{Aghanim:2016yuo,Aghanim:2018eyx,planckL05} to evaluate the low-multipoles polarization likelihood.

We note that this likelihood approach requires the data to be provided in harmonic space as power spectra $\textbf{C}^{\text{data}}$. A good estimator for the power spectrum at large scales is obtained via the quadratic maximum likelihood (QML) method. We provide a description of the method in Sec.~\ref{sec:qml}.

As a final remark, we note that the simulation-based approach is very similar in spirit to the class of methods known as Approximate Bayesian Computation (ABC). ABC methods have been initially employed in various fields as a way to bypass the evaluation of the likelihood function with the use of simulated data~\cite{pritchard1999population,beaumont2002approximate,marjoram2003markov,sisson2007sequential,toni2009approximate,beaumont2009adaptive,mckinley2009inference,beaumont2010approximate,liepe2014framework,turner2012tutorial}. Recent applications to astrophysics and cosmology include~\cite{Akeret:2015uha,Alsing:2019dvb,Alsing:2019xrx,Leclercq:2018who,Alsing:2018eau}. Similarly to ABC, the simulation-based approach approximates the likelihood by comparing simulated data with observed data using  the QML approximations to the ML points of the spectra as a distance metric. However, while the likelihood approximation in ABC is usually done by discarding mismatching simulations, the simulation-based approach described here does so by fitting the constructed histogram and then slicing.


\newpage
\section{Foreground contamination: modelling and component separation}\label{sec:foreground}
We have neglected so far the possibility that the microwave sky contains more components than the CMB. In realistic observations, one has to take into account the fact that the observed signal at a given frequency is a combination of the CMB signal plus additional emissions from so called foreground contaminants. For the range of wavelengths of interest to CMB observations (from few tens to few hundreds GHz), the most relevant contaminants are atmospheric emission (for ground-based experiments in particular) and astrophysical foreground emission. The latter includes Galactic dust, synchrotron and free-free emission and extragalactic contaminants such as clustered and Poisson CIB emission, radio point sources, molecular lines; see e.g.~\cite{2013A&A...553A..96D,2010ApJ...709..920S} for a description of the multi-component microwave sky and way to create synthetic maps, and~\cite{10.1093/ptep/ptu065} for a concise review. The temperature sky is much more composite than the polarisation sky. Nevertheless, CMB temperature signal is the dominant component over a wide range of frequencies and angular scales. On the other hand, polarised foreground can easily dominate over the cosmological signal, especially in the case of CMB $B$-modes.

To reduce the contamination from atmospheric emission, ground-based CMB experiments are usually placed in specific sites, at high altitudes (to reduce the thickness of the atmosphere above the telescope) and dry locations. Residual contamination can be further removed either by detector pair-differencing (see e.g.~\cite{Ade:2018gkx}) or by modulation of the signal (see e.g.~\cite{Takakura:2017ddx}). Balloon-borne and especially satellite missions are less or not at all concerned with atmospheric contaminations. Emission from foreground sources is instead a common issue to all CMB observatories. Of course, it is necessary to account for such foreground contaminations in a proper likelihood analysis, i.e., to decompose the observed map into the individual components. This process is called component separation; various component separation methods exist. They differ for the domain of applicability (pixel space, harmonic space, wavelet space), and for the way data are described. For example, CMB and foregrounds can be parametrized as frequency spectra (parametric methods), can be separated imposing certain conditions to the various components modelled as arbitrary templates (non-parametric), can be separated with no a-priori knowledge of the individual components (blind). A rather complete list of component separation methods can be found in the LAMBDA archive~\footnote{Component separation methods and related references can be accessed at the following url:~\url{https://lambda.gsfc.nasa.gov/toolbox/tb_comp_separation.cfm}}. Here, we cite a few examples that have been also applied to the Planck analyses: ILC~\cite{Bennett1992} (pixel space, internal linear combination of frequency components), SEVEM~\cite{Leach:2008fi} (pixel space, template-based), NILC~\cite{Delabrouille:2008qd} (needlet space, internal linear combination), COMMANDER~\cite{Eriksen:2005dr} (map domain, parametric), SMICA~\cite{Cardoso:2008qt} (harmonic space, non parametric).

After component separation, the cleaned CMB maps may still be affected by residual contamination, which have to be further taken into account. At small scales, this is achieved in various steps, see e.g., the Planck analysis~\cite{Aghanim:2015xee}. First, regions of the sky that show an excess contamination from foreground emission are masked away from the analysis. Secondly, the residual contamination in the remaining areas can be modelled at the power spectrum level via specific templates. In other words, one can exploit the fact that the harmonic shape of the foreground is different from the CMB power spectra, and the fact that the spectral dependence of the foreground emission is also different from the CMB spectral dependence. The data vector to be fed in the likelihood function would therefore contain both the CMB signal and the residual contamination, as obtained from observations of the real sky. At the same time, the theoretical spectra to be compared with the data would be given by the sum of the theoretical CMB spectra \textit{and} the foreground templates. It is crucial that multi-frequency channels are available in order to efficiently fit for both the CMB and the foreground contaminants simultaneously. 

The above prescription implicitly assumed that all the information about foreground emission in harmonic space is fully captured by their power spectrum. However, there is no reason to believe that foreground are Gaussian distributed, and indeed they are not. The assumption that is usually made is that most of the non-Gaussian contribution is removed from the maps once the foreground-saturated regions are masked away. The non-Gaussianity of the remaining contaminations can be (and actually are) neglected at the likelihood level. This assumption is demonstrated to be reasonably accurate via dedicated simulations (see e.g., discussions in~\cite{Aghanim:2015xee,Ade:2013kta}) for the current generation of cosmological surveys. At the same time, a huge effort is devoted to the study of the propagation of uncertainties in the foreground modelling and removal to the final constraints on cosmological parameters in the context of the high-sensitivity next generation CMB surveys (see e.g.~\cite{S4}). 

Foreground contaminations are of course present at large angular scales as well, and must be taken into account when realistic data are analysed. Taking the results from the Planck collaboration as an example~\cite{Aghanim:2015xee,Ade:2013kta}, the foreground treatment at large scales is somehow different from the prescription described in the case of small scales. At low multipoles in temperature, the cleaned CMB map is obtained via Gibbs sampling (see Sec.~\ref{subsec:blackwell-rao}) implemented in COMMANDER and the bright region along the Galactic plane is further masked for likelihood analysis (~7\% of the sky). In polarisation, the foreground-cleaning is implemented via template-fitting. Residual contamination can be taken into account in the construction of the noise covariance matrix to be employed in the likelihood function.

To conclude this section, we would like to enphasize that the topic of foreground modelling and component separation is extremely wide and the list of references reported in this manuscript is far from being exhaustive. The interested reader is encouraged to consider these references as a mere starting point. Further details can be accessed from the specific collaboration papers that describe the corresponding data processing, and from the overview papers of science forecasts by upcoming collaborations that describe their simulated pipelines of data reduction.


\newpage
\section{Comparing likelihood performance}\label{sec:comparison}
In this section, we compare the performance of various likelihood approaches introduced in Sec.~\ref{sec:highl} for the high-$\ell$ regime and in Sec.~\ref{sec:lowl} for the low-$\ell$ regime. In particular, we are interested in testing the different likelihood performance with respect to the ability to produce unbiased constraints on cosmological parameters. Where relevant, we also compare the different performance in terms of computational costs. In what follows, we first discuss the comparison of likelihood approaches in the small-scale regime (Sec.~\ref{sec:comparison_highl}), and then move to the presentation of the results in the large-scale regime (Sec.~\ref{sec:comparison_lowl}). 

\subsection{Small-scale regime -- Full sky}\label{sec:comparison_highl}
In this section we test the performance of the different likelihood approximations at high-$\ell$ on simulated data and compare the results with the assessment presented in Sec. \ref{subsec:compexac}, which was based on analytic arguments. We mainly want to show whether adopting a particular power spectrum likelihood approximation, when estimating cosmological parameters, may introduce a bias in the parameters recovered values or/and a misestimation of their associated error bars. In the following, we neglect the impact of foregrounds, and we consider CMB plus noise full-sky maps, for which we derive the temperature and polarization angular power spectra following Eq. \ref{eq:clall}. 

The simulated dataset consists of CMB plus noise maps. Specifically, we generate a set of $1000$ maps of the CMB sky, ${\mathbf m}=(T, Q, U)$, drawn as Gaussian random realizations of fiducial temperature and polarization power spectra, that correspond to a set of given cosmological parameters. The full-sky maps are generated at a resolution of $3.4$ arcmins ($N_{side}=1024$ in the HEALPix scheme \cite{Gorski:2004by}) and smoothed with a symmetric Gaussian beam of FWHM $10$ arcmins~\footnote{Roughly speaking, the resolution identifies the level of the discretization (i.e., the number of pixels). The beam smoothing simulates the angular resolution of the experiment, and exponentially suppresses the scales below the beam width. Note that the beam width is larger than the size of the pixel.}. To each of these maps we add a simulated realization of white isotropic noise corresponding to a noise level in polarization of $\sigma_n=1 \mu{\mathrm K}$ arcmin, which is in the ballpark of what is expected from future CMB experiments. In this way temperature anisotropy maps are signal dominated across almost all the multipoles that are relevant for primary anisotropies, up to $\ell \lesssim 2000$, compatibly with present and forthcoming measurements. 

\subsubsection{Temperature-only results}

Focusing on temperature alone, we adopt the different likelihood approximations of Sec. \ref{sec:highl} to estimate the cosmological parameters from the power spectra of the simulations. For simplicity we only fit for two parameters, specifically $A_s$ and $n_s$, that are respectively the amplitude and the spectral index of the power spectrum of primordial density fluctuations. We generate a grid of model $C_{\ell}\,$'s keeping all the other $\Lambda$CDM parameters fixed, while letting $A_s$ and $n_s$ vary in a broad range of values around the input fiducial model, $C_{f\ell}$. Given the angular power spectrum of each simulation, $\hat{C}_{\ell}$, we evaluate the likelihood of the models at each multipole, $\mathcal{L} ({C}_{\ell} )$, accounting for the noise contribution as described in Sec. \ref{sec:highl_noise}. Since here we are considering the small scale regime, the total likelihood for the set of cosmological parameters $\mathbf{\theta} = (A_s, n_s)$ is obtained by summing the log-likelihoods over the range of multipoles $\ell= [30, 2000]$: $\mathrm{ln}\mathcal{L} (\mathbf{\theta}) \propto \sum_{\ell}\mathrm{ln}\mathcal{L} ({C}_{\ell})$. 

\begin{figure}[!t]
\includegraphics[width=1.\textwidth]{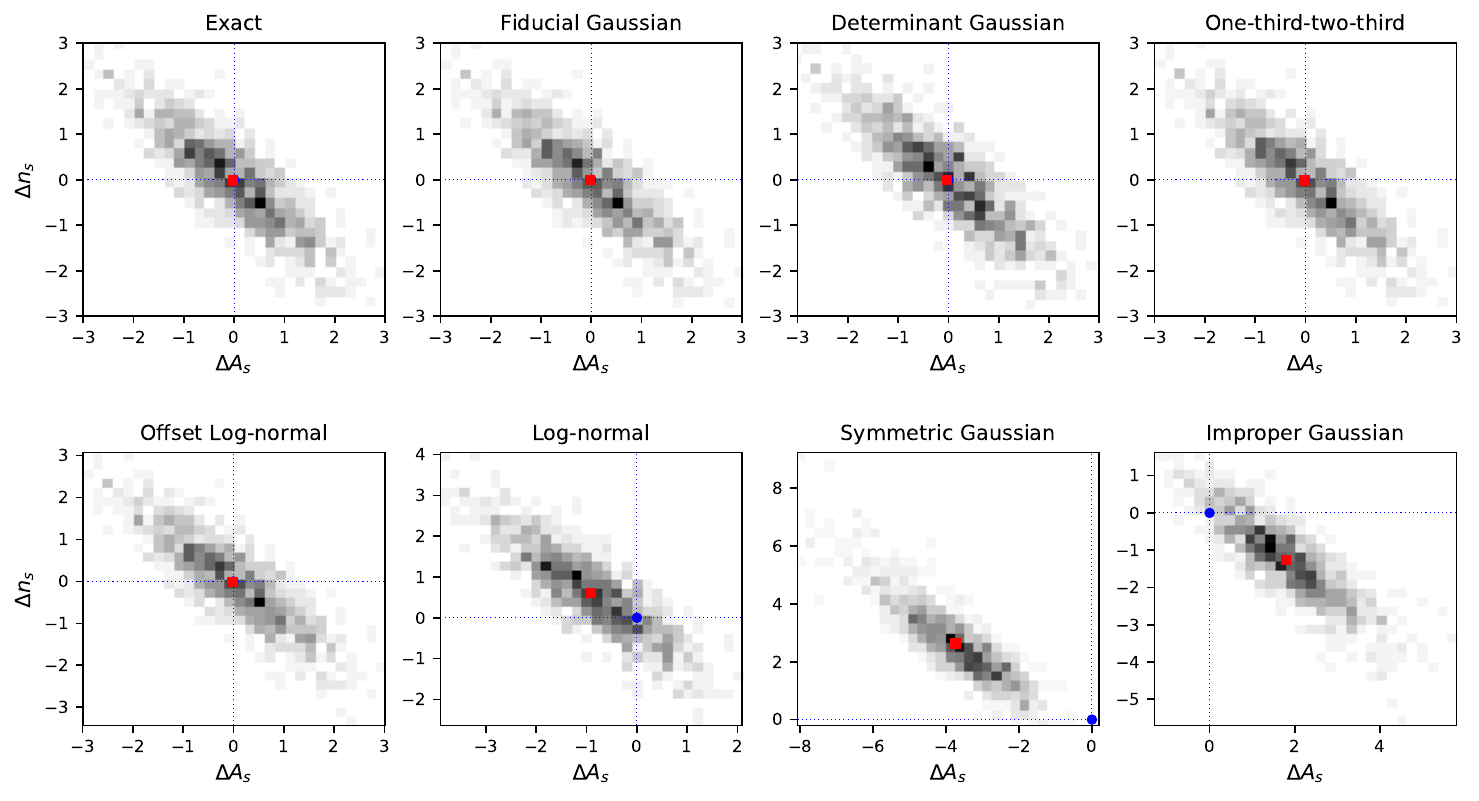}
\caption[]{\small{$2$D histograms of the biases on the parameters expectation values estimated from $1000$ simulated full-sky maps of CMB temperature plus noise. As described in more detail in the text, the biases are computed with respect to the input parameter values and normalized to the $1\,\sigma$ uncertainty associated to the parameter, $\Delta p = (\hat p -p^{\mathrm in})/ \sigma_p$, where $p$ can be either $A_s$ or $n_s$. The blue circle indicates a null bias, while the red square is the centre of the distribution. Note that for the cases plotted in the first row the two almost completely overlap.}}
\label{fig_2par_fullsky_temp}
\end{figure}

Once we have the total likelihood, assuming flat priors on parameters, we can estimate the expectation value of each parameter following Eq. \ref{eq:param_expectation}. By building histograms of the parameter values obtained from the 1000 simulated maps, we can reconstruct the posterior distribution of each parameter in a ``frequentist'' fashion. In particular, since the simulations input parameters are known, we can study the distribution of the biases in the estimated parameters, defined as the shifts with respect to the input fiducial values, normalized to the $1\,\sigma$ marginal error of the posterior distribution. For each simulation, the bias reads $\Delta p = (\hat p -p^{\mathrm in})/ \sigma_p$, where $p$ can be either $A_s$ or $n_s$. 
The distributions we derived are shown in Figure \ref{fig_2par_fullsky_temp} for the different likelihood approximations. They should be centered in zero, if there is no bias in the mean recovered parameters value, and they should have unit variance, if $\sigma_p$, as derived from the standard bayesian analysis, is consistent with the \textit{true} error. 

\begin{figure}[!t]
\centering
\includegraphics[width=0.48\textwidth]{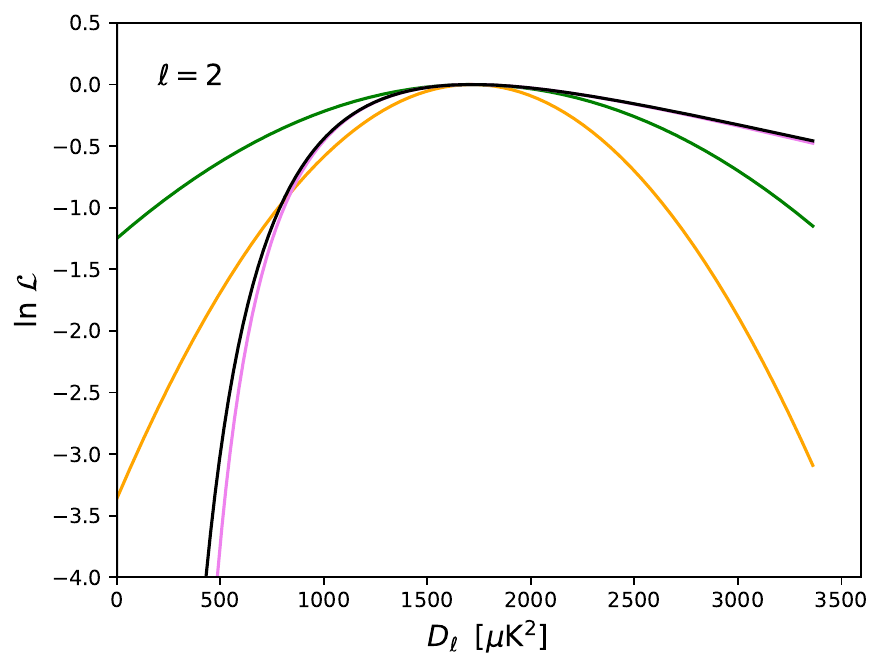}
\includegraphics[width=0.48\textwidth]{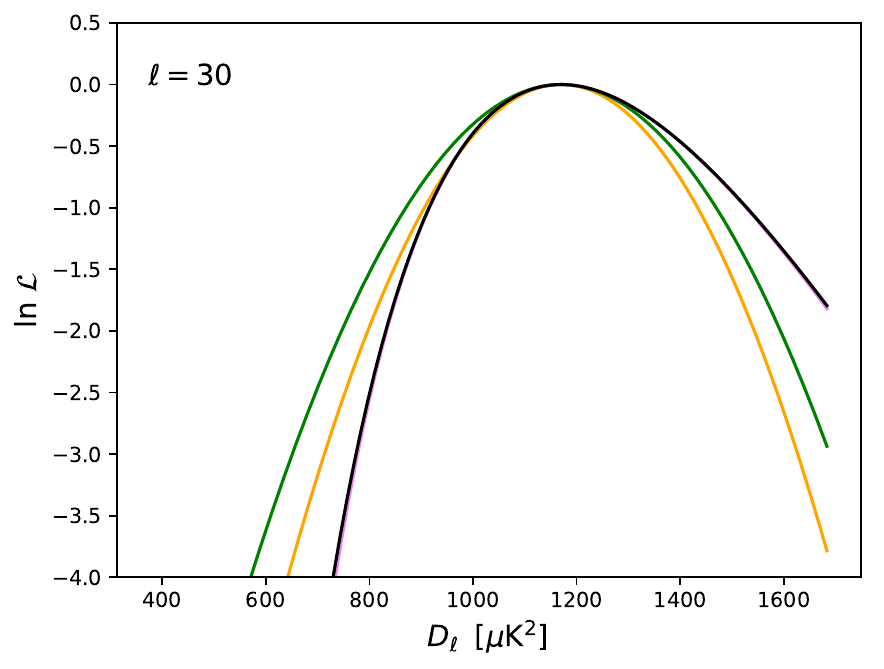}\\
\includegraphics[width=0.48\textwidth]{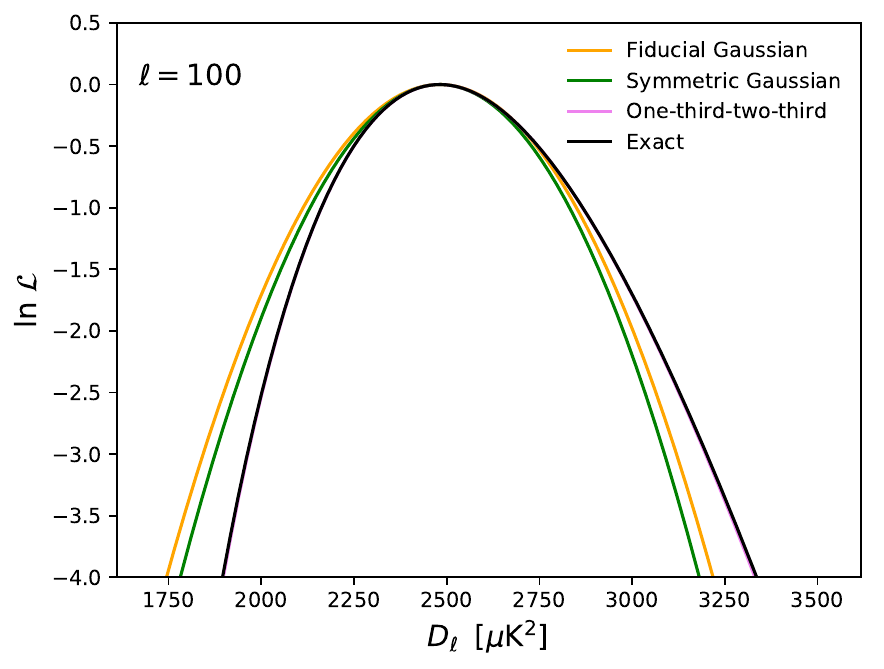}
\caption[]{\small{Comparison of some of the likelihood approximations for different multipole regimes, as marked in the plots, where the x-axis corresponds to $D_{\ell}=\ell(\ell+1)C_{\ell}/(2\pi)$}.}
\label{fig_slices}
\end{figure}

Consistently with what has been presented in Sec. \ref{subsec:compexac}, we find that the Log-Normal together with the symmetric and improper Gaussian approximations show significant bias in the recovered parameters. The bias is due to the fact that the approximations fail in capturing the correct skewness of the likelihood function. The symmetric Gaussian approximation appears to be the one with the largest associated bias and the wider distribution. For this approximation the covariance is proportional to the \textit{measured} power spectrum squared, $\hat C_{\ell}^2$. This means that the uncertainties turn out to be larger than they should be if $\hat C_{\ell}$ fluctuates upward with respect to the input fiducial model, and viceversa if $\hat C_{\ell}$ has a downward fluctuation. As a consequence, upward fluctuations are given less weight in the likelihood than downward fluctuations. This translates into an overall downward bias on the amplitude of the power spectrum, and thus on $A_s$, as it can be easily seen in Figure \ref{fig_2par_fullsky_temp}. The one-third-two-third approximation is a good match to the exact likelihood at third order, as shown in Figure \ref{fig_slices}, and it appears to provide unbiased parameter results. This is the likelihood approximation adopted by the WMAP team for the analysis of the temperature anisotropies on small angular scales \cite{Verde:2003ey}. The offset-lognormal also results in unbiased constraints (bottom left panel in Fig.~\ref{fig_2par_fullsky_temp}), as expected since the offset $a_{TT}=-1/4$ has been chosen to match the exact likelihood up to third order in the expansion, see Sec.~\ref{subsec:compexac}. Also the fiducial and determinant Gaussian approximations provide unbiased cosmological parameter estimates, despite having a wrong shape at each single $\ell$ with respect to the exact likelihood. This is because at each particular $\ell$ the shape of the likelihood can randomly be wrong upwards or downwards with respect to the \textit{true} $C_{\ell}$ (see Figure \ref{fig_slices}), so when integrating over the entire range of multipoles there is negligible bias left. An obvious advantage of the fiducial Gaussian approximation with respect to the determinant Gaussian is that the covariance matrix can be pre-computed, inverted, and then kept fixed while sampling the parameter space. Aside from speeding up the computations, this makes it easier to account for additional uncertainties, non-Gaussianities and correlations in the data by simply adding extra terms in the pre-computed covariance matrix. The fiducial Gaussian approximation is the one used for the official small-scale likelihood analysis of the Planck data \cite{Ade:2013kta}, and it has also been adopted for the cosmological analysis of data collected by the ground-based experiments ACT \cite{2013JCAP...07..025D} and SPT \cite{2011ApJ...743...28K}. Note, however, that this approximation works well if the models we are trying to fit are smooth, and if the fiducial model used to build the covariance matrix is sufficiently close to the \textit{true} model. In the tests presented here the fiducial model entering the covariance matrix actually coincides with the \textit{true} model, i.e., the power spectrum from which the simulated maps are generated. Obviously, in this case the covariance matrix is an optimal description of the simulated data. However, when dealing with real data we are not in the same fortunate position, since the \textit{true} power spectrum underlying the observed sky is unknown and is exactly what we indent to estimate. In a real situation we can only deduce a fiducial model for the covariance at the best of our knowledge from both theory and observations. We can also try to improve the accuracy of a first guess covariance matrix by iterating the cosmological parameter extraction a few times, upgrading the covariance matrix at each iteration. Note that, we did not include the Hamimeche-Lewis approximation in the tests, the reason being that on the full-sky it coincides with the exact likelihood.  

In order to perform an even more stringent validation of the likelihood approximations, we can estimate the bias of the mean of the recovered parameters from all the simulations: $(\langle{\hat p\rangle} -p^{\mathrm in})/ (\sigma_p/\sqrt{N_{sims}})$, where the shift with respect to the input is normalized by the standard deviation of the mean, and $N_{sims}=1000$. This test confirms, at high-significance level, that Log-Normal, symmetric and improper Gaussian approximations provide biased results, while the other approximations are unbiased. For the latter likelihood approximations we find that the estimated uncertainties on parameters, $\sigma_p$, agree with those derived from the exact likelihood at the sub-precent level. Furthermore, we find that the standard deviation of the values of $(\hat p -p^{\mathrm in})/ \sigma_p$ is consistent to $1$, with a precision well within the accuracy allowed by the finite number of simulations, i.e., $1/\sqrt{2\,N_{\mathrm sims}}=0.022$. This means that the uncertainties estimated from the bayesian analysis, $\sigma_p$, are consistent with the scatter of the parameters of the simulations. 

\subsubsection{Temperature and polarization joint results}

\begin{figure}[!t]
\includegraphics[width=1.\textwidth]{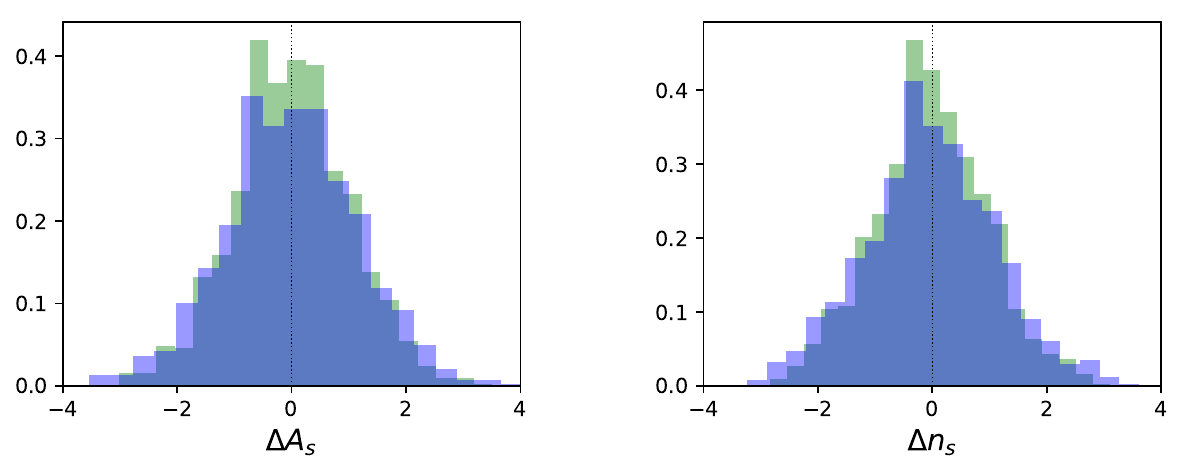}
\caption[]{\small{Histograms of the biases for the parameters expectation values estimated from the power spectra of $1000$ simulated full-sky maps of CMB temperature and polarization, plus white isotropic noise. Biases are defined with respect to the input parameter values and normalized to the $1\,\sigma$ uncertainty on the parameter, $\Delta p = (\hat p -p^{\mathrm in})/ \sigma_p$, where $p$ can be either $A_s$ or $n_s$. We adopt the fiducial Gaussian likelihood approximation, in \textit{green} the case in which the fiducial model entering the covariance matrix matches the input model of the simulations, whereas in \textit{blue} a case in which the two models differ as explained in the text.}}
\label{fig_2par_fullsky_pol}
\end{figure}

For the fiducial Gaussian approximation, we repeat the test of parameters recovery from the full-sky simulated maps also including the polarization power spectra. As we have already commented, this likelihood approximation works well provided that the fiducial power spectra used to build the covariance matrix are close enough to the \textit{true} power spectra. In order to investigate the impact on the final parameters of misestimating the covariance matrix, we change the fiducial model entering the likelihood with power spectra at the edge of the grid of models introduced above. These power spectra correspond to parameters more than $10\,\sigma$ away from the input parameters. As expected, results show some sensitivity to the choice of the model, however not for the bias, but rather for the parameters marginal error bars. In fact, using the ``wrong'' covariance matrix leaves parameters estimates unbiased. This can be noticed in Figure \ref{fig_2par_fullsky_pol}, where we show the biases on $A_s$ and $n_s$ estimated from the simulations, both using the covariance matrix with the \textit{correct} fiducial model and the one with the modified model. The latter covariance matrix, however, does not match exactly the signal in the simulated dataset, and as a consequence the estimated $1\,\sigma$ marginal error bars on the cosmological parameters do not agree with the standard deviation from the simulations, showing a $\sim 15\%$ mismatch.

\subsection{Small-scale regime -- Cut-sky tests}

\begin{figure}[!t]
\includegraphics[width=1.\textwidth]{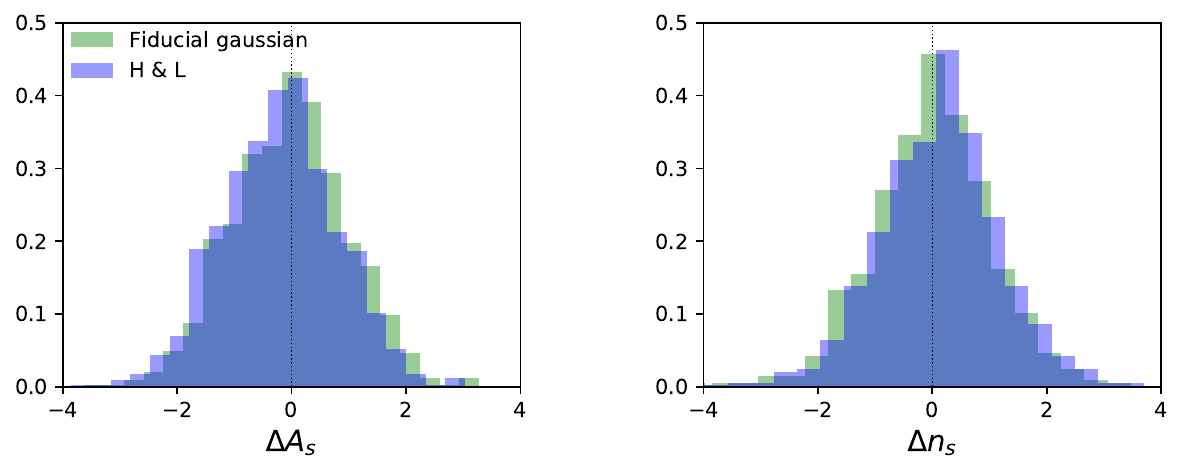}
\caption[]{\small Histograms of the biases for the parameters expectation values estimated from the power spectra of $1000$ simulated partial-sky maps of CMB temperature anisotropies plus anisotropic noise. Biases are defined analogously to Figure \ref{fig_2par_fullsky_pol}. We compare results from the fiducial gaussian and the Hamimeche-Lewis likelihood approximations.}
\label{fig_2par_cutsky}
\end{figure}

We repeat the above analysis introducing some further level of complication, specifically we assume to work with partial-sky and anisotropic noise data. We generate noise simulations which have a different level of noise variance for each pixel, thus mimicking the effect of a realistic scan strategy for which some regions of the sky are observed more often than others. 
To the simulated CMB plus noise maps we then apply a galactic sky mask, that removes the region of the galactic plane where the emission of foregrounds is expected to dominate over the CMB signal. For the sake of this test, we focus on temperature alone and we use a galactic mask that leaves for the analysis $73\%$ of sky. The mask has been apodised with a Gaussian taper that smooths sharp edges and, thus, it helps in localizing the mask power in multipole space. The angular power spectra of the cut-sky maps are extracted using an estimator that corrects for the loss of modes due to the masking and which is based on the pseudo-$C_{\ell}$ formalism. For completeness the estimator is described in Appendix \ref{sec:appendixB}. The covariance matrix associated to these power spectra has been estimated using the analytic approximation given in \cite{Efstathiou:2003dj}, and assuming as fiducial model the same model from which the simulations have been generated. 

For the parameters recovery test on the cut-sky we explore the fiducial gaussian and Hamimeche-Lewis likelihood approximations. The latter in the single field regime reduces to: 

\begin{equation}
-2\ln\mathcal{L} \simeq \sum_{\ell \ell'} \big[ g(\hat C_{\ell}/C_{\ell}) C_{f \ell} \big] \big[M^{-1}_f\big]_{\ell \ell'} \big[ g(\hat C_{\ell'}/C_{\ell'}) C_{f\ell'} \big],
\end{equation}

where the function $g(x)$ has been introduced in Eq. \ref{eq:HL8}, and $\big[M_f\big]_{\ell \ell'}$ is the covariance matrix of the $\hat C_{\ell}$ evaluated for $C_{f \ell}$.  

Following the same procedure described in the previous subsection, we fit for the cosmological parameters $A_s$ and $n_s$. We find that, also under more realistic conditions, the recovered parameters for the fiducial gaussian are unbiased well within the precision allowed by the finite number of Monte Carlo simulations. For the Hamimeche-Lewis approximation, instead, we detect a small bias at the level of $20\%$ and $11\%$ of the sigma on the parameter for $A_s$ and $n_s$ respectively, see Figure \ref{fig_2par_cutsky}. For both likelihood approximations, however, we find that the parameters marginal errors from the bayesian analysis are consistent with the standard deviation of the simulations, and thus they appear to be a good description of the \textit{true} uncertainties.   

Despite the small average bias, the Hamimeche-Lewis approximation is expected to be more robust against the choice of the fiducial model entering the likelihood, see~\cite{Hamimeche:2008ai}. In order to verify this point, when computing the likelihood and the covariance matrix of the $\hat C_{\ell}$, we use a fiducial model $C_{f \ell}$ that corresponds to parameters about $10\,\sigma$ away from those used to generate the simulations. As we have already shown in the previous subsection, in this case the fiducial Gaussian approximation provides still unbiased results, but the bayesian error bars on the parameters are not a good description of the \textit{true} uncertainties. They differ from the standard deviation of the parameters from the simulations by about $14\%$ on $A_s$ and $11\%$ on $n_s$, respectively. On the contrary, estimates with the Hamimeche-Lewis approximation still show the same level of average bias, but the uncertainties on the parameters are better characterized and they are in agreement with the standard deviation from the simulations. 

Furthermore, since it is customary to assess the goodness of the parameters fit with the chi-square statistics, we may use the value of the likelihood at the best-fit to define an effective chi-square as $\chi^2_{eff} = \mathrm{min}[-2\mathrm{ln}\mathcal{L}(\theta|\hat C_{\ell}) ] $. If we now compare the values we get for each likelihood approximation when simply varying the fiducial model, we find $\Delta \chi^2_{eff}$ of order a few hundreds for the fiducial Gaussian approximation, and $\Delta \chi^2_{eff}$ of a few tens for the Hamimeche-Lewis. As a consequence, trying to assess the goodness of the fit with the latter likelihood can surely provide more stable results, regardless of the fiducial model. However, it is worth stressing that in the present test we used a $C_{f \ell}$ that deviates significantly from the model behind the simulations, had we chosen a model closer to the simulations, also the fiducial Gaussian approximation would have provided sensible results. For further discussion of this topic refer to the appendix B of~\cite{Hamimeche:2008ai}.

\subsection{Large-scale regime}\label{sec:comparison_lowl}

In this subsection, we focus on the comparison of different likelihood approaches devoted to the analysis of large-scale CMB data. We limit the comparison of large-scale likelihoods to the polarization signal only, since this is the main target of future CMB missions \cite{Suzuki:2018cuy}. Moreover, the standard approaches used in temperature, i.e., pixel-based and Blackwell-Rao (see section \ref{subsec:blackwell-rao}) likelihoods, have been extensively characterized and validated by the Planck Collaboration, see e.g., \cite{Aghanim:2015xee,planckL05}.

We consider three approaches: pixel-based (see section \ref{subsec:pixel-based}), HL (see equations \ref{eq:HL1}-\ref{eq:HLquad}) and simulation-based (see section \ref{subsec:simulation-based}) likelihood. As we did for the small scale regime, we show whether adopting a particular likelihood approximation when estimating cosmological parameters may introduce a bias, either in the recovered values of the parameters or on their associated error bars.

We start by generating a set of $1000$ maps of the CMB sky drawn as Gaussian random realizations of a single fiducial power spectrum corresponding to a set of known cosmological parameters. The full sky maps are generated on a $N_{\textrm{side}}=16$ HEALPix grid, which roughly corresponds to a resolution of 3.7 degrees, smoothed with a cosine window function \cite{Benabed:2009af,planckL05}.  To each of these maps we add a realization of white isotropic noise corresponding to $\sigma_N = 0.01\, \mu{\mathrm K}^2$ on a pixel at our resolution. We choose this particular noise level since it is small but not completely negligible with respect to the typical peak-to-peak CMB signal in a model with a reionization optical depth $\tau=0.055$, that would be roughly $\sim 0.5 \mu{\mathrm K}$. We do not include foreground and systematics effect residuals in our simulations. We process the maps through a QML code computing the auto spectra of all our simulations.

\begin{figure}[!t]
\begin{center}
\includegraphics[width=0.5\textwidth]{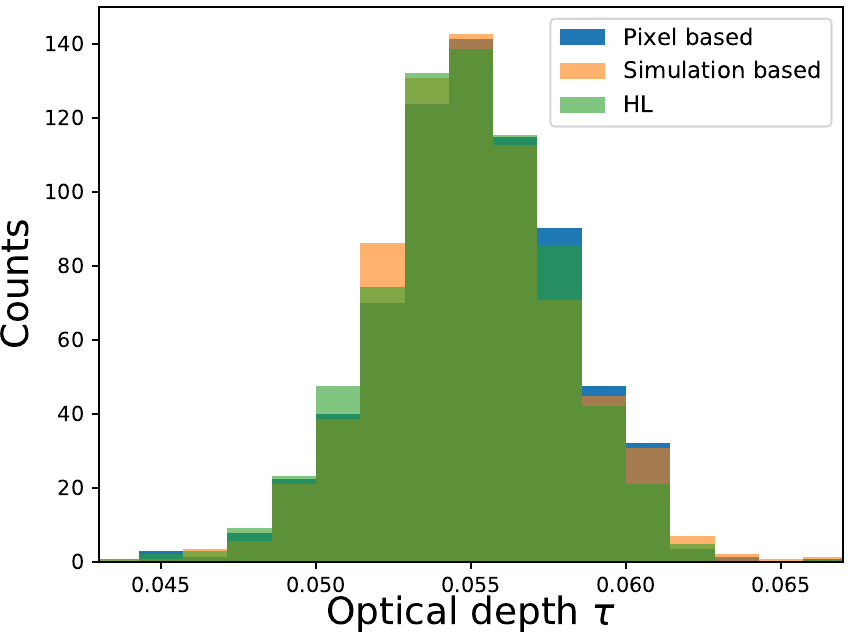}
\caption[]{\small Histograms of the recovered values of $\tau$ for the three likelihoods considered: pixel based (blue), simulation based (orange) and Hamimeche-Lewis (green).}
\label{fig_tau}
\end{center}
\end{figure}

To test the performance of the likelihood approximations, for simplicity, we only fit the reionization optical depth $\tau$ parameter keeping fixed the overall amplitude of the perturbations as parametrized through $A_{s}\exp\left(-2\tau\right)$. For each realization we compute the mean value of $\tau$. The histograms of the $\tau$ values, recovered using the three different approximations, are shown in Fig. \ref{fig_tau}. The distributions all show a bias smaller than $5\%$ of $\sigma(\tau)$, perfectly compatible with the resolution of $1000$ simulations, ( i.e., $1/\sqrt{1000}\sim3\%$).  The Hamimeche-Lewis likelihood and the simulation-based likelihood, both built on the QML estimates, perform similarly and are substantially unbiased. Likewise, the pixel-based likelihood results unbiased as it should, as also already extensively demonstrated by both the WMAP \cite{Page:2006hz} and Planck \cite{Aghanim:2015xee} collaborations. The comparison with the pixel-based likelihood allows also to validate the width of the distributions for the other two likelihood approximations, which are compatible with the pixel based at the $2\%$ level, within the resolution of the Montecarlo ( i.e., $1/\sqrt{2\times1000}\sim2\%$). 


\newpage
\section{Conclusions}\label{sec:conclusion}

CMB science is nowadays a mature yet still flourishing branch of physics. The ultimate results from the Planck satellite and from large-aperture ground-based experiments (ACT and SPT) fully characterise the CMB temperature field with great accuracy up to very small angular scales, where the contaminations from Galactic and extra-Galactic foreground emissions become dominant. In addition, they have just scratched the surface of the rich information contained in the CMB polarisation fields, also the target of small-aperture telescopes (BICEP/Keck, POLARBEAR/Simons Array, SPIDER). A large number of upcoming CMB surveys (Simons Observatory, CMB-S4, LiteBIRD) will continue the journey towards the understanding of the early phases of our Universe and its subsequent evolution. 

In the highly sophisticated process that aims to efficiently compress the information contained in the raw observed data into maps, spectra, and eventually constraints on cosmological parameters, a key ingredient is the likelihood function $\mathcal{L}(\mathbf{d}|\mathbf{\theta})$, i.e., the probability of observing a certain data collection $\mathbf{d}$ given a certain model $\mathbf{\theta}$ (see Sec.~\ref{sec:basic}). In the context of CMB analysis, $\mathbf{d}$ can be either a CMB map or CMB spectra, while $\mathbf{\theta}$ can be the set of cosmological parameters describing the cosmological model under scrutiny. In the simple case of full-sky observations, in absence of foreground contaminations and late-time-Universe effects on the CMB distribution, an exact form of the likelihood function in both real (multivariate Gaussian) and harmonic space (Wishart distribution) can be easily derived, see Sec.~\ref{sec:CMBstat}. Complications arising from realistic observations and data analysis, such as complicated noise properties, foreground obscuration, limited sky coverage, computational costs, etc, may spoil the assumptions on which the derivation of the exact likelihood functions mentioned above is based, or may limit the actual evaluation of the exact functions. Therefore, realistic analysis employ appropriate approximations of the exact likelihood functions, see Secs.~\ref{sec:highl}, ~\ref{sec:lowl}. The choice of the likelihood depends on multiple factors, such as sky-fraction retained, data resolution, computational costs, signal-to-noise properties. For example, it may (and actually did) happen that a certain likelihood approximation could work for a certain experiment given its sensitivity, and yet it could turn out to provide biased results for another, more sensitive experiment.

In this review, we have summarised the basics of CMB statistics (Sec.~\ref{sec:CMBstat}) that lead to the definition of the exact likelihoods in real and harmonic space. Then, we have moved to the descriptions of the most common likelihood approximations employed by various CMB collaborations. For the sake of simplicity, we have separated the approximations better suited to the analysis of small angular scales (higher harmonic multipoles) in Sec.~\ref{sec:highl} and those that better represent the data at large angular scales (lower multipoles) in Sec.~\ref{sec:lowl}. Although we have mostly assumed an idealised scenario of isotropic instrumental noise, Gaussian beam, unlensed CMB spectra, and absence of foreground for pedagogical reasons, we have commented about the impact of realistic deviations from the aforementioned scenario, in such a way that the reader is aware of their existence and of the extensive effort devoted to their mitigation. In particular, non-trivial modifications to the primary CMB statistics induced by the gravitational lensing of CMB photons by the evolving large-scale structure in the Universe have been discussed in Sec.~\ref{sec:lensing}. In Sec.~\ref{sec:foreground}, we have briefly discussed the main foreground contaminants in the microwave sky and mentioned methods of component separation aimed to disentangle the cosmological signal from Galactic and extra-Galactic emissions. 

In order to provide concrete examples of the different performance of various likelihood methods, we have reported results of the comparison between different likelihood approximations in Sec.~\ref{sec:comparison}. In particular, we have tested the property of the likelihood function to be unbiased, i.e., to produce a posterior distribution of cosmological parameters that matches the true cosmology (the input values, in the case of simulated data) both in terms of the mean and of the variance of the distribution. Again, we have distinguished between approximations at small scales (Sec.~\ref{sec:comparison_highl}) and at large scales (Sec.~\ref{sec:comparison_lowl}). 

In Appendix, the interested reader can find more details about some basics of statistical inference (Sec.~\ref{sec:basic}) and about the most common methods adopted to estimate CMB power spectra in the (realistic) cut-sky regime (Sec.~\ref{sec:appendixB}).

The aim of this review is to bring together and summarise a large amount of information related to a crucial aspect of CMB data analysis. The final scope is to provide a handy, yet self-consistent document to those who are approaching the field and those who are interested in learning some basic aspects of the field, and help them to navigate the vast literature produced over the past decades on these topics. If you have gone so far in your reading, we hope to have reached, at least partly, our goal.


\vspace{1cm}
\paragraph{Acknowledgments}
{The authors would like to thank D. Poletti for useful discussions. We acknowledge support from the ASI grant 2016-24-H.0 COSMOS ``Attivit\`a di studio per la comunit\`a scientifica di cosmologia''. MG acknowledges support by Argonne National Laboratory (ANL). This work was supported by the U.S. Department of Energy, Office of Science, under contract number DE-AC02-06CH11357. ML and MM acknowledge support from INFN through the InDark and Gruppo IV fundings. MM is supported by the program for young researchers ``Rita Levi Montalcini'' year 2015. LS acknowledges support from the ERC-StG ``ClustersXCosmo'' grant agreement 716762, and from the postdoctoral grant from Centre National d'\'Etudes Spatiales (CNES). LC received funding from the European Union's Horizon 2020 research and innovation programme under grant numbers 776282.


\newpage
\appendix
\section{Basic notions of probability and statistics}\label{sec:basic}

In this section we shall review some basic notions of probability and statistical inference that are used in this review.

The probability of an event $A$ occurring is quantified through a number between $0$ and $1$, and is denoted as $P(A)$. The probability of two events $A$ and $B$ both taking place is called the \emph{joint} probability of $A$ and $B$ and denoted as $P(A \cap B)$. The probability of $A$ taking place given the occurrence of $B$ (i.e., the probability of $A$ \emph{conditioned} by the observation of $B$) is denoted as $P(A|B)$. This is to be read as ''probability of $A$ given $B$''. If two events are statistically indepedent, i.e., if the occurence of one is independent on the occurrence of the other, $P(A|B) = P(A)$, and their joint probability is simply given by the product of the individual probabilities:
\begin{equation}
P(A \cap B) = P(A)\, P(B),\quad\textrm{$A$ and $B$ stat. independent.}
\end{equation}
On the other hand, if the two events are not independent, their joint probability is
\begin{equation}
P(A \cap B) = P(A|B)\, P(B) \, .
\end{equation}
Given that the two events enter symmetrically in the left-hand side of the above equation, we might as well write $P(A \cap B) = P(B|A)\, P(A)$. Equating the two conditional probabilities, we get \emph{Bayes' theorem}:
\begin{equation}
P(A|B)\, P(B) = P(B|A)\, P(A) \, .
\end{equation}
Bayes' theorem allows for the ``inversion'' of probabilities, in the sense that it relates the probability of
$A$ given $B$ to the probability of $B$ given $A$, and for this reason it plays a crucial role in Bayesian statistical inference. Note however that since Bayes' theorem follows from the relation between joint and conditional probabilities, as a mathematical relation it is valid in both the frequentist and Bayesian interpretations of probability.

In practice, one usually deals with the case in which the possible events can be mapped to numerical quantities. In particular, what we called an ``event'' might refer to a particular outcome of an experiment, for example, measuring a certain value $\bar I$ for the radiation intensity $I$ at a given frequency in a point of the sky. In this case $I$ is our random variable and $\bar I$ is a particular realization of the random variable itself. In practice, we are interested in the probability that a measurement of $I$ will yield a value
in a small interval of width $d\bar I$ around $\bar I$. 
The probability associated to this outcome is $P\left(I\in\left[\bar I, \bar I+d\bar I\right]\right) = p(\bar I) d\bar I$, where $p(\bar I$) is called the probability density function of $I$. In the following, to keep notation simple, we will usually just write $p(I)$. In a more realistic case, like for example measuring the radiation intensity over the whole sky and across different frequencies, the outcome of the experiment is represented by a vector $\mathbf{x}$. The probability density is always normalized to unity, $\int p(\mathbf x) d\mathbf x =1$. 

When performing an experiment, we are interested in using the data to perform statistical inference about the underlying physical model. This involves a hypothesis $H$ (``the theory'') that might regard, for example, the values of the parameters of a given model, or the validity of the model itself. Statistical inference can be performed in two main frameworks, Bayesian and frequentist, that we shall comment in slightly more detail in the following. In both cases, however, a central role is played by the \emph{likelihood}. The likelihood is the probability of obtaining the observed data $\mathbf{d}$ given the hypothesis $H$, regarded as a function of $H$:
\begin{equation}
\mathcal{L}(H) = P(\mathbf{d}|H) \, .
\end{equation}
If the theory is characterized by values $\btheta$ of the model parameters, then
\begin{equation}
\mathcal{L}(\btheta) = P(\mathbf{d}|\btheta) \, .
\end{equation}
and $\mathcal{L}$ is called the likelihood function.
Note that since $\mathcal{L}$ is the probability of the observed data (although expressed as a function of the theory), it is \emph{not} a probability for the hypothesis, and should not interpreted as such.

In the frequentist approach to inference, probabilities are interpreted as limiting frequencies of outcomes of a repeatable experiment. In this framework, it thus does not make sense to attach a probability to hypotheses (e.g., to speak of the probability of the ``true'' value of a parameter). Parameter inference is usually performed by providing a point estimate of the parameters of interest, together with a so-called confidence interval. Both the point estimate and the confidence intervals are functions of the data $\mathbf{d}$, built in order to have some desirable statistical properties. Being functions of random variables (the data), they are random variables themselves.
A function $\hat \theta(\mathbf{d})$ that provides a point estimate for the parameter $\theta$ is called an estimator for $\theta$. Estimators are compared and chosen on the basis of their properties. Among these, there are: i) consistency, i.e., the property that as the number of data points increases, the estimates converge in probability to the true value of the parameter; ii) bias, i.e., the distance of the expected value of the estimate to the true value of the parameter; iii) efficiency, i.e., the ratio of the variance of the estimator to the minimum possible variance, and iv) robustness, i.e., the property of yielding good performance when applied to data with a wide range of underlying statistical distributions. Many estimators can in principle be defined, and the one that is most suited for a given problem should be chosen by considering a trade-off between these (and possibly other) properties. A common choice is the maximum likelihood (ML) estimator $\hat\btheta_\mathrm{ML}$, defined as the value of $\btheta$ that maximizes $\mathcal{L}(\btheta)$:
\begin{equation}
\hat\btheta_\mathrm{ML} \equiv \argmax_{\btheta}  \mathcal{L}(\btheta) \, .
\end{equation}
In other words, the ML estimate are the values of the parameters that maximize the probability of obtaining the observed data. The ML estimator is consistent, as well as minimum variance (i.e., maximally efficient) as the number of data points goes to infinity. 

In a similar way, confidence intervals are built from the data, and there exist many recipes for doing that. We will not go into details; however, in absence of any specification, usually the term ``confidence intervals'' refers to those built from a prescription due to Neyman. This construction assures that, considering an ensemble of infinitely many repetitions of the experiment and the corresponding ensemble of confidence intervals, a fraction $X$ or larger of the intervals in the ensemble will contain the true value of the parameter, independently of the true value itself. The fraction $X$ is the coverage probability of the interval; in jargon, if for example $X=0.95$, one speaks of a ``95\% confidence interval''. Note that this does not mean that the interval constructed from the observed data has $95\%$ probability of containing the true value of the parameter. In any case, let us stress that whatever is the recipe chosen to build parameter estimates and confidence intervals, the likelihood function plays a fundamental role. Thus an accurate knowledge of the likelihood is paramount in order to perform reliable statistical inferences from the data.

In the Bayesian framework, probabilities are instead interpreted as ``degrees of belief''. Thus, other than speaking of probabilities of experimental outcomes, we can also attach a probability to quantities with respect to which we are in a condition of uncertainty, like the ``true'' value of a parameter.
For example, we can speak of the probability that the curvature parameter of the Universe $\Omega_k$ is equal to $10^{-2}$, or that the mass of the lightest neutrino is below $0.1\,\eV$.
Given this, we can use Bayes' theorem to express the probability density function of the parameters given the data as:
\begin{equation}
p(\btheta|\bd) = \frac{p(\bd|\btheta)p(\btheta)}{p(\bd)} \, .
\end{equation}
Here, $p(\btheta|\bd)$ is the posterior, $p(\bd|\btheta)$ the likelihood, $p(\btheta)$ the prior, and $p(\bd)$ the evidence. The posterior quantifies our knowledge of the model parameters after having obtained the data, while the prior takes into account the information on the parameters that we had before performing the experiment. This prior information might come from theoretical constraint, as well as from the results of previous experiments. The evidence $p(\bd)$ is the probability of the data and, by the law of total probability, is given by $p(\bd)=\int p(\bd|\btheta)p(\btheta)d\btheta = \int {\mathcal L}(\btheta)p(\btheta)d\theta$; for this reason, it is also called the marginal likelihood. The evidence is not relevant for parameter estimation since it does not depend on $\theta$, and thus only represents an irrelevant multiplicative constant in that regard. However it plays an important role in other problems, like e.g., model selection. 

The information contained in the posterior for the model parameters is usually compressed in terms of a point estimate and/or an interval for each parameter $\theta_{i}\in \btheta$. From the $N$-dimensional posterior for $\btheta$ ($N$ being the number of parameters), one can reconstruct the one-dimensional posterior for $\theta_{i}$ from the law of total probabilities, by integrating over the remaining parameters:
\begin{equation}
p(\theta_i | \bd) = \int p(\btheta|\bd)d\btheta_{-i} \, ,
\end{equation}
where $d\btheta_{-i} \equiv d\theta_1 \dots d\theta_{i-1}d\theta_{i+1}\dots d\theta_N$. This operation is called ``marginalization'' with respect
to the unwanted parameters. Common choices for the point estimate for $\theta$ (we drop the subscript $i$) include its expectation value $\langle \theta \rangle$ over the posterior distribution 
\begin{equation}\label{eq:param_expectation}
\langle \theta \rangle = \int \theta \,p(\theta|\bd)d\theta \, ,
\end{equation}
or the mode of the posterior, i.e., the value $\theta_M$ that maximizes the posterior:
\begin{equation}
\theta_{M} = \argmax_{\theta} p(\theta|\bd) \, .
\end{equation}
Note that the latter estimate numerically coincides with the frequentist ML estimate in the case of a flat prior ($p(\btheta)=\mathrm{const}$). For what concerns the construction of intervals, called credible intervals in Bayesian statistics, their defining property is that there is a probability $X$ that the true value of the parameter lies in the interval. Then, for a 95\% credible interval $\mathcal I$:
$\int_{\mathcal I} p(\theta)d\theta = 0.95$.
This is however not enough to uniquely define the interval and additional properties have to be imposed. It is common to require that the interval is symmetric around the mean, or that the probability density outside the interval is everywhere smaller than the probablity inside the interval. It is also customary, in the literature, to present results in terms of two-dimensional credible intervals for parameter pairs $\{\theta_i,\,\theta_j\}$, built in a similar way from the two-dimensional posterior $p(\theta_i,\,\theta_j|\bd)$.


\newpage
\section{Estimators in cut-sky}\label{sec:appendixB}
In this appendix, we review the main approaches to build unbiased estimators of the observed CMB power spectra. As detailed in the main text, realistic CMB experiments usually do not have access to the full sky. Actual limitations to the scanning strategy and/or exclusion of portions of observed sky obscured by foreground contaminations and excessive noise may reduce the fraction of the sky available for reconstructing the CMB signal. As we will see shortly in detail, the main effect of dealing with a cut sky is to induce a coupling between different modes, such that the effective harmonic coefficients $\tilde{a}_{\ell m}$ of the cut sky are not independent. This forbids the use of Eq.~\ref{eq:clmeas} to estimate the observed power spectrum. Different methods to estimate the power spectra in cut sky have to be considered.

\subsection{Pseudo-$C_\ell$ formalism}\label{sec:pseudocl}
The traditional method applied to the small-scale regime is the pseudo-$C_\ell$ method~\cite{Hivon:2001jp,Wandelt:2000av,Brown:MNRAS2005}, which we briefly recall here.
Let's start with a derivation of the pseudo-$C_\ell$ power spectrum. For the sake of simplicity, we will focus on the single-field temperature-only case. The limitation induced by having to deal with a cut sky can be thought as the application of a (position-dependent) weighting function $W(\hat{n})$ to the observed data. In the most simple case, the weighting function can be a discrete function $W=1$ for observed pixels, and zero otherwise. The window function can be expanded in spherical harmonics 
\begin{equation}\label{eq:Wt}
w_{\ell m}=\int d\hat{n} W(\hat{n}) Y^*_{\ell m}(\hat{n}) 
\end{equation}

with power spectrum

\begin{equation}\label{eq:wlm}
\mathcal{W}_\ell=\frac{1}{2\ell+1}\sum_m |w_{\ell m}|^2
\end{equation} 

If $\fsky$ is the fraction of the sky retained in the analysis, the i-th momentum of the window function can be defined as $w_i=(4\pi\fsky)^{-1}\int d\hat{n} W^i(\hat{n})$. The spherical harmonic expansion of the temperature anisotropy field can be written as

\begin{subequations}
\begin{eqnarray}\label{eq:almtilde}
\tilde{a}_{\ell m}&=&\int d\hat{n} \Theta(\hat{n}) W(\hat{n}) Y^*_{\ell m}(\hat{n})\\\label{eq:almtilde}
&=&\sum_{\ell' m'} a_{\ell' m'}\int d\hat{n} Y_{\ell' m'}(\hat{n}) W(\hat{n}) Y^*_{\ell m}(\hat{n})\\\label{eq:almtilde}
&=&\sum_{\ell' m'} a_{\ell' m'} K_{\ell m \ell' m'}(W)\label{eq:almtilde3},
\end{eqnarray}
\end{subequations}

where $K_{\ell m \ell' m'}$ is the coupling kernel between different modes. From Eq.~\ref{eq:almtilde3}, it is clear that $\tilde{a}_{\ell m}$ are still Gaussian variables, as they are the sum of Gaussian variables (the ``true'' $a_{\ell m}$). However, the coefficients of the temperature field in cut sky are not independent anymore, as the cut sky introduces the coupling represented by Eq.~\ref{eq:almtilde3}. The cut-sky coefficients can be used to define a pseudo-$C_\ell$ power spectrum

\begin{equation}\label{eq:pseudocl}
\tilde{C}_\ell=\frac{1}{2\ell+1}\sum_{m=-\ell}^\ell \tilde{a}_{\ell m}\tilde{a}^*_{\ell m}\;.
\end{equation}

The coupling kernel is a purely geometric factor, encoding the details of the sky cut. This geometric dependance can be further highlighted using the spherical harmonic expansion of the weighting function $W(\hat{n})$:

\begin{subequations}
\begin{eqnarray}\label{eq:klm}
K_{\ell_1 m_1 \ell_2 m_2}&=&\int d\hat{n} Y_{\ell_1 m_1}(\hat{n}) W(\hat{n}) Y^*_{\ell_2 m_2}(\hat{n})\\
&=&\sum_{\ell_3 m_3} w_{\ell_3 m_3}\int d\hat{n} Y_{\ell_1 m_1}(\hat{n}) Y_{\ell_3 m_3}(\hat{n})  Y^*_{\ell_2 m_2}(\hat{n})\\
&=&\sum_{\ell_3 m_3} w_{\ell_3 m_3} (-1)^{m_2}\left[\frac{(2\ell_1+1)(2\ell_2+1)(2\ell_3+1)}{4\pi}\right]^{1/2}\\
&&\times \left(\begin{array}{ccc}
\ell_1 &\ell_2 &\ell_3\\
0 &0 &0
\end{array}\right)\left(\begin{array}{ccc}
\ell_1 &\ell_2 &\ell_3\\
m_1 &-m_2 &m_3
\end{array}\right),\nonumber
\end{eqnarray}
\end{subequations}

where the Wigner-3j/Clebsch-Gordan symbols appear. They describe the coupling of three angular momenta such that their total sum vanishes (triangle relation).

Equation~\ref{eq:almtilde3} cannot be interpreted as an operative definition of the power spectrum. In fact, the coupling kernel is singular and therefore Eq.~\ref{eq:almtilde3} cannot be inverted to compute the true $a_{\ell m}$. As a result, a need for dedicated estimators arises in the cut-sky regime. From Eq.~\ref{eq:pseudocl}, a relation between the true power spectrum and the pseudo-power spectrum can be derived taking the ensamble average:

\begin{subequations}
\begin{eqnarray}\label{eq:pseudoav}
\langle\tilde{C}_{\ell_1}\rangle&=&\frac{1}{2\ell_1+1}\sum_{m_1=-\ell_1}^{\ell_1}\langle\tilde{a}_{\ell_1 m_1}\tilde{a}^*_{\ell_1 m_1}\rangle\\
&=&\frac{1}{2\ell_1+1}\sum_{m_1=-\ell_1}^{\ell_1}\sum_{\ell_2 m_2}\sum_{\ell_3m_3}\langle a_{\ell_2 m_2}a^*_{\ell_3 m_3}\rangle K_{\ell_1m_1\ell_2m_2}[W]K^*_{\ell_1m_1\ell_3m_3}[W]\\
&=&\frac{1}{2\ell_1+1}\sum_{m_1=-\ell_1}^{\ell_1}\sum_{\ell_2}\langle C_{\ell_2}\rangle\sum_{m_2=-\ell_2}^{\ell_2}|K_{\ell_1m_1\ell_2m_2}[W]|^2\\
&=&\sum_{\ell_2}M_{\ell_1\ell_2}\langle C_{\ell_2}\rangle \;.
\end{eqnarray}
\end{subequations}

The last line in Eq.~\ref{eq:pseudoav} has been obtained by expanding the kernel couplings in spherical harmonics and making use of the orthogonality relations of the Wigner-3j symbols. The coupling matrix $M_{\ell_1\ell_2}$ is therefore given by:

\begin{equation}\label{eq:ml1l2}
M_{\ell_1\ell_2}=\frac{2\ell_2+1}{4\pi}\sum_{\ell_3}(2\ell_3+1)\mathcal{W}_{\ell_3}\left(\begin{array}{ccc}
\ell_1 &\ell_2 &\ell_3\\
0 &0 &0
\end{array}\right)^2.
\end{equation}

Finally, the last line in Eq.~\ref{eq:pseudoav} gives the operative definition of the power spectrum estimator in the cut-sky regime:

\begin{equation}
\hat{C}_\ell=\sum_{\ell'} M_{\ell \ell'}^{-1} \tilde{C}_{\ell'}\;.
\end{equation}

For small sky cuts, the coupling matrix $M_{\ell\ell'}$ is invertible. When larger portions of the sky have to be neglected, the matrix becomes singular: some eigenvalues become negligible as a consequence of the fact that some modes end up being in the masked region of the sky. In these cases, it is possible to overcome this limitation by applying a binning scheme to the power spectrum~\cite{Hivon:2001jp}.

We have seen in the previous chapter that, in the full-sky regime, the $C_\ell$-s are distributed according to a $\chi^2$ distribution with $\nu=(2\ell+1)$ degrees of freedom and variance $\Delta C_\ell=\sqrt{2/(2\ell+1)} C_\ell$. In the cut-sky regime, the mode-mode coupling acts to effectively reduce the degrees of freedom available for each mode. The variance of the recovered power spectrum in the cut-sky regime effectively reduces~\cite{Hivon:2001jp} to $\Delta C_\ell=\sqrt{2/\nu_\ell^\mathrm{eff}} C_\ell$, where the effective number of degrees of freedom is
\begin{equation}\label{eq:nueff}
\nu_\ell^\mathrm{eff}=(2\ell+1)\fsky\frac{w_2^2}{w_4},
\end{equation}
with $w_i$ the i-th momentum of the weighting function.

\subsubsection{Pseudo-$C_\ell$ formalism for correlated fields}
The extension of the pseudo-$C_\ell$ formalism to the CMB polarisation fields follows the same steps highlighted in the case of the temperature field. First, let's us define the harmonic expansion coefficients of the polarisation $E$ and $B$-fields in the cut sky regime:
\begin{subequations}
\begin{eqnarray}\label{eq:pseudoaEB}
\tilde{a}^E_{\ell m}=\sum_{\ell'm'}\left(_+K_{\ell\ell' m m'}a^E_{\ell'm'}+i _-K_{\ell\ell' m m'}a^B_{\ell'm'}\right)\\
\tilde{a}^B_{\ell m}=\sum_{\ell'm'}\left(_+K_{\ell\ell' m m'}a^B_{\ell'm'}-i _-K_{\ell\ell' m m'}a^E_{\ell'm'}\right) 
\end{eqnarray}
\end{subequations}
and the harmonic expansion coefficients of the window function in polarisation $W_p(\hat{n})$ (analogous to that in temperature, see Eq.~\ref{eq:Wt}):
\begin{equation}\label{eq:Wp}
w^p_{\ell m}=\int d\hat{n} W_p(\hat{n}) Y^*_{\ell m}(\hat{n}) 
\end{equation}

The coupling kernels multiplying the full-sky coefficients $a^{E,B}_{\ell m}$ in Eq.~\ref{eq:pseudoaEB} are defined as
\begin{subequations}
\begin{eqnarray}\label{eq:kpm}
_+K_{\ell\ell' m m'}\equiv\frac{1}{2}\left(_{+2}K_{\ell\ell' m m'}+ _{-2}K_{\ell\ell' m m'}\right)\\
_-K_{\ell\ell' m m'}\equiv\frac{1}{2}\left(_{+2}K_{\ell\ell' m m'}- _{-2}K_{\ell\ell' m m'}\right)
\end{eqnarray}
\end{subequations}
where the spin-weighted coupling kernels for the polarisation fields $_{\pm2}K_{\ell\ell' m m'}$ appear as a function of the polarisation window function $W_p$
\begin{equation}\label{eq:Kpol}
_sK_{\ell\ell' m m'}=\int d\hat{n} W_p(\hat{n}) _sY_{\ell'm'}(\hat{n}) _sY^{*}_{\ell m}(\hat{n}),\quad s=\pm2
\end{equation}

Following the steps leading to the last line of Eq.~\ref{eq:klm}, the polarisation window function $W_p$ can be further expanded in (spin-$0$) spherical harmonics\footnote{The window function either in temperature or in polarisation is a scalar field, and therefore the relevant harmonic basis for the harmomic expansion is provided by spin-$0$ spherical harmonics $Y_{\ell m}$.} as in Eq.~\ref{eq:Wp}. Using this expansion in Eq.~\ref{eq:kpm} along with the definition of the spin-weighted coupling kernel in Eq.~\ref{eq:Kpol}, it is possible to work out the dependance of the coupling kernels $_{\pm}K_{\ell\ell' m m'}$ from the Wigner-3j symbols:

\begin{subequations}
\begin{eqnarray}\label{eq:klmP}
_{\pm}K_{\ell_1 m_1 \ell_2 m_2}&=&\int d\hat{n} W_p(\hat{n}) \left(_{+2}Y_{\ell_1 m_1}(\hat{n})  _{+2}Y^*_{\ell_2 m_2}(\hat{n})\pm_{-2}Y_{\ell_1 m_1}(\hat{n})  _{-2}Y^*_{\ell_2 m_2}(\hat{n})\right)\\
&=&\sum_{\ell_3 m_3} w^p_{\ell_3 m_3}\int d\hat{n} Y_{\ell_3 m_3}(\hat{n})  \left(_{+2}Y_{\ell_1 m_1}(\hat{n})  _{+2}Y^*_{\ell_2 m_2}(\hat{n})\pm_{-2}Y_{\ell_1 m_1}(\hat{n})  _{-2}Y^*_{\ell_2 m_2}(\hat{n})\right)\nonumber\\ 
&=&\sum_{\ell_3 m_3} w_{\ell_3 m_3} (-1)^{m_2}\left[\frac{(2\ell_1+1)(2\ell_2+1)(2\ell_3+1)}{4\pi}\right]^{1/2}\\
&&\times \left(\begin{array}{ccc}
\ell_1 &\ell_2 &\ell_3\\
m_1 &-m_2 &m_3
\end{array}\right)\left[\left(\begin{array}{ccc}
\ell_1 &\ell_2 &\ell_3\\
2 &-2 &0
\end{array}\right)\pm\left(\begin{array}{ccc}
\ell_1 &\ell_2 &\ell_3\\
-2 &2 &0
\end{array}\right)\right]\nonumber\\
&=&(-1)^{m_2} \sum_{\ell_3 m_3} \left(1\pm(-1)^L\right) w_{\ell_3 m_3} \left[\frac{(2\ell_1+1)(2\ell_2+1)(2\ell_3+1)}{4\pi}\right]^{1/2}\\
&&\times \left(\begin{array}{ccc}
\ell_1 &\ell_2 &\ell_3\\
m_1 &-m_2 &m_3
\end{array}\right)\left(\begin{array}{ccc}
\ell_1 &\ell_2 &\ell_3\\
2 &-2 &0
\end{array}\right),\,\quad L=(\ell_1+\ell_2+\ell_3)\nonumber
\end{eqnarray}
\end{subequations}

where in moving from the second-to-the-last to the last line, we have made use of the properties of the Wigner-3j symbols.

The relation between the pseudo-$C_\ell$ 
\begin{equation}\label{eq:pseudoclXY}
\tilde{C}_\ell^{XY}\equiv \frac{1}{2\ell+1}\sum_m \tilde{a}_{\ell m}^X \left(\tilde{a}_{\ell m}^Y\right)^*,\quad XY=(TT,TE,EE,BB)
\end{equation}

and the power spectra in full sky can be again given in terms of a coupling matrix $\mathcal{M}_{\ell\ell'}$ (see e.g.~\cite{Hamimeche:2008ai}):
\begin{subequations}
\begin{eqnarray}\label{eq:pseudoM}
\left(\begin{array}{c}
\langle\tilde{C}_\ell^{TT}\rangle\\
\langle\tilde{C}_\ell^{TE}\rangle\\
\langle\tilde{C}_\ell^{EE}\rangle\\
\langle\tilde{C}_\ell^{BB}\rangle\\
\end{array}\right)&=&\sum_{\ell'}\mathcal{M}_{\ell\ell'}\left(\begin{array}{c}
C_{\ell'}^{TT}\\
C_{\ell'}^{TE}\\
C_{\ell'}^{EE}\\
C_{\ell'}^{BB}\\
\end{array}\right)\\
&=&\sum_{\ell'}\left(\begin{array}{cccc}
M_{\ell\ell'}^{TT} &0 &0 &0\\
0 &M_{\ell\ell'}^{TE} &0 &0\\
0 &0 &M_{\ell\ell'}^{EE} &M_{\ell\ell'}^{EB}\\
0 &0 &M_{\ell\ell'}^{BE} &M_{\ell\ell'}^{BB}\\
\end{array}\right)\left(\begin{array}{c}
C_{\ell'}^{TT}\\
C_{\ell'}^{TE}\\
C_{\ell'}^{EE}\\
C_{\ell'}^{BB}\\
\end{array}\right)
\end{eqnarray}
\end{subequations}

The coupling matrices $M^{XY}$ for the individual power spectra can be obtained similarly to what is done for the coupling matrix $M$ in Eq.~\ref{eq:ml1l2}:
\begin{subequations}
\begin{eqnarray}\label{eq:ml1l2XY}
M_{\ell_1\ell_2}^{TT}&=&\frac{2\ell_2+1}{4\pi}\sum_{\ell_3}(2\ell_3+1)\mathcal{W}_{\ell_3}^{TT}\left(\begin{array}{ccc}
\ell_1 &\ell_2 &\ell_3\\
0 &0 &0
\end{array}\right)^2\\
M_{\ell_1\ell_2}^{TE}&=&\frac{2\ell_2+1}{8\pi}\sum_{\ell_3}(2\ell_3+1)\mathcal{W}_{\ell_3}^{TP}\left(1+(-1)^L\right)\\
&&\times \left(\begin{array}{ccc}
\ell_1 &\ell_2 &\ell_3\\
0 &0 &0
\end{array}\right)  \left(\begin{array}{ccc}
\ell_1 &\ell_2 &\ell_3\\
-2 &2 &0
\end{array}\right)\\
M_{\ell_1\ell_2}^{EE}&=&\frac{2\ell_2+1}{16\pi}\sum_{\ell_3}(2\ell_3+1)\mathcal{W}_{\ell_3}^{PP}\left(1+(-1)^L\right)^2\\
&&\times \left(\begin{array}{ccc}
\ell_1 &\ell_2 &\ell_3\\
-2 &2 &0
\end{array}\right)^2\\
M_{\ell_1\ell_2}^{EB}&=&\frac{2\ell_2+1}{16\pi}\sum_{\ell_3}(2\ell_3+1)\mathcal{W}_{\ell_3}^{PP}\left(1-(-1)^L\right)^2\\
&&\times \left(\begin{array}{ccc}
\ell_1 &\ell_2 &\ell_3\\
-2 &2 &0
\end{array}\right)^2\\
M_{\ell_1\ell_2}^{BB}&=&M_{\ell_1\ell_2}^{EE}\\
M_{\ell_1\ell_2}^{BE}&=&M_{\ell_1\ell_2}^{EB}
\end{eqnarray}
\end{subequations}

where we have made explicit the fact that, in principle, the power spectrum of the window function $\mathcal{W}$ can be different in temperature and polarisation.

The estimators for the power spectra in the cut-sky regime can be therefore obtained from the inversion of Eq.~\ref{eq:pseudoM}, either $\ell$-by-$\ell$ if the sky cut is small or in a binned version if the $\ell$-by-$\ell$ coupling matrix is not invertible in case of large cuts:
\begin{eqnarray}\label{eq:pseudoMinv}
\left(\begin{array}{c}
\hat{C}_{\ell}^{TT}\\
\hat{C}_{\ell}^{TE}\\
\hat{C}_{\ell}^{EE}\\
\hat{C}_{\ell}^{BB}\\
\end{array}\right)=\sum_{\ell'}\left(\begin{array}{cccc}
M_{\ell\ell'}^{TT} &0 &0 &0\\
0 &M_{\ell\ell'}^{TE} &0 &0\\
0 &0 &M_{\ell\ell'}^{EE} &M_{\ell\ell'}^{EB}\\
0 &0 &M_{\ell\ell'}^{BE} &M_{\ell\ell'}^{BB}\\
\end{array}\right)^{-1}\left(\begin{array}{c}
\tilde{C}_{\ell'}^{TT}\\
\tilde{C}_{\ell'}^{TE}\\
\tilde{C}_{\ell'}^{EE}\\
\tilde{C}_{\ell'}^{BB}\\
\end{array}\right)
\end{eqnarray}

As a result, it is clear that the likelihood function in cut sky cannot be written purely in terms of pseudo-$C_\ell$. In fact, one has to take into account the mode-coupling induced by the presence of disconnected regions.

\subsection{Estimating $B$-modes: the ``pure'' formalism}\label{sec:pureEB}
The pseudo-$C_\ell$ formalism can in principle be applied to estimates of the $BB$ power spectrum. However, it has been shown~\cite{Bunn:2002df} that substantial leakage from $E$-modes to $B$-modes ($E/B$ leakage) affects such estimator. Various techniques have been proposed that cure the bias induced by $E/B$ leakage, for example taking suitable linear combinations of pseudo-$C_\ell$. There still remains a non-negligible contribution of the $E/B$ leakage to the variance of the estimator. Such extra variance substantially limits the ability to detect the tiny $B$-mode signal coming from primordial sources. For example, Ref.~\cite{Challinor:2004pr} showed that pseudo-$C_\ell$ estimators can only be sensitive to values of the tensor-to-scalar ratio $r$ larger than $r>0.05$ when observations are taken in a small patch of the sky ($\fsky\sim0.01$). This value is very close to the current upper bound on the tensor-to-scalar ratio ($r<0.06$ at 95\% c.l. from BK15+Planck~\cite{Ade:2018gkx}). In addition, there is no theoretical lower bound on the predicted value of $r$, that could even be negligibly small. Therefore, efforts has been put in the construction of suitable estimators for the polarisation signal that could overcome the limitations imposed by the pseudo-$C_\ell$ approach~\cite{2009PhRvD..79l3515G,Smith:2006vq,Smith:2005gi,Bunn:2002df}.

In this section, we briefly describe the ``pure-$EB$'' approach, firstly proposed by Ref.~\cite{Bunn:2002df}. The basic idea is that, in the full-sky regime, the polarisation field $\mathbf{P}=(Q,U)^T$ can be decomposed as a linear combination of an $E$-part and a $B$-part, with the two parts being perpendicular to each other. The $E/B$ parts form a basis, in a sense that the space of the polarisation fields is the sum of the two orthogonal $E$- and $B$-subspaces. The two subspaces have the following properties:
\begin{itemize}
\item A field is $E$-like if it is curl-free;
\item A field is $B$-like if it is divergence-free;
\item An $E$-field is \textit{pure} if it orthogonal to all the possible $B$-fields;
\item A $B$-field is \textit{pure} if it orthogonal to all the possible $E$-fields.
\end{itemize}

This decomposition does not hold anymore when only a fraction of the sky is observed. In this case, it can be shown that a third subspace is needed to fully represent the polarisation field $\mathbf{P}$. This third field is called ``ambiguous'' because satisfies both the conditions for an $E$- and $B$-field simultaneously, and can therefore leak into any of the two other fields when the $E/B$ separation is attempted. Although in practice both contributions from $E$ and $B$ end up in the ambiguous modes, the dominant contribution is by far the $E$-mode. This is the reason behind the $E/B$ leakage affecting the pseudo-$C_\ell$ approach. One can understand the situation as follows: the pseudo-$C_\ell$ attempts to reconstruct the $BB$ spectrum by taking suitable linear combinations of the observed spectra in order to fit out the $EE$ spectrum. While the $BB$ spectrum obtained in such a way is \textit{on average} independent from $EE$, its variance is not.

In the cut-sky case, it is therefore preferable to estimate the $BB$ spectrum starting from the pure-$B$ field. The formalism is based on the fact that it is possible to build combinations of second derivatives of the Stokes parameters $Q$ and $U$ that depend only on $E$ or $B$. In particular,
\begin{subequations}
\begin{eqnarray}
\mathbf{P}_E&=&\mathbf{D}_E\psi_E,\quad\psi_E=-\sum_{\ell m}\left[\frac{(\ell-2)!}{(\ell+2)!}\right]^{1/2}a^E_{\ell m}Y_{\ell m}\\
\mathbf{P}_B&=&\mathbf{D}_B\psi_B,\quad\psi_B=-\sum_{\ell m}\left[\frac{(\ell-2)!}{(\ell+2)!}\right]^{1/2}a^B_{\ell m}Y_{\ell m}
\end{eqnarray}
\end{subequations}
where
\begin{equation}
\mathbf{D}_E=\frac{1}{2}\left(\begin{array}{c}
\slashed{\partial}\slashed{\partial}+\bar{\slashed{\partial}}\bar{\slashed{\partial}}\\
-\imath(\slashed{\partial}\slashed{\partial}-\bar{\slashed{\partial}}\bar{\slashed{\partial}})
\end{array}\right),\quad\mathbf{D}_B=\frac{1}{2}\left(\begin{array}{c}
\imath(\slashed{\partial}\slashed{\partial}-\bar{\slashed{\partial}}\bar{\slashed{\partial}})\\
\slashed{\partial}\slashed{\partial}+\bar{\slashed{\partial}}\bar{\slashed{\partial}}
\end{array}\right)
\end{equation}
and $\slashed{\partial},\bar{\slashed{\partial}}$ are the spin-raising and spin-lowering operators respectively (see Ref.~\cite{Zaldarriaga:1998rg} for the detailed application of such operators to the CMB polarisation study). 

The harmonic coefficients $\tilde{a}^X_{\ell m}$ of the pseudo-$C_\ell$ in polarisation can be computed by applying the operators $\mathbf{D}_X$ to the observed polarisation field:
\begin{subequations}
\begin{eqnarray}\label{eq:aepure}
\tilde{a}^E_{\ell m}&=&\int d\hat{n} [\mathbf{D}_E (W(\hat{n}) Y_{\ell m}(\hat{n})]^\dag\cdot\mathbf{P}=\int d\hat{n} W(\hat{n}) Y_{\ell m}(\hat{n}) \mathbf{D}_E^\dag\cdot\mathbf{P}\\
\tilde{a}^B_{\ell m}&=&\int d\hat{n} [\mathbf{D}_B (W(\hat{n}) Y_{\ell m}(\hat{n})]^\dag\cdot\mathbf{P}=\int d\hat{n} W(\hat{n}) Y_{\ell m}(\hat{n}) \mathbf{D}_B^\dag\cdot\mathbf{P}
\end{eqnarray}
\end{subequations}
where $W(\hat{n})$ is the window function. The passage from the first integral to the second in each line of Eq.~\ref{eq:aepure} is obtained by integrating twice by part, and it is only possible if the window function and its gradient vanish at the boundary, i.e., if the window function satisfies both the Neumann and Dirichlet conditions. In the full-sky limit ($W(x)=1$ everywhere), it is easy to show that Eq.~\ref{eq:aepure} gives the usual definition of $a^{E,B}_{\ell m}$. Comparing Eq.~\ref{eq:aepure} with the harmonic coefficients in cut sky in Eq.~\ref{eq:pseudoaEB}, it is clear that the pure harmonic coefficients only depend on the corresponding field (either $E$ or $B$): the $\mathbf{D}_X$ operator correctly avoid any $E/B$ mixing. The pure pseudo-$C_\ell$ are then given by
\begin{equation}
\tilde{C}_\ell^{XX}=\frac{1}{2\ell+1}\sum_{m=-\ell}^\ell (\tilde{a}^X_{\ell m})^* \tilde{a}^X_{\ell m},\quad X=E,B
\end{equation}

After some cumbersome algebra (see e.g., Appendix A in~\cite{2009PhRvD..79l3515G}), it can be shown that the pure power spectrum is an unbiased estimator
\begin{equation}\label{eq:clpure}
\left(\begin{array}{c}
\langle\tilde{C}_\ell^{EE}\rangle\\
\langle\tilde{C}_\ell^{BB}\rangle
\end{array}\right)=\sum_{\ell'}\left(\begin{array}{cc}
M^\mathrm{diag}_{\ell\ell'} &M^\mathrm{off}_{\ell\ell'}\\
M^\mathrm{off}_{\ell\ell'}  &M^\mathrm{diag}_{\ell\ell'}
\end{array}\right)\left(\begin{array}{c}
C_\ell^{EE}\\
C_\ell^{BB}
\end{array}\right).
\end{equation} 
Finally the observed power spectrum $\hat{C}_\ell$ can be obtained from the inversion of Eq.~\ref{eq:clpure}:
\begin{equation}\label{eq:clhatpure}
\left(\begin{array}{c}
\hat{C}_\ell^{EE}\\
\hat{C}_\ell^{BB}
\end{array}\right)=\sum_{\ell'}\left(\begin{array}{cc}
M^\mathrm{diag}_{\ell\ell'} &M^\mathrm{off}_{\ell\ell'}\\
M^\mathrm{off}_{\ell\ell'}  &M^\mathrm{diag}_{\ell\ell'}
\end{array}\right)^{-1}\left(\begin{array}{c}
\tilde{C}_\ell^{EE}\\
\tilde{C}_\ell^{BB}
\end{array}\right).
\end{equation} 

In the presence of noise, the noise bias $N_\ell$ needs to be subtracted from the pure spectrum in the right-hand side of Eq.~\ref{eq:clhatpure}. A final note concerns the boundary conditions. It has been mentioned above that, for the pure method to work, a properly apodised mask (roughly speaking, a mask with smooth borders) is needed (see e.g., discussion in Sec.III.A of Ref.~\cite{Smith:2006vq}). If the window function $W(\hat{n})$ in Eq.~\ref{eq:aepure} is properly apodised up to its first derivative, the off-diagonal terms in Eq.~\ref{eq:clpure} vanish. 

\subsection{Quadratic Maximum Likelihood Estimator}\label{sec:qml}
At large scales, the best estimator in the cut-sky regime is the Quadratic Maximum Likelihood (QML) estimator. It can be shown~\cite{Tegmark:1996qt} that the QML is not only unbiased, but also \textit{unbeatable} or optimal. The QML is unbiased in the sense that the estimator converges to the underlying power spectrum \textit{in ensemble average}. Optimality follows from the fact that the covariance of the estimator, if computed on the true model, equals the inverse of the Fisher matrix, which represents the minimum variance that can be associated to measurements of a given parameter set (Cramer-Rao inequality). In other words, no unbiased method can estimate the power spectrum with a variance lower than the QML, \textit{provided that the the fiducial model adopted to compute the covariance is the true theoretical model}. If this condition is met the estimator is maximum likelihood, as we will comment further at the end of this section.

The motivation behind the QML is the same shared by other estimators: compress the information contained in the CMB maps to reduce the dimensionality of the problem (i.e., move from a $N_\mathrm{pix}\times N_\mathrm{pix}$ problem in pixel space to a $n_\ell$ problem in harmonic space, where $n_\ell$ is the number of bandpower used to reconstruct the power spectrum), and do that in such a way that there is minimal information lost. In the case of the pseudo-$C_\ell$ estimator, the information loss is due to the mode-mode coupling (some modes are hidden in the cut region) and shows up as an increased sample variance. In the case of the QML, the estimator is lossless in the sense specified above: it provides an estimate of the underlying spectrum with minimum variance.

In what follows, we will review the algebra of the QML. For further details on the method and its applications, we refer the reader to the seminal papers of Tegmark~\cite{Tegmark:1996qt}, Tegmark\&de Olivera Costa~\cite{Tegmark:2001zv}. We will sketch the main steps in the single field regime (temperature only\footnote{see also derivation in Appendix C of F. Paci, PhD thesis \url{http://amsdottorato.unibo.it/1859/1/paci_francesco_tesi.pdf}.}) and then move to the description of the estimator in the case of correlated fields (temperature and polarisation), in a similar fashion of what has been done for the Pseudo-$C_\ell$ in App.~\ref{sec:pseudocl}.

\subsubsection{QML in single field ($T$ only)}\label{sec:qml_T}
Let's first remind that, given the data vector \textbf{x} (the observed map in real space), the covariance is
\begin{subequations}
\begin{eqnarray}\label{eq:qmlC}
\mathbf{C}&\equiv&\langle\mathbf{x} \mathbf{x}^T\rangle\\
&=&\mathbf{N}+\sum_\ell \langle|a_{\ell m}|^2\rangle Y_{\ell m} Y_{\ell m}^*\\
&=&\mathbf{N}+\sum_\ell \mathbf{P}^\ell_{ij} C_\ell
\end{eqnarray}
\end{subequations}
where $\mathbf{P}^\ell$ is the matrix of Legendre polynomial\footnote{The matrix is defined as
\begin{equation}
\textbf{P}^\ell_{ij}=\frac{2\ell+1}{4\pi}P_\ell(\cos\theta_{ij}=\hat{r}_i\cdot\hat{r}_j).
\end{equation},
where it is clear that $\mathbf{P}^\ell$ depends on the angle between two directions $\hat{r}_i$, $\hat{r}_j$ and the multipole $\ell$ through the normalisation factor.} 
at a given $\ell$, and we have used the addition theorem for the spherical harmonics. In Eq.~\ref{eq:qmlC}, we have explicitly taken into account the presence of experimental noise in the data which shows up in the noise covariance matrix \textbf{N}.

Another useful quantity is the Fisher matrix defined as the expected value of the curvature of the likelihood function around the maximum. If the likelihood is Gaussian
\begin{equation}
-2\ln\mathcal{L}\propto\ln\det\mathbf{C}+x^T\mathbf{C}^{-1}x
\end{equation}
the Fisher matrix takes the form
\begin{subequations}
\begin{eqnarray}
F_{\ell\ell'}&\equiv&\left\langle\frac{\partial^2\ln\mathcal{L}}{\partial C_\ell\partial C_{\ell'}}\right\rangle\\
&=&\frac{1}{2}\Tr\left[\mathbf{C}^{-1}\frac{\partial\mathbf{C}}{\partial C_\ell}\mathbf{C}^{-1}\frac{\partial\mathbf{C}}{\partial C_{\ell'}}\right]\\
&=&\frac{1}{2}\Tr\left[\mathbf{C}^{-1}\mathbf{P}^\ell\mathbf{C}^{-1}\mathbf{P}^{\ell'}\right]
\end{eqnarray}
\end{subequations}

We want to find an estimator that is quadratic in the data. In other words, we want to find a solution of the form
\begin{equation}\label{eq:qmlE}
\hat{C}_\ell=\mathbf{x}^T \mathbf{E}^\ell \mathbf{x}-b_\ell
\end{equation}
where $\mathbf{x}$ is the data vector, $\mathbf{E}^\ell$ is an appropriate symmetric matrix and $b_\ell$ is the noise bias. Such choice ensures that the estimator in Eq.~\ref{eq:qmlE} is unbiased
\begin{subequations}
\begin{eqnarray}
\langle\hat{C}_{\ell}\rangle&=&\langle\mathbf{x}^T \mathbf{E}^\ell \mathbf{x}\rangle-\langle b_\ell\rangle\\
&=&\Tr\left[\mathbf{N}\mathbf{E}^\ell\right]+\sum_{\ell'}\Tr\left[\mathbf{P}^{\ell'}\mathbf{E}^\ell\right]C_{\ell'}-\langle b_\ell\rangle
\end{eqnarray}
\end{subequations}
if we choose $b_\ell=\Tr\left[\mathbf{N}\mathbf{E}^\ell\right]$ and if $\Tr\left[\mathbf{P}^{\ell}\mathbf{E}^\ell\right]=1$. 

At the same time, such estimator should exhibit the minimum variance, i.e., we want to minimise the quantity
\begin{subequations}
\begin{eqnarray}\label{eq:qmlVar}
\mathbf{V}_{\ell\ell'}&=&\langle\hat{C}_\ell\hat{C}_{\ell'}\rangle-\langle\hat{C}_\ell\rangle\langle\hat{C}_{\ell'}\rangle\\
&=&2\Tr\left[\mathbf{C}\mathbf{E}^\ell\mathbf{C}\mathbf{E}^{\ell'}\right]
\end{eqnarray}
\end{subequations}
with the constraint coming from Eq.~\ref{eq:qmlE} that $\Tr\left[\mathbf{P}^{\ell}\mathbf{E}^\ell\right]=1$.
 
The constrained minimisation is obtained with the introduction of the Lagrange multiplier $\lambda$ such that the quantity
\begin{equation}
\mathrm{L}=\Tr\left[\mathbf{C}\mathbf{E}^\ell\mathbf{C}\mathbf{E}^{\ell'}-2\lambda (\mathbf{P}^{\ell}\mathbf{E}^\ell-1) \right]
\end{equation}
is minimised, i.e., its derivative with respect to $\mathbf{E}^\ell$ vanishes. It follows that \cite{Tegmark:1996qt}:
\begin{subequations}
\begin{eqnarray}
\frac{\partial \mathrm{L}}{\partial\mathbf{E}^\ell}=0&\rightarrow&\mathbf{E}^\ell=\lambda\mathbf{C}^{-1}\mathbf{P}^\ell\mathbf{C}^{-1}\\
\Tr\left[\mathbf{P}^{\ell}\mathbf{E}^\ell\right]=\Tr\left[\mathbf{P}^{\ell}\lambda\mathbf{C}^{-1}\mathbf{P}^\ell\mathbf{C}^{-1}\right]=1&\rightarrow&\lambda=\frac{1}{\Tr\left[\mathbf{P}^{\ell}\mathbf{C}^{-1}\mathbf{P}^\ell\mathbf{C}^{-1}\right]}=\frac{1}{2F_{\ell\ell}}
\end{eqnarray}
\end{subequations}
Eventually, the QML estimator is obtained as
\begin{equation}
\mathbf{E}^\ell=\frac{\mathbf{C}^{-1}\mathbf{P}^\ell\mathbf{C}^{-1}}{2F_{\ell\ell}}\rightarrow\hat{C}_\ell=\mathbf{x}^T\left(\frac{\mathbf{C}^{-1}\mathbf{P}^\ell\mathbf{C}^{-1}}{2F_{\ell\ell}}\right)\mathbf{x}.
\end{equation}

In the case of the pseudo-$C_\ell$ estimator, the observed map opportunely masked is directly converted into pseudo-power spectra, i.e., the core of the calculation is done in harmonic space (computation of the coupling kernel, inversion and computation of $\hat{C}_\ell$ out of $\tilde{C}_\ell$). In the case of the QML, the procedure lives in real space, i.e., relies on the properties of the signal $s(\hat{r})$ and noise $n(\hat{r})$ fluctuation maps, where $\hat{r}$ indicates the direction of the observed pixel in the sky.

As a final note, we would like to highlight what follows. Note that the QML is a maximum likelihood estimator when applied iteratively, i.e., it provides the $C_\ell$ that maximise the likelihood function in the limit that the fiducial model adopted to compute the covariance $\mathbf{C}$ approaches the true theoretical model~\cite{Bond:PRD571998} ($\mathbf{C}=\langle x x^T\rangle$). In such a case, it is straightforward to prove that 
\begin{subequations}
\begin{eqnarray}
2\left\langle\frac{\partial\ln\mathcal{L}}{\partial C_\ell}\right\rangle&=&\mathbf{C}^{-1}\frac{\partial\mathbf{C}}{\partial C_\ell}-\langle x x^T\rangle\mathbf{C}^{-1}\frac{\partial\mathbf{C}}{\partial C_\ell}\mathbf{C}^{-1}\\
&=&\mathbf{C}^{-1}\frac{\partial\mathbf{C}}{\partial C_\ell}-\mathbf{C}\mathbf{C}^{-1}\frac{\partial\mathbf{C}}{\partial C_\ell}\mathbf{C}^{-1}\\
&=&0.
\end{eqnarray}
\end{subequations}
The closer the fiducial model to the true spectrum, the more the $\hat{C}_\ell$ will resemble the maximum likelihood solution. Note that the iterative approach can be also adopted as an approximation to the exact likelihood~\cite{Bond:PRD571998}. If the fiducial model is a poor approximation to the true spectrum, the estimator can be proved to be still unbiased. The effect of adopting a (slightly) incorrect fiducial is a small increase in the variance of the estimator. In other words, the estimator is no longer the maximum likelihood solution, and its variance as given by Eq.~\ref{eq:qmlVar} does not equate to the Fisher matrix anymore.

\subsubsection{QML with correlated fields ($T,Q,U$)}\label{sec:qml_TQU}
When temperature and polarisation measurements are considered in a joint analysis, the QML formalism can be generalised as follows. The data vector is now a $3\times N_\mathrm{pix}$ vector
\begin{equation}
\mathbf{x}=\left(\begin{array}{c}
\mathbf{T}\\
\mathbf{Q}\\
\mathbf{U}
\end{array}\right)
\end{equation}
with covariance
\begin{equation}
\mathbf{C}\equiv\langle\mathbf{x}\mathbf{x}^T\rangle=\sum_i p_i\mathbf{P}_i
\end{equation}
where $p_i$ are the $N_b=6(\ell_\mathrm{max}-5)$ bandpowers of the 6 power spectra TT, EE, BB, TE, TB, EB \footnote{For example, $p_i$ with $i=1,...,\ell_\mathrm{max}-1$ is mapped onto $\ell(\ell+1)C_\ell^{TT}/(2\pi)$ with $\ell=2,...,\ell_\mathrm{max}$, $p_i$ with $i=\ell_\mathrm{max},...,2\ell_\mathrm{max}-2$ is mapped onto $\ell(\ell+1)C_\ell^{EE}/(2\pi)$ with $\ell=2,...,\ell_\mathrm{max}$, and so on.} plus 5 calibration parameters usually set to unity, and $\mathbf{P}_i$ are known matrices that depends only on geometrical factors and are independent from $p_i$.

We want to find a series of optimal estimators for the six power spectra that are quadratic in the data, such as
\begin{equation}
\hat{q}_i=\mathbf{x}^T\mathbf{Q}_i\mathbf{x}
\end{equation}
that is unbiased ($\langle\hat{q}_i\rangle=C_{\ell,i}$) and minimises the variance 
\begin{equation}\label{eq:qmlM}
\mathbf{M}_{ij}\equiv\langle\mathbf{q}\mathbf{q}^T\rangle-\langle\mathbf{q}\rangle\langle\mathbf{q}^T\rangle=2\Tr\left[\mathbf{Q}_i\mathbf{C}\mathbf{Q}_j\mathbf{C}\right].
\end{equation}

The request of an unbiased estimator leads to the following constraint to be satisfied while minimising the variance in Eq.~\ref{eq:qmlM}:
\begin{subequations}
\begin{eqnarray}
\langle q_i\rangle&=&\Tr\left[\mathbf{Q}_i\mathbf{C}\right]=\sum_{i'}\Tr\left[\mathbf{Q}_i\mathbf{P}_i\right]p_i\\
&=&\Tr\left[\mathbf{Q}_i\mathbf{N}\right]+\sum_{P'=1}^{6}\sum_{\ell=2}^{\ell_\mathrm{max}}W^{\ell P}_{\ell' P'}C_{\ell'}^{P'}\\
&\rightarrow&\sum_{i'}^{N_b}W^{\ell P}_{\ell' P'}=1
\end{eqnarray}
\end{subequations}
where $\Tr\left[\mathbf{Q}_i\mathbf{N}\right]$ is the noise contribution, $i=(\ell_\mathrm{max}-1)(P-1)+\ell-1$ is the index corresponding to the polarisation type $P=TT,EE,BB,TE,TB,EB$ and multipole $\ell$ (see footnote). The window functions $W^{\ell P}_{\ell' P'}=\Tr\left[\mathbf{Q}_i\mathbf{P}_i\right]$ should sum to unity for the estimator to be unbiased. They also make it explicit how the $i$-th estimator $q_i$ depends not only on the given polarisation type $P$ and multipole $\ell$, but also on the other polarisation types.

Imposing that the estimator is unbiased and minimum-variance, it can be shown that a good choice for the estimator is
\begin{equation}
\hat{q}_i=\mathbf{x}^T\mathbf{Q}_i\mathbf{x}=\mathbf{x}^T\left(\frac{1}{2}\mathcal{N}_i\sum_j\mathbf{B}_{ij}\mathbf{C}^{-1}\mathbf{P}_j\mathbf{C}^{-1}\right)\mathbf{x}
\end{equation}
with $\mathcal{N}_i$ appropriate normalisation factors such to satisfy the condition $\sum_{i'}^{N_b}W^{\ell P}_{\ell' P'}=1$, and $\mathbf{B}$ a suitable symmetric matrix. In the case of temperature-only QML, we have shown that the optimal estimator is obtained when $\mathbf{B}=\mathbf{F}^{-1}$, i.e., with the inverse of the Fisher matrix. In the multi-field regime, it can be shown that a better choice is provided by $\mathbf{B}=\mathbf{F}^{-1/2}$ (see e.g., discussion in Sec.II.C.3 of Ref.~\cite{Tegmark:2001zv}).

The key ingredient here is the full covariance matrix $\mathbf{C}$, that has to be built from a fiducial model, and has a more complicated form than in the case of temperature only. Given the data vector $\mathbf{x}_i$, the covariance is
\begin{equation}\label{eq:fullcov}
\langle\mathbf{x}_i\mathbf{x}_j^T\rangle=\mathbf{R}(\alpha_{ij})\mathbf{M}(\hat{r}_i\cdot\hat{r}_j)\mathbf{R}(-\alpha_{ij})
\end{equation}
where $\mathbf{R}(\alpha_{ij})$ is the rotation matrix around the $T$ ``direction'', i.e.,
\begin{equation}
\mathbf{R}(\alpha)=\left(\begin{array}{ccc}
1  &0  &0\\
0  &\cos2\alpha  &\sin2\alpha\\
0  &-\sin2\alpha  &\cos2\alpha
\end{array}\right)
\end{equation}
and $\mathbf{M}$ is the covariance that depends only on the angle between the two directions in the sky $\hat{r}_{i,j}$
\begin{equation}\label{eq:M}
\mathbf{M}(\hat{r}_i\cdot\hat{r}_j)=\left(\begin{array}{ccc}
\langle T_iT_j\rangle  &\langle T_iQ_j\rangle  &\langle T_iU_j\rangle\\
\langle T_iQ_j\rangle  &\langle Q_iQ_j\rangle  &\langle Q_iU_j\rangle\\
\langle T_iU_j\rangle  &\langle Q_iU_j\rangle  &\langle U_iU_j\rangle
\end{array}\right).
\end{equation}

The reason why the full covariance matrix is defined as in Eq.~\ref{eq:fullcov} and requires a rotation is given by the fact that $Q$ and $U$ are not global quantities. One can always define $\mathbf{M}$ as in Eq.~\ref{eq:M} in the reference frame of the great circle connecting the two point in $\hat{r}_{i,j}$. However, such a matrix should then be rotated into the global reference frame defined by the meridians.

The $(3\times3)$ entries in Eq.~\ref{eq:M} for any given pair of pixels $ij$ depend on the Legendre polynomial and the fiducial power spectra. As a straightforward example, the entry $\langle T_iT_j\rangle$ is the familiar expression
\begin{equation}
\langle T_iT_j\rangle=\sum_\ell\frac{2\ell+1}{4\pi}P_\ell(\hat{r}_i\cdot\hat{r}_j)C_\ell^{TT}
\end{equation}
already encountered in Sec.~\ref{sec:CMBstat_real} and in the derivation of the single-field QML estimator (Sec.~\ref{sec:qml_T}). A detailed description of the full procedure to obtain the covariance matrix, together with the expressions of the $(3\times3)$ entries, can be found in Appendix A of Ref.~\cite{Tegmark:2001zv}.


\end{document}